\documentclass{amsart}

\usepackage{psfig}
\usepackage{latexsym}
\usepackage{amssymb}
\numberwithin{equation}{section}

\theoremstyle{plain}
\newtheorem{thm}{Theorem}[section]
\newtheorem{prop}{Proposition}[section]
\newtheorem{lemma}{Lemma}[section]

\newtheorem{cor}{Corollary}[section]
\theoremstyle{definition}
\newtheorem{definition}{Definition}[section]
\newtheorem{convention}{Convention}[section]

\renewcommand{\Sp}{\ensuremath{\Sigma_{+}}}
\newcommand{\Sm}{\ensuremath{\Sigma_{-}}}
\newcommand{\No}{\ensuremath{N_{1}}}
\newcommand{\Nt}{\ensuremath{N_{2}}}
\newcommand{\Nth}{\ensuremath{N_{3}}}
\renewcommand{\O}{\ensuremath{\Omega}}
\renewcommand{\o}{\ensuremath{\omega}}
\newcommand{\g}{\ensuremath{\gamma}}
\renewcommand{\a}{\ensuremath{\alpha}}
\newcommand{\tx}{\ensuremath{\tilde{x}}}
\newcommand{\ty}{\ensuremath{\tilde{y}}}
\newcommand{\tbx}{\ensuremath{\tilde{\mathbf{x}}}}
\newcommand{\bx}{\ensuremath{\mathbf{x}}}
\renewcommand{\tt}{\ensuremath{\tilde{t}}}
\newcommand{\te}{\ensuremath{\tilde{\eta}}}
\newcommand{\txi}{\ensuremath{\tilde{\xi}}}

\begin{document}
\title{The Bianchi IX attractor}
\author{Hans Ringstr\"{o}m}
\address{Department of Mathematics\\ 
Royal Institute of Technology\\
S-100 44 Stockholm\\ 
Sweden}

\begin{abstract}
We consider the asymptotic behaviour of spatially homogeneous 
spacetimes of Bianchi type IX close to the singularity (we also 
consider some of the other Bianchi types, e. g.  Bianchi VIII in 
the stiff fluid case). The matter content is assumed to be an 
orthogonal perfect fluid with linear equation of state and zero 
cosmological constant. In terms of the variables of Wainwright and Hsu, 
we have the following results. In the stiff fluid case, the solution 
converges to a point for all the Bianchi class A types. For the other 
matter models we consider, the Bianchi IX solutions generically
converge to an attractor consisting of the closure of the vacuum type 
II orbits. Furthermore, we observe that for all the Bianchi class A
spacetimes, except those of vacuum Taub type, a curvature 
invariant is unbounded in the incomplete directions of inextendible 
causal geodesics.
\end{abstract}
\maketitle

\section{Introduction}

The last few decades, the Bianchi IX spacetimes have received
considerable attention, see for instance \cite{bogo}, \cite{hobill},
\cite{dyn} and references therein. Agreement has been reached,
at least concerning some aspects of the asymptotic behaviour as
one approaches a singularity, but the basis for the consensus
has mainly consisted of numerical studies and heuristic arguments.
The objective of this article is to provide mathematical proofs
for some aspects of the 'accepted' picture. The main result of this
paper was for example conjectured in \cite{dyn} p. 146-147, partly
on the basis of a numerical analysis. 

Why Bianchi IX? One reason is the fact that this class contains
the Taub-NUT spacetimes. These spacetimes are vacuum maximal
globally hyperbolic spacetimes that are causally geodesically
incomplete both to the future and to the past, see \cite{chris} and
\cite{law}. However, as one approaches
a singularity, in the sense of causal geodesic incompleteness, the
curvature remains bounded. In fact, one can extend the spacetime beyond
the singularities in inequivalent ways, see \cite{chris}. It is 
natural to conjecture that the behaviour exhibited
by the Taub-NUT spacetimes is non-generic, and
it is interesting to try to prove that the behaviour is
non-generic in the Bianchi IX class. In fact we prove that all
Bianchi IX initial data considered in this paper other than 
Taub-NUT yield inextendible globally hyperbolic developments
such that the curvature becomes unbounded as one approaches a 
singularity. This result is in fact more of an observation, since
the corresponding result is known in the vacuum case, see \cite{jag},
and curvature blow up is easy to prove in the non-vacuum cases
we consider.

Another reason for studying the 
Bianchi IX spacetimes is the BKL conjecture, see \cite{BKL}. 
According to this conjecture,  the 'local' approach to the singularity 
of a general solution should exhibit  oscillatory behaviour. The 
prototypes for this behaviour among the spatially homogeneous 
spacetimes are the Bianchi VIII and IX classes. Furthermore the matter
is conjectured to become unimportant as one approaches a singularity,
with some exceptions, for example the stiff fluid case. We refer to 
\cite{berger} for arguments supporting the BKL conjecture and to
\cite{Lars} for an overview of conjectures and results under symmetry 
assumptions of varying degree. In this paper we prove, under certain
restrictions on the allowed matter models, that generic Bianchi IX 
solutions exhibit oscillatory behaviour and that the matter becomes
unimportant as one approaches a singularity. What is meant by the
latter statement will be made precise below. If the matter model is 
a stiff fluid the matter will be important, and in that case we prove
that the behaviour is quiescent. This should be compared with
\cite{fuchs} concerning the structure of singularities of analytic 
solutions to Einstein's equations coupled to a scalar field or stiff fluid. 
In that paper, Andersson and Rendall prove that given a certain kind of 
solution to the so called velocity dominated system, there is a unique 
solution of Einstein's equations coupled to a stiff fluid approaching 
the velocity dominated solution asymptotically. One can then ask the 
question whether it is natural to assume that a solution has the 
asymptotics they prescribe. In Section \ref{section:avtds}, we show 
that all Bianchi VIII and IX stiff fluid solutions exhibit such 
asymptotic behaviour. 

The results presented in this paper can be divided into two parts.
The first part consists of statements about developments of 
orthogonal perfect fluid data of class A. We clarify below what we
mean by this. The results concern curvature blow up and
inextendibility of developments. The second part consists of results 
expressed in terms of the variables of Wainwright and Hsu. These 
variables describe the spacetime close to the singularity, and we 
prove that Bianchi IX solutions generically converge to a set on 
which the flow of the equation coincides with the Kasner map. 

We consider spatially homogeneous Lorentz manifolds
$(\bar{M},\bar{g})$ 
with a perfect fluid source. The stress energy
tensor is thus given by
\begin{equation}\label{eq:tdef}
T_{ab}=\mu u_{a}u_{b}+p(\bar{g}_{ab}+u_{a}u_{b}),
\end{equation}
where $u$ is a unit timelike
vectorfield, the 4-velocity of the fluid. We assume that
$p$ and $\mu$ satisfy a linear equation of state
\begin{equation}\label{eq:eqofst}
p=(\g-1)\mu,
\end{equation}
where we in this paper restrict our attention to $2/3<\g\leq 2$. 
We will also assume that $u$ is perpendicular to the hypersurfaces of
homogeneity. Einstein's equations can be written 
\begin{equation}\label{eq:einstein}
\bar{R}_{ab}-\frac{1}{2}\bar{R}\bar{g}_{ab}=T_{ab},
\end{equation}
where $\bar{R}_{ab}$ and $\bar{R}$ are the Ricci and scalar curvature
of $(\bar{M},\bar{g})$. 
In order to formulate an initial value problem in this
setting, consider a spacelike submanifold $(M,g)$ of
$(\bar{M},\bar{g})$, orthogonal to $u$. Let $e_{\a}$, $\a=0,..,3$ be a
local frame with $e_{0}=u$ and $e_{i}$, $i=1,2,3$ tangent to $M$ 
and let $k_{ij}$
be the second fundamental form of $(M,g)$. Then $g$ and $k$ must
satisfy the equations
\[
R_{g}-k_{ij}k^{ij}+(\mathrm{tr}_{g} k)^2=2\bar{R}_{00}+\bar{R}
\]
and
\[
\nabla_{i}\mathrm{tr}_{g} k-\nabla^{j}k_{ij}=\bar{R}_{0i},
\]
where $\nabla$ is the Levi-Civita connection of $g$, and $R_{g}$ is
the corresponding scalar curvature, indices are raised and lowered
by $g$.
If we specify a Riemannian metric $g$, and a symmetric covariant
2-tensor $k$, as initial data on a 3-manifold, they should thus 
in our situation satisfy
\begin{equation}\label{eq:con1}
R_{g}-k_{ij}k^{ij}+(\mathrm{tr}_{g} k)^2=2\mu
\end{equation}
and
\begin{equation}\label{eq:con2}
\nabla_{i}\mathrm{tr}_{g} k-\nabla^{j}k_{ij}=0,
\end{equation}
because of (\ref{eq:einstein}), (\ref{eq:tdef}) and the fact that 
$u$ is perpendicular to $M$. In other words, we should also specify
the initial value of $\mu$ as part of the data.

We consider only a restricted class of manifolds $M$ and initial data.
The 3-manifold $M$ is assumed to be a special type of Lie group, and
$g,\ k$ and $\mu$ are assumed to be left invariant. In order to be
more precise concerning the type of Lie groups $M=G$ we consider, let
$e_{i}$, $i=1,2,3$ be a basis of the Lie algebra with structure
constants determined by  $[e_{i},e_{j}]=\gamma_{ij}^{k}e_{k}$. If
$\gamma_{ik}^{k}=0$, then the Lie algebra and Lie group are said to 
be of class A, and  
\begin{equation}\label{eq:sconstants}
\gamma_{ij}^{k}=\epsilon_{ijm}n^{km}
\end{equation}
where the symmetric matrix $n^{ij}$ is given by
\begin{equation}\label{eq:ndef}
n^{ij}=\frac{1}{2}\gamma^{(i}_{kl}\epsilon_{}^{j)kl}.
\end{equation}

\begin{definition}\label{def:data}
\textit{Orthogonal perfect fluid data of class A} for 
Einstein's equations 
consist of the following. A  Lie group $G$ of class A, a left
invariant Riemannian metric $g$ on $G$, a left invariant symmetric 
covariant 2-tensor $k$ on $G$, and a constant $\mu_{0}\geq 0$ 
satisfying (\ref{eq:con1}) and (\ref{eq:con2}) with $\mu$ replaced 
by $\mu_{0}$.
\end{definition}

We can choose a left invariant orthonormal basis $\{ e_{i}\}$ with 
respect to $g$, so that the corresponding matrix $n^{ij}$ defined in 
(\ref{eq:ndef}) is diagonal with diagonal elements $n_{1}$, $n_{2}$ 
and $n_{3}$. By an appropriate choice of orthonormal basis, 
$n_{1}, n_{2}, n_{3}$ can be assumed to belong to one and only
one of the types given in Table \ref{table:bianchiA}. We 
assign a Bianchi type to the initial data accordingly. 
This division constitutes a classification of the class A Lie
algebras. We refer to Lemma \ref{lemma:liealg} for a  proof of 
these statements.

Let $k_{ij}=k(e_{i},e_{j})$. Then the matrices $n^{ij}$
and $k_{ij}$ commute according to (\ref{eq:con2}), so that we may
assume $k_{ij}$ to be diagonal with diagonal elements
$k_{1}$, $k_{2}$ and $k_{3}$, cf. (\ref{eq:transform}). 
\begin{definition}\label{def:generic}
Orthogonal perfect fluid data of class A 
satisfying $k_{2}=k_{3}$ and
$n_{2}=n_{3}$ or one of the permuted conditions are said to be
of \textit{Taub type}. Data with $\mu_{0}=0$ are called \textit{vacuum
data}.
\end{definition}
Observe that the Taub condition is 
independent of the choice of orthonormal basis diagonalizing 
$n$ and $k$, cf. (\ref{eq:transform}). Considering the equations
of Ellis and MacCallum (\ref{eq:dndt})-(\ref{eq:constraint1}),
one can see that if $n_{2}=n_{3}$ and $k_{2}=k_{3}$ at one point 
in time, then the equalities always hold, cf. the construction of the
spacetime carried out in the appendix. According to \cite{emac},
vacuum solutions satisfying these conditions are the Taub-NUT 
solutions. This justifies the following definition.

\begin{definition}
\textit{Taub-NUT} initial data are type IX Taub vacuum initial data.
\end{definition}

\begin{table}\label{table:bianchiA}
\caption{Bianchi class A.}
\begin{tabular}{@{}lccc}
Type & $n_{1}$ & $n_{2}$ & $n_{3}$ \\
I                   & 0 & 0 & 0 \\
II                  & + & 0 & 0 \\
V$\mathrm{I}_{0}$   & 0 & + & $-$ \\
VI$\mathrm{I}_{0}$  & 0 & + & + \\
VIII                & $-$ & + & + \\
IX                  & + & + & + \\
\end{tabular}
\end{table}

\begin{definition}\label{def:development}
By an \textit{orthogonal perfect fluid development} of orthogonal 
perfect fluid data of class A, we will mean the following. A
connected 4-dimensional Lorentz manifold $(\bar{M},\bar{g})$ and a
2-tensor $T$, as in (\ref{eq:tdef}), on $(\bar{M},\bar{g})$,
such that there is an embedding $i:G\rightarrow \bar{M}$ with
$i^{*}(\bar{g})=g$, $i^{*}(\bar{k})=k$ and $i^{*}(\mu)=\mu_{0}$,
where $\bar{k}$ is the second fundamental form of $i(G)$ in
$(\bar{M},\bar{g})$.
\end{definition}

In the appendix, we construct globally hyperbolic orthogonal
perfect fluid developments, given initial data, and we refer to
them as class A developments, cf. Definition \ref{def:utveckling}.
We also assign a type to such a development according to the
type of the initial data. Let us make a division of the initial
data according to their global behaviour.

\begin{thm}\label{thm:classification}
Consider a class A development with $1\leq \g\leq 2$.
\begin{enumerate}
\item If the initial data are not of type IX, but satisfy 
  $\mathrm{tr}_{g}k=0$, then $\mu_{0}=0$ and the 
  development is causally geodesically complete. Only types 
  I and VI$I_{0}$ permit this possibility.
\item If the initial data are of type I, II,  V$I_{0}$, VI$I_{0}$ 
  or VIII, and 
  satisfy $\mathrm{tr}_{g}k<0$, then the development is future
  causally geodesically complete and past causally
  geodesically incomplete. Such initial data we will refer to
  as expanding.
\item Bianchi IX initial data yield
  developments that are past and future causally geodesically
  incomplete. Such data are called recollapsing.
\end{enumerate}
\end{thm}

A proof is to be found in the appendix, but observe that this
theorem is not new. As far as class A developments are concerned, 
we will restrict our attention to equations of state with 
$1\leq \g\leq 2$. The reason is that there is cause to doubt the
well posedness of the initial value problem for $2/3<\g<1$, 
cf. \cite{friedrich} p. 85 and p. 88.
Furthermore, in the Bianchi IX case we use results from \cite{law}
concerning recollapse, see Lemma \ref{lemma:recollapse}. In order
to be allowed to do that, we need the above mentioned condition on $\g$.
What is meant by inextendibility is explained in the following.
\begin{definition}
Consider a connected Lorentz manifold $(M,g)$.
If there is a connected $C^{2}$ Lorentz manifold $(\hat{M},\hat{g})$
of the same dimension, and a map $i:M\rightarrow \hat{M}$, with
$i(M)\neq \hat{M}$, which is an isometry onto its image, then $(M,g)$ 
is said to be $C^{2}$-\textit{extendible} and $(\hat{M},\hat{g})$ is
called a $C^{2}$-\textit{extension} of $(M,g)$. A Lorentz manifold
which is not $C^{2}$-extendible is said to be
$C^{2}$-\textit{inextendible}. 
\end{definition}
\textit{Remark}. There is an analogous definition of smooth
extensions. Unless otherwise mentioned, manifolds are assumed to 
be smooth, and maps between manifolds are assumed to be as regular as
possible.

We will use the \textit{Kretschmann scalar}, 
\begin{equation}\label{eq:k1}
\kappa=\bar{R}_{\alpha\beta\gamma\delta}
\bar{R}^{\alpha\beta\gamma\delta},
\end{equation}
as our main measure of whether curvature blows up or not, but in
the non-vacuum case it is natural to consider the Ricci tensor 
contracted with itself $\bar{R}_{\a\beta}\bar{R}^{\a\beta}$. 
The next theorem states the main conclusion concerning
developments.
\begin{thm}\label{thm:main}
For class A developments with $1\leq \g\leq 2$, we have the following 
division.
\begin{enumerate}
\item Consider expanding initial data of type I, II or VI$I_{0}$ with 
  $1\leq \g <2$ which are not of Taub vacuum type. Then the
  Kretschmann scalar is unbounded along all
  inextendible causal geodesics in the incomplete direction. 
\item Consider non-Taub-NUT recollapsing initial data with $1\leq \g < 2$.
  Then the Kretschmann scalar is unbounded along all inextendible
  causal geodesics in both incomplete directions. 
\item Expanding and recollapsing data with $\g=2$ and $\mu_{0}>0$. 
  Then the Kretschmann scalar is unbounded along all inextendible
  causal geodesics in all incomplete directions.
\item Expanding and recollapsing data with $\mu_{0}>0$. Then
  $\bar{R}_{\a\beta}\bar{R}^{\a\beta}$ is unbounded along all
  inextendible causal geodesics in all incomplete directions.
\end{enumerate}
In all cases mentioned above the class A
development is $C^{2}$-inextendible.
\end{thm}
\textit{Remark}. Observe that the Bianchi VIII vacuum case was handled
in \cite{jag}, and the Bianchi V$\mathrm{I}_{0}$ vacuum case in
\cite{rendall}. The above theorem thus isolates the vacuum Taub type
solutions as the only ones among the Bianchi class A spacetimes
that do not exhibit curvature blow up, given our particular matter
model.

We now turn to the results that are expressed in terms of the
variables of Wainwright and Hsu. The equations and some of
their properties are to be found in Section \ref{section:whsu}.
The appendix contains a derivation. It is natural to divide the 
matter models into two categories; the non-stiff fluid case and 
the stiff fluid case ($\g=2$). 

\begin{figure}[hbt]
  \centerline{\psfig{figure=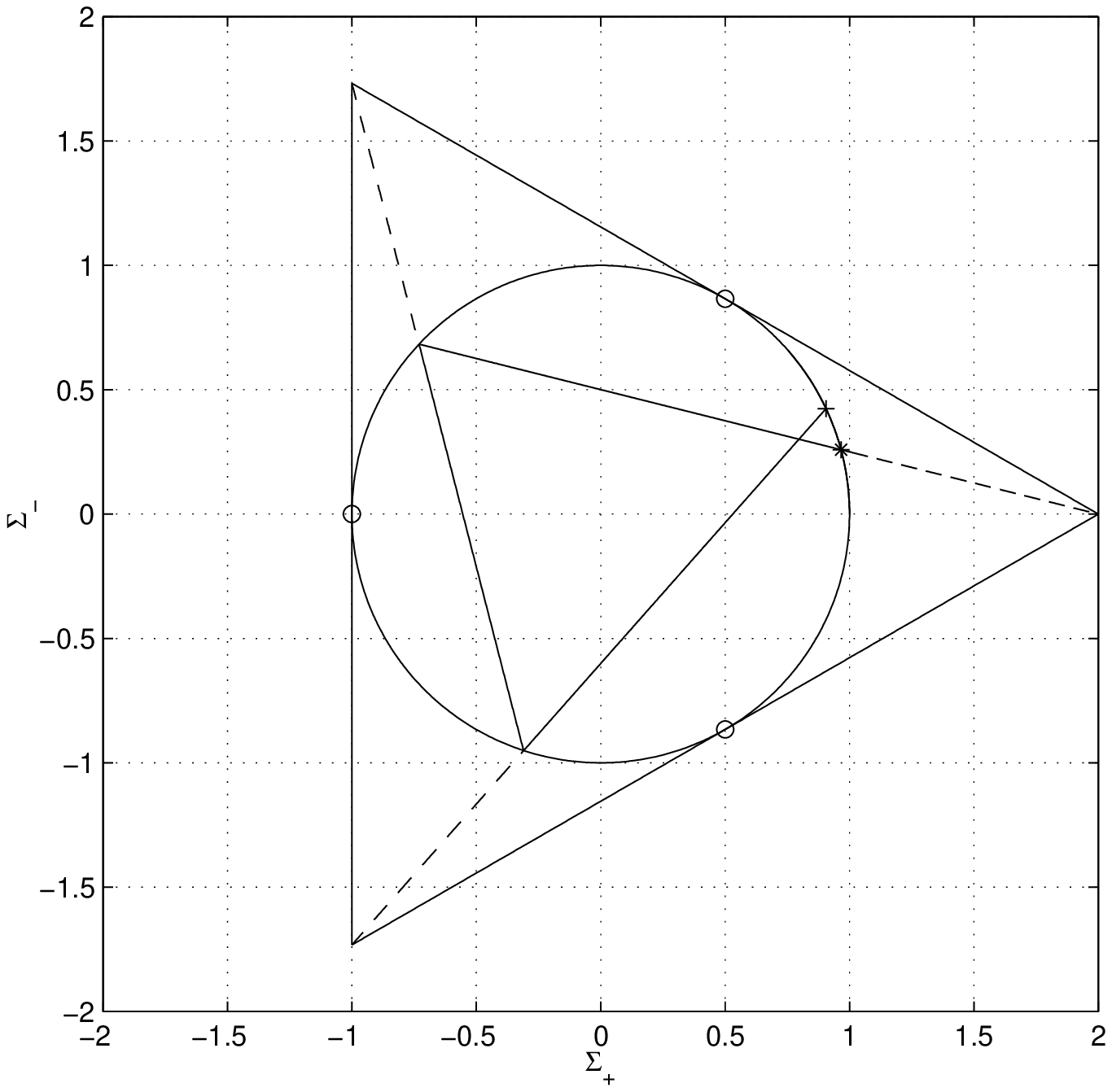,height=7cm}}
  \caption{The Kasner map.}\label{fig:bkl}
\end{figure}

Let us begin with the non-stiff fluid case, including the vacuum case. 
We confine our attention to Bianchi IX solutions. The existence
interval stretches back to $-\infty$ which corresponds to the
singularity. There are some fixed points to which certain solutions 
converge, and data which lead to such solutions  together 
with data of Taub type will be considered to be non-generic. The Kasner
map, which is supposed to be an approximation of the Bianchi IX
dynamics as one approaches a singularity, is illustrated in Figure 
\ref{fig:bkl}. The circle in the $\Sp\Sm$-plane appearing in the
figure is called the Kasner circle, and we have depicted two bounces 
of the Kasner map. The starting point is marked by a star, and the end 
point by a plus sign. Given a point $x$ on the Kasner circle, the Kasner
map yields a new point $y$ on the Kasner circle by taking the corner 
of the triangle closest to $x$, drawing a straight line from the 
corner through $x$, and then letting $y$ be the second point of 
intersection between the line and the Kasner circle. One solid line  
corresponds to the closure of a vacuum type II orbit of the equations 
of Wainwright and Hsu. Actually, it is the projection of the closure of
such an orbit to the $\Sp\Sm$-plane. A vacuum type II solution has 
one $N_{i}$  non-zero and the other zero, and the three different 
$N_{i}$ correspond to the three corners of the triangle; the 
rightmost corner corresponds to $\No\neq 0$ and the corner on the top 
left corresponds to $\Nth\neq0$. The constraint
(\ref{eq:constraint}) for the vacuum type II solutions is given by
\[
\Sp^2+\Sm^2+\frac{3}{4}N_{i}^2=1.
\]
The closure of this set is given a name in the following definition.
\begin{definition}\label{def:attractor}
The set 
\[ 
\mathcal{A}=\{(\O,\Sp,\Sm,\No,\Nt,\Nth):
\O+|\No\Nt|+|\Nt\Nth|+|\Nth\No|=0\}\cap M,
\]
where $M$ is defined by (\ref{eq:constraint}), is called 
\textit{the Bianchi attractor}. 
\end{definition}
The main result of this paper is that for generic Bianchi
IX data, the solution converges to the attractor. That is
\begin{equation}\label{eq:main}
\lim_{\tau\rightarrow -\infty}(\O+\No\Nt+\Nt\Nth+\Nth\No)=0.
\end{equation}
This conclusion supports the statement that the Kasner map approximates
the dynamics, and also the statement that the matter content 
loses significance close to the singularity. Let us introduce some
terminology. 

\begin{definition}\label{def:als}
Let $f\in C^{\infty}(\mathbb{R}^{n},\mathbb{R}^{n})$, and consider 
a solution $x$ to the equation
\[
\frac{dx}{dt}=f\circ x,\ x(0)=x_{0},
\]
with maximal existence interval $(t_{-},t_{+})$. 
We call a point $x_{*}$ an $\alpha$-\textit{limit point} of the 
solution $x$, if there is a sequence $t_{k}\rightarrow
t_{-}$ with $x(t_{k})\rightarrow x_{*}$. The $\alpha$-\textit{limit
set} of $x$ is the set of its $\alpha$-limit points. The
$\o$-limit set is defined similarly by replacing $t_{-}$ with
$t_{+}$.
\end{definition}
\textit{Remark}. If $t_{-}>-\infty$ then the $\a$-limit set is
empty, cf. \cite{jag}.

Thus, the $\a$-limit set of a generic solution is contained in the 
attractor. The desired statement is that the $\a$-limit set coincides
with the attractor, but the best result we have achieved in this direction
is that there must at least be three $\a$-limit points on the Kasner
circle. This worst case situation corresponds to the solution
converging to a periodic orbit of the Kasner map with period three.
Observe that we have not proven anything concerning Bianchi VIII
solutions.

Let us sketch the proof. It is natural to divide it into two parts.
The first part consists of proving the existence
of an $\a$-limit point on the Kasner circle. We achieve this in the
following steps. First we analyze the $\a$-limit sets of the 
Bianchi types I, II and VI$\mathrm{I}_{0}$. An analysis of types 
I of II can also be found in Ellis and Wainwright \cite{dyn}. Then 
we prove the existence of an $\a$-limit point for a generic
Bianchi IX solution. To go from the existence
of an $\a$-limit point to an $\a$-limit point on the 
Kasner circle, we use the analysis of the lower Bianchi types. 
In the second part, we prove (\ref{eq:main}). Let $d$ be the
function appearing in that equation. We assume that $d$ does not
converge to zero in order to reach a contradiction. The existence 
of an $\a$-limit point on the Kasner circle proves that there is
a sequence $\tau_{k}\rightarrow -\infty$ such that $d(\tau_{k})
\rightarrow 0$. If $d$ does not converge to zero there is a $\delta>0$,
and a sequence $s_{k}\rightarrow -\infty$ such that 
$d(s_{k})\geq \delta$. We can assume $s_{k}\leq \tau_{k}$ and conclude
that $d$ on the whole has to grow (going backwards) in the 
interval $[s_{k},\tau_{k}]$.
What can be said about this growth? In Section \ref{section:ocon},
we prove that we can control the density parameter $\O$ in this 
process, assuming $\delta$ is small enough, which is not a restriction.
As a consequence $\O$ can be assumed to be arbitrarily small during 
the growth. Some further arguments, given in Section
\ref{section:attractor}, show that we can assume the growth to occur
in the product $\Nt\Nth$, using the symmetries of the equations. 
Furthermore, one can assume the $\Sp\Sm$-variables to 
be arbitrarily close to $(\Sp,\Sm)=(-1,0)$,
and that some expressions dominate others. For instance 
$1+\Sp$ can be assumed to be arbitrarily much smaller than $\Nt\Nth$.
This control introduces a natural concept of order of magnitude.
The behaviour of the product $\Nt\Nth$ will be oscillatory; it 
will look roughly like a sine wave. The point is to prove that 
the product decays during a period of its oscillation; that would
lead to a contradiction. The variation during a period can be
expressed in terms of an integral, and we use the order of magnitude 
concept to prove an estimate showing that this integral has the right
sign. 

\begin{figure}[hbt]
  \centerline{\psfig{figure=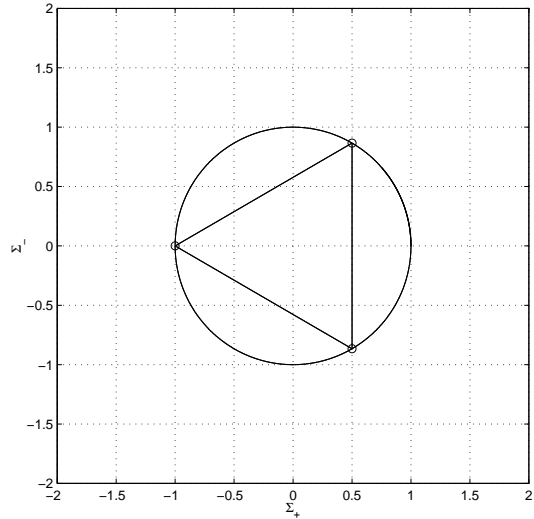,height=7cm}}
  \caption{The triangle mentioned in the text.}\label{fig:triangle}
\end{figure}

Now consider the stiff fluid case with positive density parameter. 
In this case we 
will consider Bianchi VIII and IX solutions. The analysis is similar 
for the other cases and a description of the results is to be found 
in Section \ref{section:conclusions}. Again the singularity corresponds
to $-\infty$. The density parameter $\O$ converges to a non-zero
value, all the $N_{i}$ converge to zero, and in the $\Sp\Sm$-plane
the solution converges to a point inside the triangle shown in
Figure \ref{fig:triangle}.

In Section \ref{section:whsu}, we formulate the equations of Wainwright
and Hsu and briefly describe their origin and some of their
properties. Section \ref{section:eprop} contains some elementary
properties of solutions. We give the existence intervals of solutions
to the equations, and prove that the $\O\Sp\Sm$-variables are contained
in a compact set to the past for Bianchi IX solutions. As in the 
vacuum case, we also prove that $(\Sp,\Sm)$ can converge to $(-1,0)$ 
only if the solution is of Taub type, although this is no longer a 
characterization. In Section \ref{section:critp}, we mention some 
critical points and 
make more precise the statement that solutions converging to these
points are non-generic. Included in this section are also two
technical lemmas relevant to the analysis. The monotonicity principle
is explained in Section \ref{section:monotone}. It is fundamental
to the analysis of the $\a$-limit sets of the solutions. We present
two applications; the fact that all $\a$-limit points of Bianchi
IX solutions are of type I, II or VI$\mathrm{I}_{0}$ and an analysis
of the vacuum type II orbits. The last application is not complicated,
but illustrates the arguments involved as well as demonstrating how 
the map depicted in Figure \ref{fig:bkl} can be viewed as a sequence 
of type II orbits. Section \ref{section:shear} deals with situations
such that one has control over the shear variables and the density
parameter. Specifically, it gives a geometric interpretation of some 
of the equations in $\O\Sp\Sm$-space. As an application, we prove that
if a Bianchi IX solution has an $\a$-limit point on the Kasner circle
then all the points obtained by applying the Kasner map to this point
belong to the $\a$-limit set of the solution. The stiff fluid case 
is handled in Section \ref{section:stiff}. In this case the
$\a$-limit set consists of a point regardless of type.
Sections \ref{section:typeI}-\ref{section:typeVII0} deal with the 
lower order Bianchi types needed in order
to analyze Bianchi IX. An analysis of types I of II can also be found
in Ellis and Wainwright \cite{dyn}.
Section \ref{section:taub} gives the possibilities for a Taub type 
Bianchi IX solution. The technical Section \ref{section:osc} is 
needed in order to prove the existence of
an $\alpha$-limit point for Bianchi IX solutions, and also to prove 
that the set of vacuum type II points is an attractor. It is used for 
approximating the  solution in situations where the behaviour
is oscillatory. Section
\ref{section:typeIX} proves the existence of an $\alpha$-limit point
for a Bianchi IX solution and the existence of an $\alpha$-limit 
point on the Kasner circle for generic Bianchi IX solutions.
 In Section \ref{section:ocon}, we prove that if
one has control over the sum $|\No\Nt|+|\Nt\Nth|+|\Nth\No|$ in some
time interval $[\tau_{1},\tau_{2}]$, and control over $\O$ in
$\tau_{2}$ then one has control over $\O$ in the entire interval. 
This rather technical observation is essential in the proof that
generic solutions converge to the attractor.  The heart of this
paper is Section \ref{section:attractor} which contains a proof 
of (\ref{eq:main}). It also contains arguments that will be used
in Section \ref{section:regular} to analyze the regularity of
the set of non-generic points.
In Section \ref{section:uniform}, we observe that
the convergence to the attractor is uniform, and in Section
\ref{section:nsale} we prove the existence of at least three
non-special $\a$-limit points on the Kasner circle. We formulate
the main conclusions and prove Theorem \ref{thm:main} in Section
\ref{section:conclusions}. In Section \ref{section:avtds}, we relate
our results concerning stiff fluid solutions to those of 
\cite{fuchs}. The appendices contain results relating
solutions to the equations of Wainwright and Hsu with properties
of the class A developments and some curvature computations.

\section{Equations of Wainwright and Hsu}
\label{section:whsu}
The essence of this paper is an analysis of the asymptotic 
behaviour of solutions to the equations of Wainwright and
Hsu (\ref{eq:whsu})-(\ref{eq:constraint}). One important
property of these equations is that they describe all the 
Bianchi class A types at the same time. Another important
property is that it seems that the variables remain in 
a compact set as one approaches a singularity. In the Bianchi
IX case, this follows from the analysis presented in this paper. Let 
us give a rough description of the origin of the variables. In the 
situations we consider, there is a foliation of the Lorentz manifold
by homogeneous spacelike hypersurfaces diffeomorphic to a Lie group 
$G$ of class A. One can define an orthonormal basis $e_{\a}$, 
$\a=0,...,3$, such that $e_{i}$, $i=1,2,3$, span the tangent space
of the spacelike hypersurfaces of homogeneity, and $e_{0}=\partial_{t}$
for a suitable globally defined time coordinate $t$. It is possible
to  associate 
a matrix $n_{ij}$ with the spacelike vectors $e_{i}$, as in 
(\ref{eq:ndef}), and assume it to be diagonal with diagonal 
components $n_{i}$. One changes the time coordinate by
$dt/d\tau=3/\theta$, where $\theta$ is minus the trace of the second
fundamental form of the spacelike hypersurface corresponding to $t$.
The $N_{i}(\tau)$ below are the $n_{i}(\tau)$ divided by
$\theta(\tau)$, the $\Sp$ and $\Sm$ correspond to the traceless
part of the second fundamental form of the spacelike hypersurface
corresponding to $\tau$, similarly normalized, and finally
$\O=3\mu/\theta^2$.  We will refer to 
$\Sp$ and $\Sm$ as the \textit{shear variables}, and to $\O$ as 
the \textit{density parameter}. 
The question then arises to what extent this makes sense, since
$\theta$ could become zero. An answer is given in the appendix. 
For all the Bianchi types except IX, this procedure is essentially
harmless, and the variables of Wainwright and Hsu capture the entire
Lorentz manifold. In the Bianchi IX case, there is however a point
at which $\theta=0$, at least if $1\leq \g\leq 2$, see the appendix,
and the variables are only valid for half a 
development in that case. As far as the analysis of the asymptotics
are concerned, this is however not important. A derivation of the 
equations is given in the appendix. They are
\begin{eqnarray}
\No'&=&(q-4\Sp)\No  \nonumber \\
\Nt'&= &(q+2\Sp +2\sqrt{3}\Sm)\Nt  \nonumber \\
\Nth'&=& (q+2\Sp -2\sqrt{3}\Sm)\Nth  \label{eq:whsu}\\
\Sp'&=& -(2-q)\Sp-3S_{+}  \nonumber \\
\Sm'&=& -(2-q)\Sm-3S_{-} \nonumber \\
\O'&=& [2q-(3\g-2)]\O. \nonumber
\end{eqnarray}
The prime denotes derivative with respect to a time coordinate 
$\tau$, and
\begin{eqnarray}
q & = & \frac{1}{2}(3\g-2)\O+2(\Sp^2+\Sm^2) \nonumber \\
S_{+} & = & \frac{1}{2}[(\Nt-\Nth)^2-\No(2\No-\Nt-\Nth)]
\label{eq:whsudef}\\
S_{-} & = & \frac{\sqrt{3}}{2}(\Nth-\Nt)(\No-\Nt-\Nth). \nonumber
\end{eqnarray}
The constraint is
\begin{equation}
\O+\Sp^2+\Sm^2+\frac{3}{4}[\No^2+\Nt^2+\Nth^2-2(\No\Nt+\Nt\Nth+
\Nth\No)]=1.
\label{eq:constraint}
\end{equation}
We demand that $2/3<\g\leq 2$ and $\O\geq 0$. 
The equations (\ref{eq:whsu})-(\ref{eq:constraint}) have certain
symmetries, described in Wainwright and Hsu \cite{whsu}. By permuting
$\No,\Nt,\Nth$ arbitrarily, we get new solutions, if we at the same 
time carry out appropriate combinations of rotations by integer 
multiples of $2\pi/3$, and reflections in the $(\Sp,\Sm)$-plane. 
Explicitly, the transformations
\[
(\tilde{N}_{1},\tilde{N}_{2},\tilde{N}_{3})=(\Nth,\No,\Nt),\
(\tilde{\Sigma}_{+},\tilde{\Sigma}_{-})=(-\frac{1}{2}\Sp+
\frac{1}{2}\sqrt{3}\Sm,-\frac{1}{2}\sqrt{3}\Sp-\frac{1}{2}\Sm)
\]
and
\[
(\tilde{N}_{1},\tilde{N}_{2},\tilde{N}_{3})=(\No,\Nth,\Nt),\
(\tilde{\Sigma}_{+},\tilde{\Sigma}_{-})=(\Sp,-\Sm)
\]
yield new solutions. Below, we refer to rotations
by integer multiples of $2\pi/3$ as rotations.
Changing the sign of all the $N_{i}$ at the same time does not change
the equations. Classify points $(\O,\Sp,\Sm,\No,\Nt,\Nth)$ according 
to the values of $\No,\Nt,\Nth$ in the same way as in Table
\ref{table:bianchiA}. Since the sets 
$N_{i}>0$, $N_{i}<0$ and $N_{i}=0$ are invariant under the flow of 
the equations, we may classify solutions to 
(\ref{eq:whsu})-(\ref{eq:constraint}) accordingly.

\begin{definition}
The \textit{Kasner circle} is defined by the conditions $N_{i}=\O=0$ 
and the constraint (\ref{eq:constraint}). There are three points on 
this circle
called \textit{special}: $(\Sp,\Sm)=(-1,0)$ and $(1/2,\pm\sqrt{3}/2)$.
\end{definition}

The following reformulation of $\Sp'$ is written down for future 
reference,
\begin{equation}\label{eq:sppr0}
\Sp'=-(2-2\O-2\Sp^2-2\Sm^2)(\Sp+1)-\frac{3}{2}(2-\g)\O\Sp+\frac{9}{2}
\No(\No-\Nt-\Nth).
\end{equation}

\section{Elementary properties of solutions}\label{section:eprop}
Here we collect some miscellaneous observations that will be of
importance. Most of them are similar to results obtained in
\cite{jag}. The $\a$-limit set defined in Definition \ref{def:als}
plays an important role in this paper, and here we mention some 
of its properties. 

\begin{lemma}\label{lemma:als}
Let $f$ and $x$ be as in Definition \ref{def:als}.
The $\a$-limit set of $x$ is closed and invariant under the flow of
$f$. If there is a $T$ such that $x(t)$ is contained in a compact 
set for $t\leq T$, then the $\a$-limit set of $x$ is connected.
\end{lemma}

\textit{Proof}. See e. g. \cite{irwin}. $\Box$

\begin{definition}
A solution to (\ref{eq:whsu})-(\ref{eq:constraint}) satisfying
$\Nt=\Nth$ and $\Sm=0$, or one of the conditions found by applying the
symmetries, is said to be of \textit{Taub type}.
\end{definition}
\textit{Remark}. The set defined by $\Nt=\Nth$ and $\Sm=0$ is
invariant under the flow of (\ref{eq:whsu}). 

\begin{lemma}\label{lemma:existence}
The existence intervals for all solutions to 
(\ref{eq:whsu})-(\ref{eq:constraint}) except Bianchi IX are 
$(-\infty,\infty)$. For Bianchi IX solutions we have past global
existence.
\end{lemma}
\textit{Proof}. As in the vacuum case, see \cite{jag}. $\Box$

By observations made in the appendix, $-\infty$ corresponds to the
singularity.

\begin{lemma}\label{lemma:B9N}
Let $2/3<\g\leq 2$.
Consider a solution of type IX. The image $(\Sp,\Sm,\O)((-\infty,0])$ 
is contained in a compact set whose size depends on the initial 
data. Further, if at a point in time $\Nth\geq\Nt\geq\No$ and 
$\Nth\geq 2$, then $\Nt\geq\Nth/10$.
\end{lemma}
\textit{Proof}. As in the vacuum case, see \cite{jag}. $\Box$

That $(\Sp,\Sm,\O)$ is contained in a compact set for all the other
types follows from the constraint. The second part of this
lemma will be important in the proof of the existence of an
$\a$-limit point. One consequence is that one $N_{i}$ may not
become unbounded alone. 

The final observation is relevant in proving curvature blow up.
One can define a normalized version (\ref{eq:nk}) of the Kretschmann
scalar (\ref{eq:k1}), and it can be expressed as a polynomial in the
variables of Wainwright and Hsu. One way of proving that 
a specific solution exhibits curvature blow up is to prove that it
has an $\a$-limit point at which the normalized Kretschmann scalar
is non-zero. We refer to the appendix for the details. It turns out 
that this polynomial is zero when $\Nt=\Nth$, $\No=0$, $\Sm=0$,
$\Sp=-1$ and $\O=0$. The same is true of the points obtained by
applying the symmetries. It is then natural to ask the question: 
for which solutions does $(\Sp,\Sm)$ converge to $(-1,0)$?

\begin{prop}\label{prop:limchar}
A solution to (\ref{eq:whsu})-(\ref{eq:constraint}) with
$2/3<\g<2$ satisfies
\[
\lim_{\tau\rightarrow -\infty}(\Sp (\tau),\Sm (\tau))=(-1,0),
\]
only if it is contained in the invariant set
$\Sm=0$ and $\Nt=\Nth$.
\end{prop}
\textit{Remark}. The proposition does not apply 
to the stiff fluid case. The analogous statements for the points
$(\Sp,\Sm)=(1/2,\pm \sqrt{3}/2)$ are true by an application of the 
symmetries. We may not replace the implication with an
equivalence, cf. Proposition \ref{prop:typeII}.

\textit{Proof}. The argument is essentially the same as in the 
vacuum case, see \cite{jag}. We only need to observe that $\O$ will decay
exponentially when $(\Sp,\Sm)$ is close to $(-1,0)$. $\Box$

\section{Critical points}\label{section:critp}

\begin{definition}
The critical point $F$ 
is defined by $\O=1$ and all other variables zero. In the case
$2/3<\g<2$, we define the critical point
$P_{1}^{+}(II)$ to be the type II point with $\Sm=0$, $\No>0$, 
$\Sp=(3\g-2)/8$ and $\O=1-(3\g-2)/16$. The critical points 
$P_{i}^{+}(II)$, $i=2,3$ are found by applying the symmetries.
\end{definition}

It will turn out that there are solutions which converge to these
points as $\tau\rightarrow -\infty$. The main objective of this section
is to prove that the set of such solutions is small. Observe that 
only non-vacuum solutions can converge these critical points.

\begin{definition}
Let $\mathcal{I}_{\mathrm{VII}_{0}}$ denote initial data to
(\ref{eq:whsu})-(\ref{eq:constraint}) of type
VI$\mathrm{I}_{0}$ with $\O>0$, and correspondingly for the other
types. Let $\mathcal{P}_{\mathrm{VII}_{0}}$ be the elements of
$\mathcal{I}_{\mathrm{VII}_{0}}$ such that the corresponding solutions
converge to one of $P_{i}^{+}(II)$ as $\tau\rightarrow -\infty$
and similarly for Bianchi II and IX. Finally, let
$\mathcal{F}_{\mathrm{VII}_{0}}$ be the elements of
$\mathcal{I}_{\mathrm{VII}_{0}}$ such that the corresponding solutions
converge to $F$ as $\tau\rightarrow -\infty$,
and similarly for the other types. 
\end{definition}

\textit{Remark}. The sets $\mathcal{F}_{\mathrm{II}}$ and so
on depend on $\g$, but we omit this reference.
 
Observe that $\mathcal{I}_{\mathrm{I}}$,
$\mathcal{I}_{\mathrm{II}}$, $\mathcal{I}_{\mathrm{VII}_{0}}$ and
$\mathcal{I}_{\mathrm{IX}}$ are submanifolds of $\mathbb{R}^{6}$
of dimensions 2, 3, 4 and 5 respectively. They are diffeomorphic with 
open sets in a suitable $\mathbb{R}^{n}$;
project $\O$ to zero. We will prove that $\mathcal{P}_{\mathrm{II}}$ 
consists of points and that $\mathcal{F}_{\mathrm{I}}$ is the point $F$.
Let $2/3<\g<2$ be fixed.
In Theorem \ref{thm:regular}, we will be able to prove that 
the sets $\mathcal{F}_{\mathrm{II}},
\mathcal{F}_{\mathrm{VII}_{0}}$, $\mathcal{F}_{\mathrm{IX}}$,
$\mathcal{P}_{\mathrm{VII}_{0}}$ and
$\mathcal{P}_{\mathrm{IX}}$ are $C^{1}$ submanifolds of $\mathbb{R}^{6}$
of dimensions $1$, $2$, $3$, $1$ and $2$ respectively.
This justifies the following definition.

\begin{definition}
Let $2/3<\g<2$.
A solution to (\ref{eq:whsu})-(\ref{eq:constraint}) is said to be
\textit{generic} if it is not of Taub type, and if it does not belong 
to $\mathcal{F}_{\mathrm{I}}, \mathcal{F}_{\mathrm{II}},
\mathcal{F}_{\mathrm{VII}_{0}}$, $\mathcal{F}_{\mathrm{IX}}$,
$\mathcal{P}_{\mathrm{II}}$, $\mathcal{P}_{\mathrm{VII}_{0}}$ or 
$\mathcal{P}_{\mathrm{IX}}$.
\end{definition}

We will need the following two lemmas in the sequel. 

\begin{lemma}\label{lemma:palp}
Consider a solution $x$ to (\ref{eq:whsu})-(\ref{eq:constraint})
such that $x$ has $P^{+}_{1}(II)$ as an $\a$-limit point but
does not converge to it. Then $x$ has an $\a$-limit point  of
type II, which is not $P^{+}_{1}(II)$.
\end{lemma}

\textit{Remark}. There is no solution satisfying the conditions of
this lemma, but we will need it to establish that fact.

\textit{Proof}. 
Consider the solution to belong to $\mathbb{R}^{6}$, and let the
point $x_{0}$ represent $P^{+}_{1}(II)$. There is an $\epsilon>0$ 
such that for each $T$, there is a $\tau\leq T$ such that $x(\tau)$
does not belong to the open ball $B_{\epsilon}(x_{0})$. In $x_{0}$
one can compute that
\[
q+2\Sp\pm 2\sqrt{3}\Sm>0.
\]
Let $\epsilon$ be so small that these expressions are positive in 
$B_{\epsilon}(x_{0})$. Let $\tau_{k}\rightarrow -\infty$ be 
a sequence such that $x(\tau_{k})\rightarrow x_{0}$, and let
$s_{k}\leq \tau_{k}$ be a sequence such that $x(s_{k})\in
\partial B_{\epsilon}(x_{0})$ and $x((s_{k},\tau_{k}])\subseteq
B_{\epsilon}(x_{0})$. Since $x(s_{k})$ is contained in
a compact set, there is a convergent subsequence yielding an 
$\a$-limit point which is not $P^{+}_{1}(II)$. Since $\Nt$ and
$\Nth$ converge to zero in $\tau_{k}$ and decay in absolute value 
from $\tau_{k}$ to $s_{k}$, the $\a$-limit point has to be of type II
($\No$ has to be non-zero for the new $\a$-limit point if $\epsilon$
is small enough).
$\Box$

\begin{lemma}\label{lemma:falp}
Consider a solution $x$ to (\ref{eq:whsu})-(\ref{eq:constraint})
such that $x$ has $F$ as an $\a$-limit point, but which does
not converge to $F$. Then $x$ has an $\a$-limit point of type I
which is not $F$.
\end{lemma}

\textit{Remark}. The same remark as that made in connection
with Lemma \ref{lemma:palp} holds concerning this lemma.

\textit{Proof}. The idea is the same as the previous lemma.
We need only observe that $q-4\Sp,\ q+2\Sp+2\sqrt{3}\Sm$ and
$q+2\Sp-2\sqrt{3}\Sm$ are positive in $F$. $\Box$

\section{The monotonicity principle}\label{section:monotone}
The following lemma will be a basic tool in the analysis of the
asymptotics, we will refer to it as \textit{the monotonicity
principle}.
\begin{lemma}\label{lemma:monotone}
Consider
\begin{equation}\label{eq:dxdteqf}
\frac{dx}{dt}=f\circ x
\end{equation}
where $f\in C^{\infty}(\mathbb{R}^{n},\mathbb{R}^{n})$.
Let $U$ be an open subset of $\mathbb{R}^{n}$, and $M$ a
closed subset invariant 
under the flow of the vectorfield $f$. Let $G:U\rightarrow \mathbb{R}$ 
be a continuous function such that $G(x(t))$ is strictly monotone
for any solution $x(t)$ of (\ref{eq:dxdteqf}), as long as 
$x(t)\in U\cap M$. Then no solution of (\ref{eq:dxdteqf}) whose
image is contained in $U\cap M$ has an $\alpha$- or $\omega$-limit 
point in $U$.
\end{lemma}
\textit{Remark}. Observe that one can use $M=\mathbb{R}^{n}$. We will
mainly choose $M$ to be the closed invariant subset of 
$\mathbb{R}^{6}$ defined by (\ref{eq:constraint}). If one $N_{i}$ is
zero and two are non-zero, we consider the number of variables to
be four etc.

\textit{Proof}. Suppose $p\in U$ is an $\a$-limit point
of a solution $x$ contained in $U\cap M$. Then $G\circ x$ is 
strictly monotone. There is a sequence $t_{n}\rightarrow t_{-}$
such that $x(t_{n})\rightarrow p$ by our supposition. Thus 
$G(x(t_{n}))\rightarrow G(p)$, but $G\circ x$ is monotone
so that $G(x(t))\rightarrow G(p)$. Thus $G(q)=G(p)$ for all
$\a$-limit points $q$ of $x$. Since $M$ is closed $p\in M$.
The solution $\bar{x}$ of (\ref{eq:dxdteqf}), with initial value $p$,
is contained in $M$ by the invariance property of $M$, and it
consists of $\a$-limit points of $x$ so that $G(\bar{x}(t))=G(p)$
which is constant. Furthermore, on an open set containing zero it 
takes values in $U$ contradicting the assumptions of the lemma.
$\Box$

Let us give an example of an application. 

\begin{lemma}\label{lemma:n1n2n3tz}
Consider a solution to (\ref{eq:whsu})-(\ref{eq:constraint}) of
type VIII or IX. If it has an $\a$-limit point, then
\[
\lim_{\tau\rightarrow -\infty}(\No\Nt\Nth)(\tau)=0.
\]
\end{lemma}

\textit{Proof}. 
Let $U$ of Lemma \ref{lemma:monotone} be defined by the union of the
sets $N_{i}\neq 0$, $i=1,2,3$, $M$ by the constraint
(\ref{eq:constraint}), and $G$ by the function $\No\Nt\Nth$. 
Compute
\begin{equation}\label{eq:nnnder}
(\No\Nt\Nth)'=3q\No\Nt\Nth.
\end{equation}
Consider a solution $x$ of (\ref{eq:whsu})-(\ref{eq:constraint}). 
We need to prove that $G\circ x$ is strictly monotone as long as 
$x(\tau)\in U\cap M$. By (\ref{eq:nnnder}) the only problem that 
could occur is $q=0$. However, $q=0$ implies $|\Sp'|+|\Sm'|>0$
by (\ref{eq:whsu})-(\ref{eq:constraint}) so that $G\circ x$ has
the desired property. If the sequence $\tau_{k}\rightarrow -\infty$ 
yields the $\a$-limit point we assume exists, then we conclude that
\[
(\No\Nt\Nth)(\tau_{k})\rightarrow 0.
\]
Since $\No\Nt\Nth$ is monotone, we conclude that it converges to 
zero. $\Box$

One important consequence of this observation is the fact that 
all $\a$-limit points of Bianchi VIII and IX solutions are of one of
the lower Bianchi types. Since the $\a$-limit set is invariant under 
the flow, it is thus of interest to know something about the 
$\a$-limit sets of the lower Bianchi types, if one wants to prove
the existence of an $\a$-limit point on the Kasner circle. 

Let us now analyze the vacuum type II orbits and define the Kasner map.

\begin{prop}\label{prop:b2}
A Bianchi II vacuum solution of (\ref{eq:whsu})-(\ref{eq:constraint})
with $\No>0$ and $\Nt=\Nth=0$ satisfies
\begin{equation}\label{eq:nol}
\lim_{\tau\rightarrow \pm\infty}\No=0.
\end{equation}
The $\o$-limit set is a point in $\mathcal{K}_{1}$ and the
$\a$-limit set is a point on the Kasner circle, in the complement 
of the closure of $\mathcal{K}_{1}$.
\end{prop}
\textit{Remark}. What is meant by $\mathcal{K}_{1}$ is explained
in Definition \ref{def:k123}.

\textit{Proof}. Using the constraint (\ref{eq:constraint}) we deduce
that
\[
\Sp'=\frac{3}{2}\No^{2}(2-\Sp).
\]
We wish to apply the monotonicity principle. There are three 
variables. Let $U$ be defined by $\No>0$, $M$ be defined by
(\ref{eq:constraint}), and $G(\Sp,\Sm,\No)=\Sp$. 
We conclude that (\ref{eq:nol}) is true as follows.
Let $\tau_{n}\rightarrow \infty$. A subsequence yields an  
$\omega$-limit point by (\ref{eq:constraint}). The monotonicity
principle yields $\No(\tau_{n_{k}})\rightarrow 0$ for the
subsequence. The argument for the $\a$-limit set is similar, and
equation (\ref{eq:nol}) follows.
Combining this with the constraint, we deduce
\[
\lim_{\tau\rightarrow \pm\infty}q=2.
\]
Using the monotonicity of $\Sp$, we conclude that $(\Sp,\Sm)$ has
to converge. As for the $\a$-limit set, convergence to
$\mathcal{K}_{1}$ is not allowed since $\No'<0$ close to 
$\mathcal{K}_{1}$. Convergence to one of the special points in
the closure of $\mathcal{K}_{1}$ is also forbidden, since 
Proposition \ref{prop:limchar} would imply $\No=0$ for the
solution in that case. Assume now that $(\Sp,\Sm)\rightarrow
(\sigma_{+},\sigma_{-})$ as $\tau\rightarrow \infty$. Compute
\begin{equation}\label{eq:quo}
\left(\frac{\Sm}{2-\Sp}\right)'=0.
\end{equation}
We get
\begin{equation}\label{eq:b2l}
\frac{\Sm}{2-\Sp}=\frac{\sigma_{-}}{2-\sigma_{+}}
\end{equation}
for arbitrary $(\Sp,\Sm)$ belonging to the solution. Since
$\No'=(q-4\Sp)\No$ and $\No\rightarrow 0$, we have to have
$\sigma_{+}\geq 1/2$. If $\sigma_{+}=1/2$, then $\sigma_{-}=
\pm\sqrt{3}/2$. The two corresponding lines in the $\Sp\Sm$-plane,
obtained by substituting $(\sigma_{+},\sigma_{-})$ into
(\ref{eq:b2l}), do not intersect any points interior to the Kasner
circle. Therefore $\sigma_{+}=1/2$ is not an allowed limit point,
and the proposition follows. $\Box$

Observe that by (\ref{eq:quo}), the projection of the solution to
the $\Sp\Sm$-plane is a straight line. The orbits when $\Nt>0$ and
when $\Nth>0$ are obtained by applying the symmetries.
Figure \ref{fig:bkl} shows a sequence of vacuum type II orbits
projected to the $\Sp\Sm$-plane. The first line, starting at the
star, has $\No>0$, the second $\Nth>0$ and the third $\Nt>0$.

\begin{definition}
If $x_{0}$ is a non-special point on the Kasner circle, then the
\textit{Kasner map} applied to $x_{0}$ is defined to be the point 
$x_{1}$ on the Kasner circle, with the property that there is a vacuum 
type II orbit with $x_{0}$ as an $\o$-limit point and $x_{1}$ as 
an $\a$-limit point. 
\end{definition}

\section{Dependence on the shear variables}\label{section:shear}

\begin{figure}[hbt]
  \centerline{\psfig{figure=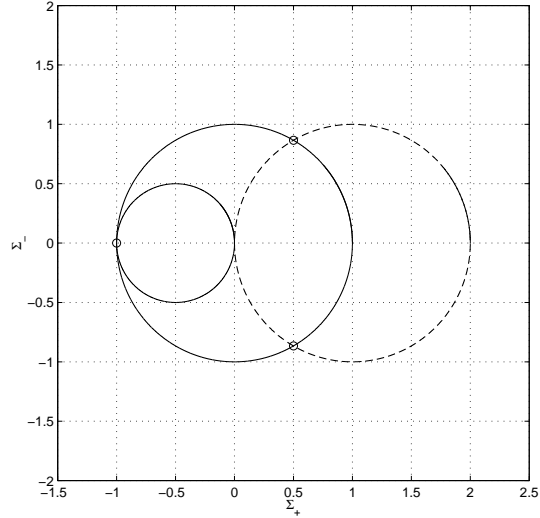,height=7cm}}
  \caption{The circles mentioned in the text.}\label{fig:circles}
\end{figure}

In several arguments, we will have control over the shear variables
and the density parameter in some time interval, and it is of interest
to know how the remaining variables behave in such situations. 
Consider for instance the expression multiplying $\No$ in the
formula for $\No'$, see (\ref{eq:whsu}). It is given by 
$q-4\Sp$ and equals zero when
\begin{equation}\label{eq:paraboloid}
\frac{1}{4}(3\g-2)\O+(1-\Sp)^2+\Sm^2=1.
\end{equation}
The set of points in $\O\Sp\Sm$-space satisfying this equation is
a paraboloid, and the intersection with $\O=0$  is the dashed 
circle shown in Figure
\ref{fig:circles}. If $(\O,\Sp,\Sm)$ belongs to  the interior of 
the paraboloid (\ref{eq:paraboloid}) with $\O\geq 0$, then $|\No|'$ 
will be negative, so that $|\No|$ increases as we go backward. Outside
of the paraboloid, $|\No|$ decreases. The situation is similar for
$\Nt$ and $\Nth$. Observe that the circle obtained by 
letting $\O=0$ in (\ref{eq:paraboloid}) intersects the Kasner circle
in two special points. The same is true of the rotated circles
corresponding to $\Nt$ and $\Nth$. It will be convenient to introduce 
notation for the points on the Kasner circle at which $|N_{i}|'$
is negative.

\begin{definition}\label{def:k123}
We let  $\mathcal{K}_{1},\ \mathcal{K}_{2}$ and $\mathcal{K}_{3}$ be
the subsets of the Kasner circle where $q-4\Sp<0,\
q+2\Sp+2\sqrt{3}\Sm<0$ and $q+2\Sp-2\sqrt{3}\Sm<0$ respectively.
\end{definition}
\textit{Remark}. On the Kasner circle, $\O=0$ so that
$q=2(\Sp^2+\Sm^2)=2$ under the conditions of this definition.

It also of interest to know when the derivatives of $\Nt\Nth$ and 
similar products are zero. Since $(\Nt\Nth)'=(2q+4\Sp)\Nt\Nth$,
we consider the set on which $q+2\Sp$ equals zero. This set is
a paraboloid and is given by
\[
\frac{1}{4}(3\g-2)\O+(\Sp+\frac{1}{2})^2+\Sm^2=\frac{1}{4}.
\]
The intersection with the plane $\O=0$ is the circle
with radius $1/2$ shown in Figure \ref{fig:circles}. Again, inside
the paraboloid $|\Nt\Nth|$ increases as we go backward, and outside it
decreases. There are corresponding paraboloids for the products 
$\No\Nt$ and $\No\Nth$. Observe that in the non-vacuum case, it is 
harmless to introduce $\omega=\O^{1/2}$ and then the paraboloids 
become half spheres.

\begin{prop}\label{prop:BKL}
Consider a Bianchi IX solution to
(\ref{eq:whsu})-(\ref{eq:constraint}) with $2/3<\g<2$.
If the solution has a non-special $\alpha$-limit point $x$ on the 
Kasner circle, then the closure of the vacuum type II orbit with $x$ 
as an $\o$-limit point belongs to the $\alpha$-limit set. 
\end{prop}
\textit{Remark}. The same conclusion holds for a Bianchi type 
VI$\mathrm{I}_{0}$ solution with $\No=0$, if it has an $\alpha$-limit 
point in $\mathcal{K}_{2}$ or $\mathcal{K}_{3}$. 

\textit{Proof}. Assume the limit point lies in $\mathcal{K}_{1}$ 
with $(\Sp,\Sm)=(\sigma_{+},\sigma_{-})$.
There is a sequence $\tau_{k}\rightarrow -\infty$, such that the
solution evaluated at $\tau_{k}$ converges to the point on the 
Kasner circle. There is a ball $B_{\eta}(\sigma_{+},\sigma_{-})$ 
in the $\Sp\Sm$-plane, centered at this point, such that $|\Nt|,\ |\Nth|,\
|\No\Nt|,\ |\No\Nth|$ and $\O$ all decay exponentially, at least as
$e^{\xi\tau}$ for some fixed $\xi>0$, and $\No$ increases 
exponentially, at least as $e^{-\xi\tau}$, in the closure of this 
ball. There is a $K$ such that $(\Sp(\tau_{k}),\Sm(\tau_{k}))\in 
B_{\eta}(\sigma_{+},\sigma_{-})$ 
for all $k\geq K$. For each time we enter the ball, we must leave it,
since if we stay in it to the past, $\No$ will grow to infinity 
whereas
$\Nt$ and $\Nth$ will decay to zero, in violation of the constraint.
Thus for each $\tau_{k}$, $k\geq K$, there is a $t_{k}\leq \tau_{k}$
corresponding to the first time we leave the ball, starting at 
$\tau_{k}$ and going backward. We may compute
\[
(\frac{\Sm}{2-\Sp})'=h
\]
where 
\[
|h(\tau)|\leq \epsilon_{k}e^{\xi(\tau-\tau_{k})}
\]
in $[t_{k},\tau_{k}]$ and $\epsilon_{k}\rightarrow 0$. Thus
\[
\frac{\Sm(\tau_{k})}{2-\Sp(\tau_{k})}-
\frac{\Sm(t_{k})}{2-\Sp(t_{k})}=\int_{t_{k}}^{\tau_{k}}hd\tau.
\]
But 
\[
|\int_{t_{k}}^{\tau_{k}}hd\tau|\leq \frac{\epsilon_{k}}{\xi},
\]
and in consequence
\[
\frac{\Sm(\tau_{k})}{2-\Sp(\tau_{k})}-
\frac{\Sm(t_{k})}{2-\Sp(t_{k})}\rightarrow 0.
\]
We thus get a type II vacuum limit point with $\No>0$, to which 
we may apply the flow, and deduce the conclusion of the lemma.
The statement made in the remark follows in the same way. Observe
that the only important thing was that the limit point was in 
$\mathcal{K}_{1}$ \textit{and} $\No$ was non-zero for the solution.
$\Box$

\section{The stiff fluid case}\label{section:stiff}
In this section we will assume $\O>0$ and $\g=2$ for all solutions we
consider. We begin by explaining the origin of the triangle
shown in Figure \ref{fig:triangle}. Then we analyze the type II
orbits. They yield an analogue of the Kasner map, connecting
two points inside the Kasner circle, and we state an analogue
of Proposition \ref{prop:BKL} for this map. We then prove that
$\O$ is bounded away from zero to the past. Only in the case of
Bianchi IX is an argument required, but this result is the
central part of the analysis of the stiff fluid case. A peculiarity
of the equations then yields the conclusion
that $|\No\Nt|+|\Nt\Nth|+|\Nth\No|$ converges to zero exponentially.
This proves that any solution is contained in a compact set to the
past, and that all $\a$-limit points are of type I or II. Another
consequence is that $\O$ has to converge to a non-zero value;
this requires a proof in the Bianchi IX case. Next one concludes that
all $N_{i}$ converge to zero, since if that were not the case, there
would be an $\a$-limit point of type II to which one could apply the
flow, obtaining $\a$-limit points with different $\O$:s. Then if
a Bianchi IX solution had an $\a$-limit point outside the triangle,
one could apply the 'Kasner' map to such a point, obtaining an
$\a$-limit point with some $N_{i}>0$. Finally, some technical 
arguments finish the analysis.

In the case of a stiff fluid, that is $\g=2$, it is convenient to 
introduce
\[
\o=\O^{1/2}.
\]
We then have, since $3\g-2=4$,
\begin{equation}\label{eq:sopr}
\o'=-(2-q)\o.
\end{equation}
The expression $\O+\Sp^2+\Sm^2$ turns into 
$\omega^2+\Sp^2+\Sm^2$, and the $\o,\Sp,\Sm$-coordinates of the type
I points obey
\begin{equation}\label{eq:typeIcon}
\omega^2+\Sp^2+\Sm^2=1,\ \omega\geq 0.
\end{equation}
In the stiff fluid case, all the type I points are fixed points, and
they play a role similar to that of the Kasner circle in the vacuum
case.

Let us make some observations. 
If $\No\neq 0$, then $\No'=0$ is equivalent to $q-4\Sp=0$. Dividing
by $2$ and completing squares, we see that this condition is equivalent
to
\begin{equation}\label{eq:nosp}
\o^2+(1-\Sp)^2+\Sm^2=1,\ \omega\geq 0.
\end{equation}
By applying the symmetries, the conditions $N_{i}'=0,\
N_{i}\neq 0$ are consequently all fulfilled precisely on
half spheres of radii $1$.
Since $|\No|'<0$ corresponds to an increase in $|\No|$ as we go
backward, $|\No|$ increases exponentially as we are inside the
half sphere (\ref{eq:nosp}) and decreases exponentially as we are 
outside it. If one takes the intersection of (\ref{eq:typeIcon})
and (\ref{eq:nosp}), one gets the subset $\Sp=1/2$ of 
(\ref{eq:typeIcon}). The corresponding intersections for $\Nt$ and
$\Nth$ yield two more lines in the $\Sp\Sm$-plane. Together they yield
the triangle in Figure \ref{fig:triangle}. Consequently, if
$(\o,\Sp,\Sm)$ is close to (\ref{eq:typeIcon}) and $(\Sp,\Sm)$ is in
the interior of the triangle, then all the $N_{i}$ decay exponentially
as $\tau\rightarrow -\infty$. 

Let $\mathcal{M}_{1}$ be the subset $\o\Sp\Sm$-space obeying 
(\ref{eq:typeIcon}) with
$\o>0$ and $\Sp>1/2$ and $\mathcal{M}_{2}$, $\mathcal{M}_{3}$ be the 
corresponding sets for $\Nt$ and $\Nth$. We also let $\mathcal{L}_{1}$
be the subset of the intersection between (\ref{eq:typeIcon})
and (\ref{eq:nosp}) with $\o>0$ and correspondingly $\Nt$ and
$\Nth$ yield $\mathcal{L}_{2}$ and $\mathcal{L}_{3}$.

\begin{lemma}\label{lemma:typeIIs}
Consider a solution to (\ref{eq:whsu})-(\ref{eq:constraint})
with $\g=2$ such that $\No>0$, $\o>0$ and $\Nt=\Nth=0$. Then 
\begin{equation}\label{eq:notz}
\lim_{\tau\rightarrow \pm\infty}\No(\tau)=0
\end{equation}
and $(\o,\Sp,\Sm)$ converges to a point, satisfying
(\ref{eq:typeIcon}) and $\o>0$, in the complement of 
$\mathcal{L}_{1}\cup \mathcal{M}_{1}$, as $\tau\rightarrow -\infty$. 
In $\o\Sp\Sm$-space, the orbit of the solution
is a straight line connecting two points satisfying (\ref{eq:typeIcon}).
If $\o>0$, it is strictly increasing along the solution, going
backwards in time.
\end{lemma}

\textit{Proof}. Since $q<2$ for the entire solution, we can apply
the monotonicity principle with $U$ defined by $q<2$, $G$
defined by $\Sp$ and $M$ by the constraint (\ref{eq:constraint}).
If $q$ does not converge to 2 as $\tau\rightarrow -\infty$, we
get an $\alpha$-limit point with $q<2$. We have a contradiction. 
This argument also yields the conclusion that $\No\rightarrow 0$ as
$\tau\rightarrow \infty$. Equation (\ref{eq:notz})
follows. Observe that 
\begin{equation}\label{eq:sspr}
\Sp'=\frac{3}{2}\No^2(2-\Sp),\ \Sm'=-\frac{3}{2}\No^2\Sm
\end{equation}
and
\begin{equation}\label{eq:sopr2}
\o'=-\frac{3}{2}\No^2\o.
\end{equation}
Consequently, $\Sp$, $\Sm$ and $\o$ are all monotone so that they
converge, both as $\tau\rightarrow \infty$ and as $\tau\rightarrow 
-\infty$. It also follows from (\ref{eq:sspr}) and (\ref{eq:sopr2})
that the quotients $(2-\Sp)/\o$ and $\Sm/\o$ are constant. Thus the
orbit in $\o\Sp\Sm$-space describes a straight line connecting two 
points satisfying (\ref{eq:typeIcon}). As $\tau\rightarrow -\infty$, 
the solution cannot converge to a point in $
\mathcal{L}_{1}\cup \mathcal{M}_{1}$ for the following reason. Assume 
it does. Since $\Sp$ decreases as $\tau$ decreases, see
(\ref{eq:sspr}), we must
have $\Sp\geq 1/2$ for the entire solution, since $\Sp$ by assumption
converges to a value $\geq 1/2$. But then $\No'<0$ for the entire 
solution by (\ref{eq:whsu}) and (\ref{eq:constraint}). Thus $\No$
increases as we go backward, contradicting the fact that
$\No\rightarrow 0$. $\Box$

The next thing we wish to prove is that if a solution has an 
$\alpha$-limit point $x$ in the set $\mathcal{M}_{1}$,
and $\No\neq 0$ for the solution, then we can apply the
'Kasner' map to that point. What we mean by that is that an entire
type II orbit with $x$ as an $\o$-limit point belongs to the 
$\alpha$-limit set of the original solution. From this one can draw
quite strong conclusions. Observe for instance that by (\ref{eq:sopr}),
$\o$ is monotone for a Bianchi VIII solution to 
(\ref{eq:whsu})-(\ref{eq:constraint}). Thus $\o$ converges as
$\tau\rightarrow -\infty$ since it is bounded. If the Bianchi VIII
solution has an $\alpha$-limit point of type I outside the triangle,
we can apply the Kasner map to it to obtain $\alpha$-limit points with
different $\o$. But that is impossible.

\begin{lemma}\label{lemma:sBKL}
Consider a solution to (\ref{eq:whsu})-(\ref{eq:constraint}) with
$\g=2$ such that $\No\neq 0$. Then if the solution has an 
$\alpha$-limit point $x\in \mathcal{M}_{1}$,
the orbit of a type II solution with $x$ as an $\o$-limit point
belongs to the $\alpha$-limit set of the solution.
\end{lemma}

\textit{Proof}. The proof is analogous to the proof of
Proposition \ref{prop:BKL}. $\Box$

Consider a solution such that $\o>0$. We want to exclude the
possibility that $\o\rightarrow 0$ as $\tau\rightarrow -\infty$. 
Considering (\ref{eq:sopr}), we see that the
only possibility for $\o$ to decrease is if $q>2$. In that context,
the following lemma is relevant.

\begin{lemma}\label{lemma:sB9N}
Consider a Bianchi IX solution to
(\ref{eq:whsu})-(\ref{eq:constraint}) with $\g=2$. There is an 
$\alpha_{0}$ such that if $\a\leq\a_{0}$ and 
\[
(\No\Nt\Nth)(\tau)\leq \a,
\]
then
\[
q(\tau)-2\leq 4\a^{1/3}.
\]
\end{lemma}

\textit{Proof}. By a permutation of the variables, we can assume 
$\No\leq\Nt\leq\Nth$ in $\tau$. Observe that 
\[
q-2\leq 3\No(\Nt+\Nth)
\]
by the constraint (\ref{eq:constraint}). If $\Nth\leq \a^{1/2}$
in $\tau$, we get $q-2\leq 6\a\leq 4\a^{1/3}$ if $\a_{0}$ is small
enough. If $\Nth\geq \alpha^{1/2}$ in $\tau$, we get
\[
\No\Nt\leq \alpha^{1/2}.
\]
Assume, in order to reach a contradiction,  $(\No\Nth)(\tau)\geq
\a^{1/3}$. Then $\Nt(\tau)\leq \a^{2/3}$, so that  $\No(\tau)\leq
\a^{2/3}$ and $\Nth(\tau)\geq \a^{-1/3}$. By Lemma \ref{lemma:B9N} 
we get a contradiction if $\a_{0}$ is small enough. Thus
\[
q-2\leq 3(\No\Nt+\No\Nth)(\tau)\leq 3(\a^{1/3}+\a^{1/2})\leq 4\a^{1/3}
\]
if $\a_{0}$ is small enough. $\Box$

For all solutions except those of Bianchi IX type, $\o$ is monotone
increasing as $\tau$ decreases. Thus, $\o$ is greater than zero on the
$\a$-limit set of any  non-vacuum solution which is not of type IX.
It turns out that the same is true for a Bianchi IX solution.

\begin{lemma}\label{lemma:onz}
Consider a Bianchi IX solution to
(\ref{eq:whsu})-(\ref{eq:constraint}) with $\g=2$ such that $\o>0$.
Then there is an $\epsilon>0$ such that $\o(\tau)\geq \epsilon$ for
all $\tau\leq 0$.
\end{lemma}

\textit{Proof}. 
Assume all the $N_{i}$ are positive. The function
\[
\phi=\frac{(\No\Nt\Nth)^{1/3}}{\o}
\]
satisfies $\phi'=2\phi$. Thus, for $\tau\leq 0$,
\[
(\No\Nt\Nth)^{1/3}(\tau)=\o(\tau)\phi(0)e^{2\tau}\leq C e^{2\tau},
\]
because of Lemma \ref{lemma:B9N}. For $\tau\leq T\leq 0$, we can thus
apply Lemma \ref{lemma:sB9N}, so that for $\tau\leq T$,
\[
\int_{\tau}^{0}(q(s)-2)ds=
\int_{\tau}^{T}(q(s)-2)ds+\int_{T}^{0}(q(s)-2)ds\leq
4C\int_{\tau}^{T}e^{2s}ds+
\]
\[
+\int_{T}^{0}(q(s)-2)ds\leq
2Ce^{2T}+\int_{T}^{0}(q(s)-2)ds\leq C'<\infty.
\]
Consequently,
\[
\o(\tau)=\o(0)\exp(-\int_{\tau}^{0}(q(s)-2)ds)\geq
\o(0) e^{-C'},
\]
and the lemma follows. $\Box$

The next lemma will be used to prove that $\o$ converges for
a Bianchi IX solution.

\begin{lemma}\label{lemma:n2b}
Consider a solution to (\ref{eq:whsu})-(\ref{eq:constraint}) 
with $\g=2$ and $\o>0$. Then there is an $\a>0$ and a $T$ such that 
\[
|\No\Nt|+|\Nt\Nth|+|\Nth\No|\leq e^{\a\tau}
\]
for all $\tau\leq T$. 
\end{lemma}

\textit{Proof}. Consider $g=|\Nt\Nth|/\o$. Then
\[
g'=(2\o^2+2(1+\Sp)^2+2\Sm^2)g.
\]
Since $\o(\tau)\geq \epsilon$ for all $\tau\leq 0$, we conclude that
\[
g(\tau)\leq g(0)\exp (2\epsilon^2\tau)
\]
so that
\[
|(\Nt\Nth)(\tau)|\leq g(0)\o(\tau)\exp (2\epsilon^2\tau).
\]
There are similar estimates for the other products.  By Lemma
\ref{lemma:B9N}, we know that $\o$ is bounded in $(-\infty,0]$
so that by choosing $\a=\epsilon^2$ and $T$ negative enough the
lemma follows. $\Box$

\begin{cor}\label{cor:comp}
Consider a solution to (\ref{eq:whsu})-(\ref{eq:constraint}) 
with $\g=2$ and $\o>0$. Then $(\o,\Sp,\Sm,\No,\Nt,\Nth)(-\infty,0]$
is contained in a compact set and all the $\a$-limit points are
of type I or II.
\end{cor}

\begin{lemma}\label{lemma:oconv}
Consider a solution to (\ref{eq:whsu})-(\ref{eq:constraint}) 
with $\g=2$ and $\o>0$. Then
\[
\lim_{\tau\rightarrow -\infty} \o(\tau)=\o_{0}>0.
\]
\end{lemma}

\textit{Proof}. 
Since this follows from the monotonicity of $\o$ in all cases except
Bianchi IX, see (\ref{eq:sopr}), we assume that the solution is of
type IX. Let $\tau_{k}\rightarrow -\infty$ be a sequence such
that $\o(\tau_{k})\rightarrow \o_{1}>0$. This is possible since $\o$
is constrained to belong to a compact set for $\tau\leq 0$ by
Lemma \ref{lemma:B9N}, and since $\o$ is bounded away from zero to
the past by Lemma \ref{lemma:onz}.
Assume $\o$ does not converge to $\o_{1}$. Then there is a sequence
$s_{k}\rightarrow -\infty$ such that $\o(s_{k})\rightarrow \o_{2}$
where we can assume $\o_{2}>\o_{1}$. We can also assume $\tau_{k}\leq
s_{k}$. Then
\[
\o(s_{k})=\exp (\int_{\tau_{k}}^{s_{k}}(q-2)ds)\o(\tau_{k}).
\]
Since 
\[
q-2\leq 3(\No\Nt+\Nt\Nth+\Nth\No)\leq 3e^{\a\tau}
\]
for $\tau\leq T$ by Lemma \ref{lemma:n2b} and the constraint
(\ref{eq:constraint}), we have, assuming $s_{k}\leq T$,
\[
\int_{\tau_{k}}^{s_{k}}(q-2)ds\leq
3\int_{\tau_{k}}^{s_{k}}e^{\a\tau}d\tau\leq\frac{3}{\a}e^{\a s_{k}}.
\]
Thus
\[
\o(s_{k})\leq \exp(\frac{3}{\a}e^{\a s_{k}})\o(\tau_{k})\rightarrow
\o_{1},
\]
so that $\o_{2}\leq\o_{1}$ contradicting our assumption. $\Box$

\begin{cor}\label{cor:ntz}
Consider a solution to (\ref{eq:whsu})-(\ref{eq:constraint}) 
with $\g=2$ and $\o>0$. Then
\[
\lim_{\tau\rightarrow -\infty}N_{i}(\tau)=0
\]
for $i=1,2,3$. 
\end{cor}

\textit{Proof}. Assume $\No$ does not converge to zero. Then
there is a type II $\a$-limit point with $\No$ and $\o$ non-zero by
Corollary \ref{cor:comp} and Lemma \ref{lemma:oconv}.
If we apply the flow, we get $\a$-limit points with different
$\o$ in contradiction to Lemma \ref{lemma:oconv}. $\Box$

\begin{lemma}
Consider a solution to (\ref{eq:whsu})-(\ref{eq:constraint}) 
with $\g=2$ and $\o>0$. If it has an $\a$-limit point of type
I inside the triangle, the solution converges to that point.
\end{lemma}

\textit{Proof}. Let $x$ be the limit point. Let $B$ be a ball of
radius $\epsilon$ in $\o\Sp\Sm$-space, with center given by the
$\o,\Sp,\Sm$-coordinates of $x$. Let $\tau_{k}\rightarrow -\infty$
be a sequence that yields $x$. Assume the solution leaves $B$ to the
past of every $\tau_{k}$. Then there is a sequence $s_{k}\rightarrow
-\infty$, such that the $\o,\Sp,\Sm$-coordinates of the solution 
evaluated in $s_{k}$ converges to a point on the boundary of $B$,
$s_{k}\leq \tau_{k}$, and the $\o,\Sp,\Sm$-coordinates of the solution
are contained in $B$ during $[s_{k},\tau_{k}]$, $k$ large enough. 

Since all expressions in the $N_{i}$ decay exponentially as
$e^{\a\tau}$, for some $\a>0$, as long as the $\o,\Sp,\Sm$-coordinates 
are in $B$ ($\epsilon$ small enough), we have
\[
|\Sp'|+|\Sm'|+|\o'|\leq \a_{k}e^{\a(\tau-\tau_{k})}
\]
for $\tau\in [s_{k},\tau_{k}]$ where $\a_{k}\rightarrow 0$.
We get
\[
|\Sp(\tau_{k})-\Sp(s_{k})|\leq \frac{\a_{k}}{\a}\rightarrow 0,
\]
and similarly for $\Sm$ and $\o$. The assumption that we always leave
$B$ consequently yields a contradiction. We must thus converge to 
the given $\a$-limit point. $\Box$

\begin{prop}\label{prop:notmi}
Consider a solution to (\ref{eq:whsu})-(\ref{eq:constraint}) 
with $\g=2$ and $\o>0$. If $N_{i}$ is non-zero for the solution,
it converges to a type I point in the complement of $\mathcal{M}_{i}$
with $\o>0$.
\end{prop}

\textit{Proof}. If there is an $\a$-limit point on $\mathcal{M}_{i}$,
we can use Lemma \ref{lemma:sBKL} to obtain a contradiction to 
Lemma \ref{lemma:oconv}. If there is an $\a$-limit point in 
$\mathcal{M}_{k}$ and $N_{k}$ is zero for the solution, the solution
converges to that point by an argument similar to the one given in the
previous lemma. What remains is the possibility that all the
$\a$-limit points are on the $\mathcal{L}_{k}$. Since $\o$
converges, the possible points projected to the $\Sp\Sm$-plane
are the intersection between a triangle and a circle. Since the
$\a$-limit set is connected, we conclude that the solution must
converge to a point on one of the $\mathcal{L}_{k}$. $\Box$

\begin{prop}\label{prop:notli}
Consider a solution to (\ref{eq:whsu})-(\ref{eq:constraint}) 
with $\g=2$ and $\o>0$. If $N_{i}$ is non-zero for the solution,
the solution cannot converge to a point in $\mathcal{L}_{i}$.
\end{prop}

\textit{Proof}. Assume $i=1$. Then $\mathcal{L}_{i}$ is the
subset of (\ref{eq:typeIcon}) consisting of points with $\Sp=1/2$
and $\o>0$. Since $\Nt,\ \Nth,\ \Nt\Nth,\ \Nt\No$ and $\Nth\No$
converge to zero faster than $\No^{2}$, $\Sp'$ will in the end be
positive, cf. (\ref{eq:sspr}),
so that there is a $T$ such that $\Sp(\tau)\geq 1/2$ for $\tau\leq T$.
Since $\No$ will dominate in the end, we can also assume $q(\tau)<2$
for $\tau\leq T$. By (\ref{eq:whsu}) we conclude that $|\No|$
increases backward as $\tau\leq T$ contradicting Corollary
\ref{cor:ntz}. $\Box$

Adding up the last two propositions, we conclude that the
$\Sp\Sm$-variables of Bianchi VIII and IX solutions converge to 
a point interior to the triangle of Figure \ref{fig:triangle},
and $\O$ to the value then determined by the constraint
(\ref{eq:constraint}). In the Bianchi VI$\mathrm{I}_{0}$ case,
a side of the triangle disappears, increasing the set of 
points to which $\Sp,\Sm$ may converge. We sum up the conclusions
in Section \ref{section:conclusions}.

\section{Type I solutions}\label{section:typeI}
Consider type I solutions ($N_{i}=0$). The point $F$ and the points 
on the Kasner circle are fixed points. Consider a solution with 
$0<\O(\tau_{0})<1$. Using the constraint, we may express the time 
derivative of $\O$ in terms of $\O$. Solving the resulting equation 
yields 
\[
\lim_{\tau\rightarrow -\infty}\O(\tau)=0,\
\lim_{\tau\rightarrow \infty}\O(\tau)=1.
\]
By (\ref{eq:whsu}) $(\Sp,\Sm)$ moves radially. 
\begin{prop}\label{prop:typeI}
For a type I solution, with $2/3<\g<2$, which is not F, we have
\[
\lim_{\tau\rightarrow -\infty}(\Sp,\Sm,\O)(\tau)=
(\sigma_{+}/|\sigma|,\sigma_{-}/|\sigma|,0),
\]
where $(\sigma_{+},\sigma_{-})$ is the initial value of $(\Sp,\Sm)$,
and $|\sigma|$ is the Euclidean norm of the initial value.
\end{prop}

\section{Type II solutions}\label{section:typeII}

\begin{prop}\label{prop:typeII}
Consider a type II solution with $\No>0$ and $2/3<\g<2$. If the
initial value for $\Sm$ is non-zero, the $\a$-limit set is
a point in $\mathcal{K}_{2}\cup\mathcal{K}_{3}$.  
If the initial value for $\Sm$ is zero, either the solution
is the special point $P_{1}^{+}(II)$, it is contained in 
$\mathcal{F}_{\mathrm{II}}$, or 
\begin{equation}\label{eq:t2tl}
\lim_{\tau\rightarrow -\infty}(\O,\Sp,\No)(\tau)=(0,-1,0).
\end{equation}
\end{prop}

\textit{Proof}. Let the initial data be given by
$(\sigma_{+},\sigma_{-},\O_{0})$. The vacuum case was handled in 
Proposition \ref{prop:b2}, so we will assume $\O_{0}>0$.

Consider first the case $\sigma_{-}\neq 0$. Compute
\[
q-2=-\frac{3}{2}(2-\g)\O-\frac{3}{2}\No^{2}.
\]
Thus, $\Sm$ decreases if it is negative, and increases if it is
positive, as we go backward in time, by (\ref{eq:whsu}). Thus, 
both $\No$ and $\O$ must converge to $0$ as $\tau\rightarrow
-\infty$, since the variables are constrained to belong to a 
compact set, and because of the monotonicity principle. Since $\Sm$ is 
monotonous and the $\alpha$-limit set is connected, see Lemma
\ref{lemma:als},  $(\Sp,\Sm)$ must 
converge to a point, say $(s_{+},s_{-})$  on the Kasner circle. We 
must have $s_{-}\neq 0$, and
\[
2s_{+}^{2}+2s_{-}^{2}-4s_{+}\geq 0,
\]
since $\No$ converges to $0$. There are two special
points in this set, but we may not converge to them, since that
would imply $\No=0$ for the entire solution by Proposition
\ref{prop:limchar}. The first part of the proposition follows.

Consider the case $\sigma_{-}=0$. There is a fixed point 
$P_{1}^{+}(II)$. 
Eliminating $\O$ from (\ref{eq:whsu})-(\ref{eq:constraint}), we are
left with the two variables $\No$ and $\Sp$. The linearization has
negative eigenvalues at $P_{1}^{+}(II)$, so that no solution which does
not equal $P_{1}^{+}(II)$ can have it as an $\a$-limit point, cf.
\cite{hart} pp. 228-234. 
There is also a set of solutions converging to the 
fixed point $F$. Consider now the complement of the above. The
function
\[
Z_{7}=\frac{\No^{2m}\O^{1-m}}{(1-v\Sp)^2},
\]
where $v=(3\g-2)/8$ and $m=3v(2-\g)/8(1-v^2)$, found by Uggla satisfies
\[
Z_{7}'=\frac{3(2-\g)}{1-v\Sp}\frac{1}{1-v^{2}}(\Sp-v)^2 Z_{7}.
\]
Apply the monotonicity principle. Let $G=Z_{7}$ and $U$ be defined
as the subset of $\O\Sp\No$-space consisting of points
different from $P_{1}^{+}(II)$, which have $\O>0$, $\No>0$ and
$|\Sp|<1$. Let $M$ be defined by the constraint. 
If $\Sp=v$ then $Z_{7}'=0$, but if we are not at $P_{1}^{+}(II)$,
$\Sp=v$ implies $\Sp'\neq 0$. Thus, $G\circ x$ is strictly monotone as 
long as $x$ is contained in $U\cap M$. Since the solution cannot have
$P_{1}^{+}(II)$ as an $\a$-limit point, we must thus
have $\No=0$ or $\O=0$ in the $\alpha$-limit set. Observe that
\begin{equation}\label{eq:spprII}
\Sp'=\frac{3}{2}\No^2(2-\Sp)-\frac{3}{2}(2-\g)\O\Sp.
\end{equation}
Thus, if the solution attains a point $\Sp\leq 0$, then (\ref{eq:t2tl})
holds. We will now prove that this is the only possibility. 

a. Assume we have an $\alpha$-limit point with $\No>0$ and $\O=0$. 
Then we may apply the flow to that limit point to get $\Sp=-1$ as a 
limit point, but then the solution must attain $\Sp\leq 0$. 

b. If $\O>0$ but $\No=0$, then we may assume $\Sp\neq 0$ since we are
not on $\mathcal{F}_{\mathrm{II}}$, cf. Lemma \ref{lemma:falp}. 
Apply the flow to arrive at $\Sp=-1$ or
$\Sp=1$. The former alternative has been dealt with, and the latter 
case allows us to construct an $\alpha$-limit point with $\No>0$ and
$\O=0$, since $\No$ increases exponentially, and $\O$ decreases 
exponentially, in a neighbourhood of the point on the Kasner circle
with $\Sp=1$, cf. Proposition \ref{prop:BKL}. 

c. The situation $\O=\No=0$ can be handled as above. $\Box$

We make one more observation that will be relevant in analyzing the
regularity of $\mathcal{F}_{\mathrm{II}}$. 

\begin{lemma}\label{lemma:ffaII}
The closure of $\mathcal{F}_{\mathrm{II}}$ does not intersect
$\mathcal{A}$.
\end{lemma}

\textit{Proof}. Assume there is a sequence $x_{k}\in
\mathcal{F}_{\mathrm{II}}$ such that the distance from
$x_{k}$ to $\mathcal{A}$ goes to zero. We can assume that all
the $x_{k}$ have $\No>0$ by choosing a suitable subsequence
and then applying the symmetries. We can also assume that
$x_{k}\rightarrow x\in\mathcal{A}$. Since $\Sm=0$ for all
the $x_{k}$ by Proposition \ref{prop:typeII}, the same 
holds for $x$. Observe that no element of $\mathcal{F}_{
\mathrm{II}}$ can have $\Sp\leq 0$, because of (\ref{eq:spprII}).
If $\No$ corresponding to $x$ is zero, we then conclude that
$x$ is defined by $\Sp=1$ and all the other variables zero.
Applying the flow to the past to the points $x_{k}$ will
then yield a sequence $y_{k}\in\mathcal{F}_{\mathrm{II}}$
such that $y_{k}$ converges to a type II vacuum point with
$\No>0$ and $\Sm=0$, cf. the proof of Proposition \ref{prop:BKL}.
Thus, we can assume that the limit point $x\in\mathcal{A}$
has $\No>0$. Applying the flow to $x$ yields the point
$\Sp=-1$ on the Kasner circle by Proposition \ref{prop:b2}.
By the continuity of the flow, we can apply the flow to 
$x_{k}$ to obtain elements in $\mathcal{F}_{\mathrm{II}}$
with $\Sp<0$ which is impossible. $\Box$

\section{Type VI$\mathrm{I}_{0}$ solutions}\label{section:typeVII0}
When speaking of Bianchi VI$\mathrm{I}_{0}$ solutions, we will always 
assume $\No=0$ and $\Nt,\Nth>0$. Consider first the case $\Nt=\Nth$
and $\Sm=0$

\begin{prop}\label{prop:typeVII0t}
Consider a type VI$I_{0}$ solution with $\No=0$ and $2/3<\g<2$. If
$\Nt=\Nth$ and $\Sm=0$, one of the following possibilities 
occurs
\begin{enumerate}
  \item The solution converges to $\Sp=1$ on the
        Kasner circle.
  \item The solution converges to $F$.
  \item $\lim_{\tau\rightarrow -\infty}\Sp=-1,\ 
         \lim_{\tau\rightarrow -\infty}\Nt=n_{2}>0,\
         \lim_{\tau\rightarrow -\infty}\O=0$.
\end{enumerate}
\end{prop}

\textit{Proof}. Since
\[
\Sp'=-\frac{3}{2}(2-\g)\O\Sp
\]
if $\Nt=\Nth$, the conclusions of the lemma follow, except for
the statement that $\Nt$ converges to a non-zero value if $\Sp$ 
converges
to $-1$. However, $\O$ will decay to zero exponentially close to the 
Kasner circle, and by the constraint, $1+\Sp$ will behave as $\O$ 
close to $\Sp=-1$. Thus, $q+2\Sp$ will be integrable. $\Box$

\begin{figure}[hbt]
  \centerline{\psfig{figure=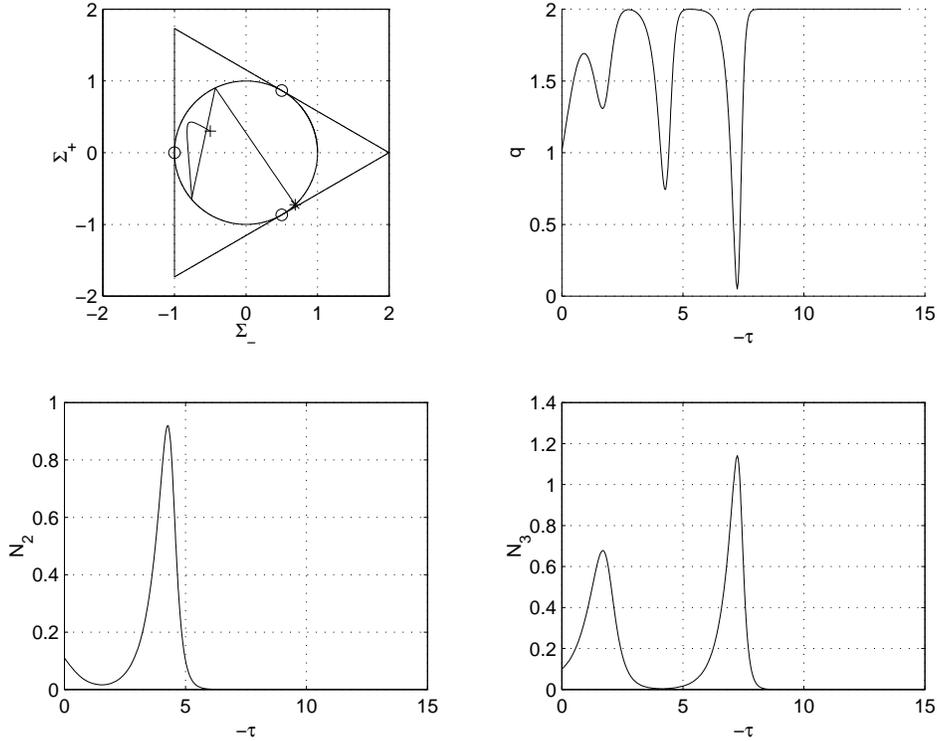,height=10cm}}
  \caption{Illustration of a Bianchi VI$\mathrm{I}_{0}$ solution.}
\label{fig:bsz}
\end{figure}

Before we state a proposition concerning the behaviour of 
generic Bianchi VI$\mathrm{I}_{0}$ solutions, let us give
an intuitive picture. Figure \ref{fig:bsz} shows a simulation
with $\g=1$, where the plus sign represents the starting point, and
the star the end point, going backward. $\O$ will decay to zero 
quite rapidly, and the same holds for the product $\Nt\Nth$. 
In that sense, the solution will asymptotically behave like a 
sequence of type II vacuum orbits. If both $\Nt$ and $\Nth$ are small,
and we are close to the section $\mathcal{K}_{2}$ on the Kasner
circle, then $\Nt$ will increase exponentially, and $\Nth$ will
decay exponentially, yielding in the end roughly a type II orbit with
$\Nt>0$. If this orbit ends in at a point in $\mathcal{K}_{3}$, then
the game begins anew, and we get roughly a type II orbit with $\Nth>0$.
Observe however that if we get close to $\mathcal{K}_{1}$, there is 
nothing to make us bounce away, since $\No$ is zero. The simulation
illustrates this behaviour. Consider the figure of the solution
projected to the $\Sp\Sm$-plane. The three points that appear to 
be on the Kasner circle are close to $\mathcal{K}_{2}$,
$\mathcal{K}_{3}$ and $\mathcal{K}_{1}$ respectively. Observe how this
correlates with the graphs of $\Nt$, $\Nth$ and $q$.

\begin{prop}\label{prop:typeVII0}
Generic Bianchi VI$I_{0}$ solutions with $\No=0$ and
$2/3<\g<2$ converge to a point in $\mathcal{K}_{1}$.
\end{prop}

We divide the proof into lemmas. First we prove that the 
past dynamics are contained in a compact set.

\begin{lemma}\label{lemma:bsz}
For a generic Bianchi VI$I_{0}$ solution with $\No=0$ and
$2/3<\g<2$, $(\Nt,\Nth)(-\infty,0]$ is contained in
a compact set.
\end{lemma}

\textit{Proof}.
For a generic solution,
\[
Z_{-1}=\frac{\frac{4}{3}\Sm^2+(\Nt-\Nth)^2}{\Nt\Nth}
\]
is never zero. Compute
\begin{equation}\label{eq:zmopr1}
Z_{-1}'=-\frac{16}{3}\frac{\Sm^2(1+\Sp)}{\frac{4}{3}\Sm^2+
(\Nt-\Nth)^2} Z_{-1}.
\end{equation}
The proof that the
past dynamics are contained in a compact set is as in Rendall
\cite{rendall}. Let $\tau\leq 0$. Then
\[
Z_{-1}(\tau)\geq Z_{-1}(0),
\]
so that
\[
(\Nt\Nth)(\tau)\leq \frac{4}{3Z_{-1}(0)}.
\]
Combining this fact with the constraint, we see that all the variables
are contained in a compact set during $(-\infty,0]$. $\Box$

We now prove that $\Nt\Nth\rightarrow 0$. The reason being the desire
to reduce the problem by proving that all the limit points are of
type I or II, and then use our knowledge about what happens when we
apply the flow to such points. 

\begin{lemma}\label{lemma:n23tz}
Generic Bianchi VI$I_{0}$ solutions with $\No=0$ and
$2/3<\g<2$ satisfy
\[
\lim_{\tau\rightarrow -\infty}(\Nt\Nth)(\tau)=0.
\]
\end{lemma}
\textit{Proof}. Assume the contrary. Then we can use
Lemma \ref{lemma:bsz} to construct an $\alpha$-limit point
$(\omega,\sigma_{+},\sigma_{-},0,n_{2},n_{3})$ where
$n_{2}n_{3}>0$. We apply the monotonicity principle in
order to arrive at a contradiction. With notation as in
Lemma \ref{lemma:monotone}, let $U$ be defined
by $\Nt>0,\ \Nth>0$ and $\Sm^{2}+(\Nt-\Nth)^{2}>0$. Let
$G$ be defined by $Z_{-1}$, and $M$ by the constraint
(\ref{eq:constraint}). We have to show that $G$ evaluated
on a solution is strictly monotone as long as the  solution is 
contained in $U\cap M$. Consider (\ref{eq:zmopr1}). By
the constraint (\ref{eq:constraint}),  $\Sm^{2}+(\Nt-\Nth)^{2}>0$
implies $\Sp>-1$. Furthermore, $Z_{-1}>0$ on $U$. If
$Z_{-1}'=0$ in $U\cap M$, we thus have $\Sm=0$, but then
$\Sm'\neq 0$ since $\Sm^{2}+(\Nt-\Nth)^{2}>0$ and 
$\Nt+\Nth>0$. The $\alpha$-limit point we have constructed
cannot belong to $U$. On the other hand, $n_{2},n_{3}>0$ and
since $Z_{-1}$ increases as we go backward, $\sigma_{-}^{2}+
(n_{2}-n_{3})^{2}$ cannot be zero. We have a contradiction. 
$\Box$ 

\textit{Proof of Proposition \ref{prop:typeVII0}}. Compute
\begin{equation}\label{eq:sppr}
\Sp'=-(2-2\O-2\Sp^2-2\Sm^2)(1+\Sp)-\frac{3}{2}(2-\g)\O\Sp
\end{equation}
by (\ref{eq:sppr0}).
Assume we are not on $\mathcal{P}_{\mathrm{VII}_{0}}$ or
$\mathcal{F}_{\mathrm{VII}_{0}}$. 
Let us first prove that there is an $\alpha$-limit point on the Kasner
circle. Assume $F$ is an $\alpha$-limit point. Then we may construct 
a type I limit point which is not $F$, and thus a limit point on the 
Kasner 
circle, cf. Lemma \ref{lemma:falp} and Proposition \ref{prop:typeI}. 
By Lemma \ref{lemma:n23tz}, we may then assume that there is 
a limit point of type I or II, which is not $P_{2}^{+}(II)$ or 
$P_{3}^{+}(II)$, and does not lie in $\mathcal{F}_{\mathrm{I}}$ or 
$\mathcal{F}_{\mathrm{II}}$, cf. Lemma \ref{lemma:palp}. Thus, we get 
a limit point on the Kasner circle by Proposition \ref{prop:typeI} 
and Proposition \ref{prop:typeII}. 

Next, we prove that there has to be an $\a$-limit point which lies 
in the closure of
$\mathcal{K}_{1}$. If the $\a$-limit point we have constructed is in
$\mathcal{K}_{2}$ or $\mathcal{K}_{3}$, we can apply the
Kasner map according to the remark following Proposition 
\ref{prop:BKL}. After a finite number of Kasner iterates we will end
up in the desired set. If the $\a$-limit point we obtained
has $\Sp=-1$, we may construct a limit point
with $1+\Sp=\epsilon>0$  by Proposition \ref{prop:limchar}.
We can also assume that $\O=0$ for this point, since $\O$ decays
exponentially going backward when $\Sp$ is close to $-1$.
By Lemma \ref{lemma:n23tz}, this limit point will be a type I or II 
vacuum point, and by applying 
the flow we get a non special limit point on the Kasner circle. As 
above, we then get an $\a$-limit point in the desired set. Let the
$\Sp\Sm$-variables of one $\a$-limit point in the closure of 
$\mathcal{K}_{1}$ be $(\sigma_{+},\sigma_{-})$. 

By (\ref{eq:sppr}), we conclude that once $\Sp$ has become greater 
than $0$, it becomes monotone so that it has to converge. Moreover,
we see by the same equation that $\O$ then has to converge to 
zero, and $\Sp^2+\Sm^2$ has to converge to $1$. Since the $\a$-limit
set is connected, by Lemma \ref{lemma:als} and Lemma \ref{lemma:bsz},
we conclude that $(\Sp,\Sm)$ has to converge to 
$(\sigma_{+},\sigma_{-})$. By Proposition \ref{prop:limchar},
$(\sigma_{+},\sigma_{-})$ cannot equal $(1/2,\pm \sqrt{3}/2)$,
since otherwise $\Nt$ or $\Nth$ would be zero for the entire 
solution. Consequently, $\sigma_{+}>1/2$, and we conclude that
$\Nt$ and $\Nth$ have to converge to zero. The proposition
follows. $\Box$

\section{Taub type IX solutions}\label{section:taub}
Consider the Taub type solutions: $\Sm=0$ and $\Nt=\Nth$. We 
prove that except for the cases when the solution belongs to
$\mathcal{F}_{\mathrm{IX}}$ or $\mathcal{P}_{\mathrm{IX}}$, 
$(\Sp,\Sm)$ converges to $(-1,0)$.

\begin{figure}[hbt]
  \centerline{\psfig{figure=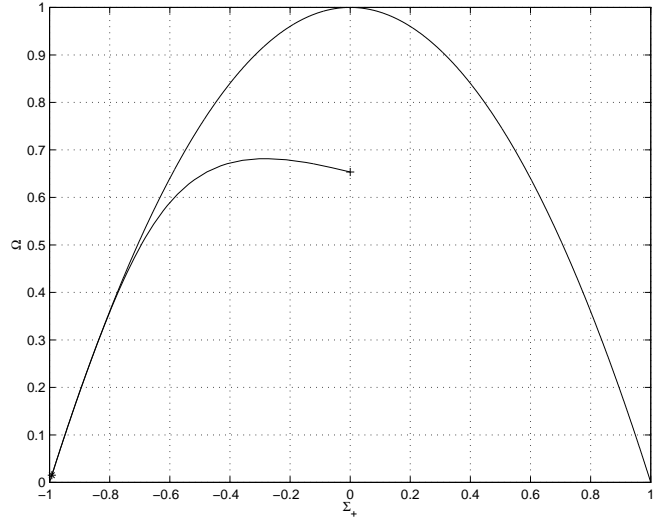,height=7cm}}
  \caption{Part of a Taub type IX solution projected to the $\Sp\O$-plane.}
\label{fig:taub}
\end{figure}

\begin{lemma}\label{lemma:taubconv}
Consider a type IX solution with $\Sm=0$, $\Nt=\Nth$ and
$2/3<\g<2$. Then $\Sp(\tau_{0})\leq
0$ and $\O(\tau_{0})<1$ imply
\[
\lim_{\tau\rightarrow -\infty}(\O,\Sp,\Sm,
\No,\Nt,\Nth)(\tau)=(0,-1,0,0,n_{2},n_{2}),
\]
where $0<n_{2}<\infty$.
\end{lemma}
\textit{Proof}. We prove that the flow will take us to the boundary
of the parabola $\O+\Sp^2=1$ with $\Sp<0$, and that we will then slide
down the side on the outside to reach $\Sp=-1$, see Figure
\ref{fig:taub}. The plus sign in the figure represents the starting
point, and the star the end point.

1. \textit{Let us first assume} $\Sp(\tau_{0})\leq 0$,
$\O(\tau_{0})<1$ \textit{and} $\O(\tau_{0})+\Sp^2(\tau_{0})\geq 1$.
Consider
\[
\mathcal{C}=\{\tau\leq \tau_{0} : t\in [\tau,\tau_{0}]\Rightarrow
\Sp(t)\leq 0,\ \O(t)\leq \O(\tau_{0}),\ \O(t)+\Sp^{2}(t)\geq 1\}.
\]
We prove that $\mathcal{C}$ is not bounded from below. Assume the 
contrary. Let $t$ be the infimum of $\mathcal{C}$, which exists
since $\mathcal{C}$ is non-empty and bounded from below. Since
$t\in\mathcal{C}$, $\Sp(t)<0$. Let $t'<t$ be such that $\Sp<0$ in
$[t',t]$. Observe that
\begin{equation}\label{eq:opr}
\O'=[(3\g-2)(\O+\Sp^2-1)+3(2-\g)\Sp^2]\O.
\end{equation}
By the constraint,
\begin{equation}\label{eq:taubc}
\O+\Sp^2-1=\frac{3}{4}\No^2(4\frac{\Nt}{\No}-1).
\end{equation}
Since $\Sp<0$ in $[t',t]$, $\Nt/\No$ increases as we go backward in 
that interval, because of
\[
(\frac{\Nt}{\No})'=6\Sp\frac{\Nt}{\No}.
\]
Consequently $\O+\Sp^2\geq 1$ in $[t',t]$, by (\ref{eq:taubc}), so that
$\O$ decreases in the interval by (\ref{eq:opr}). Thus $t'\in
\mathcal{C}$, contradicting the fact that $t$ is the infimum of
$\mathcal{C}$. 

Let $\tau\leq \tau_{0}$. Then $\Sp(\tau)\leq -\sqrt{1-\O(\tau_{0})}$.
By (\ref{eq:opr}), we then conclude
$\O\rightarrow 0$. By (\ref{eq:whsu}), we also conclude that
$\No\Nt\rightarrow 0$ and $\No\rightarrow 0$. By (\ref{eq:taubc}), we
have $\Sp\rightarrow -1$. Using the constraint (\ref{eq:taubc})
and (\ref{eq:whsudef}), we conclude that $q+2\Sp$ is integrable, so
that $\Nt=\Nth$ will converge to a finite non-zero value. 

2. \textit{Assume now} $\Sp(\tau_{0})\leq 0$, 
$\O(\tau_{0})<1$ \textit{and} $\O(\tau_{0})+\Sp^2(\tau_{0})<
1$. Observe that
\begin{equation}\label{eq:sppr2}
\Sp'=(1-\O-\Sp^2)(4-2\Sp)-\frac{3}{2}(2-\g)\O\Sp+9\No\Nt.
\end{equation}
As long as $\O+\Sp^2<1$, $\Sp$ 
decreases as we go backward in time by (\ref{eq:sppr2}). Then
$\Nt/\No$ will increase exponentially until $\O+\Sp^2=1$, by the
constraint, and $\Sp<0$. $\Box$

\begin{lemma}\label{lemma:taubcomp}
Consider a type IX solution with $\Sm=0$, $\Nt=\Nth$ and $2/3<\g<2$.
It is contained in a compact set for $\tau\leq 0$ and
$\No\Nt\rightarrow 0$.
\end{lemma}

\textit{Proof}. Note that $\No$ must be bounded for $\tau\leq 0$, as 
follows from Lemma \ref{lemma:B9N}, the fact that $\Nt=\Nth$, and the 
fact that $\No\Nt\Nth$ decreases backward in time.
To prove the first statement, assume the contrary. 
Then there is a sequence
$\tau_{k} \rightarrow -\infty$ such that $\Nt(\tau_{k})\rightarrow
\infty$. We can assume $\Nt'(\tau_{k})\leq0$, and thus
\begin{equation}\label{eq:ntpr}
\frac{1}{2}(3\g-2)\O+2\Sp^2+2\Sp\leq0
\end{equation}
in $\tau_{k}$. Since $\No\Nt^2$ is decreasing as we go
backward, $\No$ and $\No\Nt$ evaluated at $\tau_{k}$ must go to 
zero. Thus $\O+\Sp^2-1$ will become arbitrarily small
in $\tau_{k}$ by (\ref{eq:taubc}). If $\O(\tau_{k})\geq 1 $ for all 
$k$, we get 
\[
\Sp(\tau_{k})\leq-\frac{1}{4}(3\g-2)
\]
by (\ref{eq:ntpr}), so that
\[
\Sp^2(\tau_{k})+\O(\tau_{k})\geq 1+\frac{1}{16}(3\g-2)^2,
\]
which is a contradiction. In other words, there is a $k$ such that
$\Sp(\tau_{k})\leq 0$, by (\ref{eq:ntpr}), and $\O(\tau_{k})<1$. We 
can then use Lemma \ref{lemma:taubconv} to arrive at a contradiction 
to the assumption that the solution is not contained in a compact set.

To prove the second part of the lemma, observe that $\No\Nt^2$
converges to zero, as follows from the existence of an $\alpha$-
limit point and Lemma \ref{lemma:n1n2n3tz}. Thus
\[
\No\Nt=\No^{1/2}[\No\Nt^2]^{1/2}\leq C[\No\Nt^2]^{1/2}\rightarrow 0.
\]
$\Box$

\begin{prop}\label{prop:taub9}
For a type IX solution with $\Sm=0$, $\Nt=\Nth$ and $2/3<\g<2$, either
the solution is contained in $\mathcal{F}_{\mathrm{IX}}$ or 
$\mathcal{P}_{\mathrm{IX}}$, or 
\[
\lim_{\tau\rightarrow -\infty}(\O,\Sp,\Sm,
\No,\Nt,\Nth)(\tau)=(0,-1,0,0,n_{2},n_{2})
\]
where $0<n_{2}<\infty$.
\end{prop}
\textit{Remark}. Compare with Proposition \ref{prop:limchar}. Observe
also that when $\Sp$ for the solution converges to $-1$, we approach
$\Sp=-1,\ \O=0$ from outside the parabola $\O+\Sp^2=1$, as follows from
the proof of Lemma \ref{lemma:taubconv}.

\textit{Proof}. Consider a solution which is not contained in 
$\mathcal{F}_{\mathrm{IX}}$ or $\mathcal{P}_{\mathrm{IX}}$.
By Lemma \ref{lemma:taubcomp}, there is an $\alpha$-limit point with
$\No\Nt=0$. We can assume it is not $P_{1}^{+}(II)$. We have the
following possibilities.

1. \textit{It is contained in} $\mathcal{F}_{\mathrm{I}}\cup
\mathcal{F}_{\mathrm{II}}\cup \mathcal{F}_{\mathrm{VII}_{0}}$.
Then $F$ is an $\alpha$-limit point. Since the solution is not 
contained in $\mathcal{F}_{\mathrm{IX}}$, we get a type I 
limit point which is not $F$, by Lemma \ref{lemma:falp}, and thus 
either $\Sp=-1$ or $\Sp=1$ as limit points, by Proposition 
\ref{prop:typeI}. The first alternative implies convergence to 
$\Sp=-1$, by Lemma \ref{lemma:taubconv}. If we have a type I 
$\a$-limit point with $\Sp=1$, we can apply the Kasner map by Proposition 
\ref{prop:BKL} in order to obtain a type I limit point with $\Sp=-1$. 

2. \textit{The limit point is of type I}. This possibility can be
dealt with as above.
 
3. \textit{It is of type II}. We can assume that it is not
$P_{1}^{+}(II)$, by Lemma \ref{lemma:palp}, and that it is not 
contained in $\mathcal{F}_{\mathrm{II}}$. Thus we get $\Sp=-1$ on the 
Kasner circle as an $\alpha$-limit point, by Proposition 
\ref{prop:typeII}, and thus as above convergence to $\Sp=-1$. 

4. \textit{The limit point is of type VI}$I_{0}$. We can assume
$\Sp\neq 0$. If $\Sp<0$, we can apply Lemma \ref{lemma:taubconv} again,
and if $\Sp>0$, we get $\Sp=1$ on the Kasner circle as an 
$\alpha$-limit point, by Proposition \ref{prop:typeVII0t}, a case 
which can be dealt with as above. $\Box$

\section{Oscillatory behaviour}\label{section:osc}

It will be necessary to consider Bianchi IX solutions to 
(\ref{eq:whsu})-(\ref{eq:constraint}) under circumstances such
that the behaviour is oscillatory. This section provides the
technical tools needed.

Let $g$ be a function,
\begin{equation}\label{eq:Adef}
A=\left( \begin{array}{rr}
                0 & g \\
                -g & 0
            \end{array}
     \right),
\end{equation}
and  $\tbx=(\tx,\ty)^{t}$ satisfy
\[
\tbx'=A\tbx+\epsilon,
\]
where $\epsilon$ is some vector valued function.

\begin{lemma}\label{lemma:oscest}
Let $\phi_{0}$ be such that
$(\sin(\phi_{0}),\cos(\phi_{0}))$ and $(\tx (\tau_{0}),
\ty (\tau_{0}))$ are parallel. Define 
\begin{equation}
\xi (\tau)=\int_{\tau_{0}}^{\tau}g(s)ds+\phi_{0}\label{eq:xi}
\end{equation}
and
\begin{equation}
\bx(\tau)=\left(\begin{array}{c}
    x(\tau)\\
    y(\tau)
          \end{array}
    \right)=
\left(\begin{array}{c}
 \sin\xi(\tau)\\
 \cos\xi(\tau)
      \end{array}
\right).
\end{equation}
Then 
\begin{equation}\label{eq:errorest}
\| \tbx(\tau)-\bx(\tau)\|\leq
|1-(\tx^{2}(\tau_{0})+\ty^{2}(\tau_{0}))^{1/2}|+
|\int_{\tau_{0}}^{\tau}\|\epsilon (s)\|ds|.
\end{equation}
\end{lemma}
\textit{Proof}. Let
\[
\Phi=\left( \begin{array}{rr}
          y & -x \\
          x & y
         \end{array}
  \right).
\]
We have $[A,\Phi]=0$, $\Phi'=-A\Phi$ and $\bx'=A\bx$. We get
\[
(\Phi(\tbx-\bx))'=-A\Phi(\tbx-\bx)+\Phi(A(\tbx-\bx)+\epsilon)=
\Phi\epsilon.
\]
Thus
\[
(\tbx-\bx)(\tau)=\Phi^{-1}(\tau)\Phi(\tau_{0})(\tbx-\bx)(\tau_{0})
+\Phi^{-1}(\tau)\int_{\tau_{0}}^{\tau}\Phi(s)\epsilon(s)ds.
\]
But $\Phi$ takes values in $ SO(2)$ and the lemma follows. $\Box$

In order to prove the existence of an $\alpha$-limit point for 
Bianchi IX solutions, and that, generically, there is a limit
point on the Kasner circle, we need the following lemma.

\begin{lemma}\label{lemma:sppos}
Consider a Bianchi IX solution with $2/3<\g<2$.
Assume there is a sequence $\tau_{k}\rightarrow -\infty$ such that
$q(\tau_{k})\rightarrow 0$, and $\Nt(\tau_{k}),\ \Nth(\tau_{k})
\rightarrow \infty$, then for each $T$, there is a $\tau\leq T$
such that $\Sp(\tau)\geq 0$.
\end{lemma}
\textit{Proof}. Observe that by (\ref{eq:sppr0}), $q=0$ and 
$\Nt+\Nth\geq \No$ implies
$\Sp'\leq -2$. However, the only term appearing in the constraint 
which does not go to zero in $\tau_{k}$ is $(\Nt-\Nth)^2$, since the 
product $\No\Nt\Nth$ decreases as we go backward. Thus 
$|\Sm'(\tau_{k})| \rightarrow \infty$, and the behaviour is 
oscillatory. It is clear that $\Sp'$ could become positive during 
the oscillations, but only when $|\Sm|$ is big, so that we on the 
whole should move in the positive direction.

Assume there is a $T$ such that $\Sp(\tau)<0$ for all $\tau\leq T$.

We begin by examining the behaviour of different expressions in
the sets
\[
\mathcal{D}_{k}=\cup_{n=k}^{\infty}[\tau_{n}-1,\tau_{n}]
\]
and
\[
\mathcal{D}=\cup_{n=1}^{\infty}[\tau_{n}-1,\tau_{n}].
\]
Observe that by the fact that $(\O,\Sp,\Sm)$ are constrained to 
belong to a compact set during $(-\infty,0]$, according to 
Lemma \ref{lemma:B9N}, $\Nt$ and $\Nth$ go to infinity 
uniformly in $\mathcal{D}$ (by which we will mean the following):
\[
\forall M\ \exists K: k\geq K\Rightarrow N_{i}(\tau)\geq M\
\forall \tau\in\mathcal{D}_{k},\ i=2,3.
\] 
Thus $\No$ and $\No(\Nt+\Nth)$ go to zero uniformly in
$\mathcal{D}$. By (\ref{eq:whsu}), $\O$ also converges to zero 
uniformly in $\mathcal{D}$. Due to the 
constraint, we get a bound on $\Sm^2+\frac{3}{4}(\Nt-\Nth)^2$
in $\mathcal{D}$. Consider (\ref{eq:sppr0}). The last two terms
go to zero uniformly. If the first term is not negative,
$1-\O-\Sp^{2}-\Sm^{2}\leq 0$. By the constraint, it will then be
bounded by an expression that converges to zero uniformly in
$\mathcal{D}$. Thus,
for every $\delta>0$ there is a $K$ such that $k\geq K$ implies
$\Sp'\leq \delta$ in $\mathcal{D}_{k}$. Combining this with
the fact that $q(\tau_{k})\rightarrow 0$, and the assumption that
$\Sp(\tau)<0$ for $\tau\leq T$, we conclude that $\Sp$ converges
uniformly to zero in $\mathcal{D}$. 

Next, we use Lemma \ref{lemma:oscest} in order to approximate the
oscillatory behaviour. Define the functions
\begin{eqnarray*}
\tx & = & \frac{\Sm}{(1-\Sp^2)^{1/2}} \\
\ty & = & \frac{\sqrt{3}}{2}\frac{\Nt-\Nth}{(1-\Sp^2)^{1/2}}.
\end{eqnarray*}
We can apply Lemma \ref{lemma:oscest} with 
\[
g= -3(\Nt+\Nth)-2(1+\Sp)\tx\ty=g_{1}+g_{2}
\]
and $\epsilon_{x}$, $\epsilon_{y}$ given by (\ref{eq:exdef}) and
(\ref{eq:eydef}), cf. Lemma \ref{lemma:oscillation}. 
By the above, we conclude that $\tx$ and $\ty$ are
uniformly bounded on $\mathcal{D}_{k}$, if $k$ is great enough, and
that $\|\epsilon\|$ converges to zero uniformly on $\mathcal{D}$.
Let $\bx_{k}$ be the expression given by Lemma \ref{lemma:oscest},
with $\tau_{0}$ replaced by $\tau_{k}$ and $\phi_{0}$ by a suitable
$\phi_{k}$. Let $\delta>0$. By the 
above and $q(\tau_{k})\rightarrow 0$, we get
\begin{equation}\label{eq:xmtx}
\|(\tbx-\bx_{k})(\tau)\|\leq \delta,
\end{equation}
if $\tau\in [\tau_{k}-1,\tau_{k}]$, and $k$ is great enough.
In $[\tau_{k}-1,\tau_{k}]$, we thus have
\begin{equation}\label{eq:spwell}
\Sp' = -2+2x_{k}^{2}(1-\Sp^{2})+\rho_{k},
\end{equation}
where the error $\rho_{k}$ can be assumed to be arbitrarily small by
choosing $k$ great enough, cf. (\ref{eq:sppr0}). 

Let 
\[
\xi_{k}(\tau)=\int_{\tau_{k}}^{\tau}g(s)ds+\phi_{k}
\]
be as in (\ref{eq:xi}). Since $\Nt+\Nth$ goes to infinity
uniformly, $[\tau_{k}-1,\tau_{k}]$ can be assumed to contain
an arbitrary number of periods of $\xi_{k}$, if $k$ is great 
enough. Thus, we can assume the existence of $\tau_{1,k},
\tau_{2,k}\in [\tau_{k}-1,\tau_{k}]$, such that $\tau_{2,k}-
\tau_{1,k}\geq 1/2$ and $\xi_{k}(\tau_{1,k})-\xi_{k}(\tau_{2,k})$
is an integer multiple of $\pi$. Let
$[\tau_{1},\tau_{2}]\subseteq [\tau_{1,k},
\tau_{2,k}]$ satisfy $\xi_{k}(\tau_{1})-\xi_{k}(\tau_{2})=\pi$. 
We can assume $\tau_{2}-\tau_{1}$ to be arbitrarily small by
choosing $k$ great enough. Considering (\ref{eq:whsu}), and using
the fact that $q$ is bounded, we conclude that $\Nt+\Nth$ cannot 
change by more than a factor arbitrarily close to one during
$[\tau_{1},\tau_{2}]$. Since the expression involving $\Nt+\Nth$
dominates $g$, we conclude that
\[
-\frac{3}{4}g(\tau_{\mathrm{max}})\leq -g(\tau_{\mathrm{min}}),
\]
where $\tau_{\mathrm{max}}$ and $\tau_{\mathrm{min}}$ correspond
to the maximum and the minimum of $-g$ in $[\tau_{1},\tau_{2}]$. 
Estimate
\[
\int_{\tau_{1}}^{\tau_{2}} 2x_{k}^{2}(1-\Sp^{2})ds=
\int_{\xi_{k}(\tau_{1})}^{\xi_{k}(\tau_{2})}
\frac{2x_{k}^{2}(1-\Sp^{2})}{g}d\eta=
-\int_{\xi_{k}(\tau_{1})-\pi}^{\xi_{k}(\tau_{1})}
\frac{2x_{k}^{2}(1-\Sp^{2})}{g}d\eta\leq
\]
\[
\leq -\frac{1}{g(\tau_{\mathrm{min}})}
\int_{\xi_{k}(\tau_{1})-\pi}^{\xi_{k}(\tau_{1})} 2\sin^{2} (\eta)
d\eta=-\frac{\pi}{g(\tau_{\mathrm{min}})}.
\]
We get 
\[
\tau_{2}-\tau_{1}=\int_{\xi_{k}(\tau_{1})}^{\xi_{k}(\tau_{2})}
\frac{1}{g}d\eta\geq
-\frac{\pi}{g(\tau_{\mathrm{max}})}\geq
\frac{3}{4}\int_{\tau_{1}}^{\tau_{2}} 2x_{k}^{2}(1-\Sp^{2})ds.
\]
Consequently, (\ref{eq:spwell}) yields
\[
\Sp(\tau_{2})-\Sp(\tau_{1})=-2(\tau_{2}-\tau_{1})+
\int_{\tau_{1}}^{\tau_{2}} 2x_{k}^{2}(1-\Sp^{2})ds+
\int_{\tau_{1}}^{\tau_{2}}\rho_{k} d\tau\leq
\]
\[
\leq-\frac{2}{3}(\tau_{2}-\tau_{1})+
\int_{\tau_{1}}^{\tau_{2}}\rho_{k} d\tau.
\]
Since $\xi_{k}(\tau_{1,k})-\xi_{k}(\tau_{2,k})$
corresponds to an integer multiple of $\pi$, we conclude that
\[
\Sp(\tau_{2,k})-\Sp(\tau_{1,k})\leq-\frac{2}{3}(\tau_{2,k}-
\tau_{1,k})+
\int_{\tau_{1,k}}^{\tau_{2,k}}\rho_{k} d\tau\leq
-\frac{1}{3}+\int_{\tau_{1,k}}^{\tau_{2,k}}\rho_{k} d\tau.
\]
However, the expressions on the far left can be assumed to be
arbitrarily small, and the integral of $\rho_{k}$ can be assumed to 
be arbitrarily small. We have a contradiction. $\Box$

\section{Bianchi IX solutions}\label{section:typeIX}
We first prove that there is an $\a$-limit point. If we assume
that there is no $\a$-limit point, we get the conclusion that 
the Euclidean norm $\|N\|$ of the vector $(\No,\Nt,\Nth)$ has to
converge to infinity, since $(\O,\Sp,\Sm)$ is constrained to
belong to a compact set to the past by Lemma \ref{lemma:B9N}.
In fact, Lemma \ref{lemma:B9N} yields more; it implies that
two $N_{i}$ have to be large at any given time. Since the product
$\No\Nt\Nth$ decays as we go backward, the third $N_{i}$ has to
be small. Sooner or later, the two $N_{i}$ which are large and the
one which is small have to be fixed, since a 'changing of roles'
would require two $N_{i}$ to be small, and thereby also the third
by Lemma \ref{lemma:B9N}, contradicting the fact that 
$\|N\|\rightarrow \infty$. Therefore, one can assume that two $N_{i}$
converge to infinity, and that the third converges to zero. More
precisely we have.

\begin{lemma}\label{lemma:aesetup}
Consider a Bianchi IX solution. 
If $\|N\|\rightarrow \infty$, we can, by applying the symmetries to
the equations, assume that $\Nt,\ \Nth
\rightarrow \infty$ and $\No,\ \No(\Nt+\Nth)\rightarrow 0$.
\end{lemma}
\textit{Proof}. As in the vacuum case, see \cite{jag}. $\Box$

\begin{lemma}\label{lemma:aexists}
A Bianchi IX solution with $2/3<\g<2$ has an $\alpha$-limit point. 
\end{lemma}

\textit{Proof}. If the solution is of Taub type, we already know 
that it is true so assume not. We assume $\Nt,\ \Nth\rightarrow 
\infty$, since if this does not occur, there is an $\alpha$-limit 
point by Lemma \ref{lemma:B9N} and Lemma \ref{lemma:aesetup}. By
(\ref{eq:sppr0})
we have $\Sp'<0$ if $\Sp=0$ using the constraint (assuming
$\Nt+\Nth>3\No$). Thus, there is a $T$ such that if $\Sp$ attains
zero in $\tau\leq T$, it will be non-negative to the past, and thus 
$\Nt\Nth$ will 
be bounded to the past since $\Sp$ has to be negative for the product 
to grow. If there is a sequence $\tau_{k}\rightarrow -\infty$ such
that $q(\tau_{k})\rightarrow 0$, we can apply Lemma \ref{lemma:sppos}
to arrive at a contradiction. Thus there is an $S$ such that
\begin{equation}\label{eq:qgte}
q(\tau)\geq\epsilon>0
\end{equation}
for all $\tau\leq S$. 

Consider
\begin{equation}\label{eq:zmodef}
Z_{-1}=\frac{\frac{4}{3}\Sm^2+(\Nt-\Nth)^2}{\Nt\Nth}.
\end{equation}
The reason we consider this function is that the derivative is in a
sense almost negative, so that it almost increases as we go backward.
On the other hand, it converges to zero as $\tau\rightarrow -\infty$
by our assumptions. The lemma follows from the resulting
contradiction. We have 
\begin{equation}\label{eq:zmopr}
Z_{-1}'=\frac{h}{\Nt\Nth}=\frac{-\frac{16}{3}\Sm^2(1+\Sp)+
4\sqrt{3}\Sm(\Nt-\Nth)\No}{\Nt\Nth}.
\end{equation}
Letting
\[
f=\frac{4}{3}\Sm^2+(\Nt-\Nth)^2,
\]
we have, using the constraint,
\[
h\leq 4\Sm^2\No(\Nt+\Nth)+2\sqrt{3}\No f\leq
\No\Nt\Nth f
\]
for, say, $\tau\leq T'\leq S$. Thus 
\begin{equation}\label{eq:zmpr}
Z_{-1}'\leq \No\Nt\Nth Z_{-1}
\end{equation}
for all $\tau\leq T'$. Since $q\geq \epsilon>0$ for all 
$\tau\leq T'\leq S$ by (\ref{eq:qgte}), we get
\[
(\No\Nt\Nth)(\tau)\leq
(\No\Nt\Nth)(T')\exp [3\epsilon (\tau-T')]
\]
for $\tau\leq T'$.
Inserting this inequality in (\ref{eq:zmpr}) we can integrate
to obtain
\[
Z_{-1}(\tau)\geq Z_{-1}(T')\exp (-\frac{1}{3\epsilon}(\No\Nt\Nth)(T')
)>0
\]
for $\tau\leq T'$. But $Z_{-1}(\tau)\rightarrow 0$ as $\tau\rightarrow
-\infty$ by our assumption, and we have a contradiction. $\Box$

\begin{cor}\label{cor:parabcon}
Consider a Bianchi IX solution with $2/3<\g<2$. For all $\epsilon>0$,
there is a $T$ such that 
\[
\O+\Sp^2+\Sm^2\leq 1+\epsilon
\]
for all $\tau\leq T$. Furthermore
\[
\lim_{\tau\rightarrow -\infty}(\No\Nt\Nth)(\tau)=0.
\]
\end{cor}
\textit{Proof}. As in the vacuum case, see \cite{jag}. The second
 part follows from Lemma \ref{lemma:n1n2n3tz} and Lemma
 \ref{lemma:aexists}. $\Box$

\begin{prop}\label{prop:alke}
A generic Bianchi IX solution with $2/3<\g<2$ has an 
$\alpha$-limit point on the Kasner circle.
\end{prop}

\textit{Proof}. Observe that by Lemma \ref{lemma:aexists} and
Corollary \ref{cor:parabcon}, there is an $\a$-limit point of type
I, II or VI$\mathrm{I}_{0}$.

1. \textit{First we prove that we can assume the
$\alpha$-limit point to be a type VI$I_{0}$ point with
$\No=0,\ 0<\Nt=\Nth,\ \O=0,\ \Sm=0$ and $\Sp=-1$}.

a. If there is an $\alpha$-limit point in $\mathcal{F}_{\mathrm{I}}$, 
$\mathcal{F}_{\mathrm{II}}$ or $\mathcal{F}_{\mathrm{VII}_{0}}$,
$F$ is a limit point, but then there is an
$\alpha$-limit point on the Kasner circle, by Lemma \ref{lemma:falp}
and Proposition \ref{prop:typeI}.

b. Assume there is an $\alpha$-limit point in
$\mathcal{P}_{\mathrm{VII}_{0}}$, or that one of $P_{i}^{+}(II)$ is an 
$\a$-limit point. Then there is a limit point of type II which is not
$P_{i}^{+}(II)$, by Lemma \ref{lemma:palp}, and we can assume it
does not belong to
$\mathcal{F}_{\mathrm{II}}$. We thus get an $\alpha$-limit point on 
the Kasner circle by Proposition \ref{prop:typeII}. 

c. Consider the complement of the above. We have an $\alpha$-limit 
point of type I, II or VI$\mathrm{I}_{0}$ which is generic or 
possibly of Taub type. If the limit point is of type I or II, we get
an $\alpha$-limit point on the Kasner circle by Proposition 
\ref{prop:typeI} and Proposition \ref{prop:typeII}. If the limit 
point is a non-Taub type VI$\mathrm{I}_{0}$ point, we get an 
$\alpha$-limit point on the Kasner
circle by Proposition \ref{prop:typeVII0}. Assume it is of Taub type 
with $\Sm=0$, $\Nt=\Nth$. By Proposition \ref{prop:typeVII0t}, we can 
assume that we have an $\alpha$-limit point of the type mentioned.

2. We construct an $\alpha$-limit point on the Kasner circle
given an $\alpha$-limit point as in 1. 
Since the solution is not of Taub type,
we must leave a neighbourhood of the point $(\Sp,\Sm)=(-1,0)$. If
$\Nt$ and $\Nth$ evaluated at the times we leave do not go to 
infinity, we are done. The reason is that we
can choose the neighbourhood to be so small that $\O$ and $\No$
decrease exponentially in it, see (\ref{eq:whsu}). If $\Nt(t_{k})$ or 
$\Nth(t_{k})$ is bounded, we  get a vacuum Bianchi VI$\mathrm{I}_{0}$ 
$\alpha$-limit point which is not of Taub-type by choosing a suitable 
subsequence (if we get a type I or II point we are done, see the above
arguments). By Proposition \ref{prop:typeVII0}, we then get an 
$\alpha$-limit point on the Kasner circle. Thus, we can assume the 
existence of a sequence $t_{k}\rightarrow -\infty$ such that 
$\Nt(t_{k})$ and $\Nth(t_{k})$ go to infinity. 

There are two problems we have to confront. First of all $\Nt$ and
$\Nth$ have to decay from their values in $t_{k}$ in order for us to
get an $\alpha$-limit point. Secondly, and more importantly, we need
to see to it that we do not get an $\alpha$-limit point of the same
type we started with. Let us divide the situation into two cases.

a. Assume that for each $t_{k}$ there is an $s_{k}\leq t_{k}$ 
such that $\Sp(s_{k})=0$. Observe that when $\Sp=0$, we have
\[
\Sp'\leq \frac{1}{2}\No(9\No-3\Nt-3\Nth)
\]
by the constraint (\ref{eq:constraint}), and (\ref{eq:sppr0}). 
Thus, we can assume that we have $3\No\geq \Nt+\Nth$ in $s_{k}$, since 
there is an $\alpha$-limit point with 
$\Sp=-1$. Thus there must be an $r_{k}\leq t_{k}$ such that, at
$r_{k}$, either  $\No=\Nt<\Nth$, $\No=\Nth<\Nt$ or 
$\No<\Nt$, $\No<\Nth$ and $3\No\geq \Nt+\Nth$. One of these
possibilities must occur an infinite number of times. The first two
possibilities yield a type I or II limit point, and the last a type I 
limit point because, of the fact that $\No\Nt\Nth\rightarrow 0$ and 
Lemma \ref{lemma:B9N}. As above, we get an $\a$-limit point on the 
Kasner circle.

b. Assume there is a $T$ such that $\Sp(\tau)< 0$ for all 
$\tau\leq T$. Then $\No\rightarrow 0$, since $\No(t_{k})\rightarrow 0$,
and $\Sp<0$ implies that $\No$ is monotone. 
Assume there is a sequence $\tau_{k}
\rightarrow -\infty$ such that $\Nt$ or $\Nth$ evaluated at it goes 
to zero. Then we get an $\alpha$ limit point of type I or II, a 
situation we may deal with as above. Thus we may assume 
$N_{i}\geq \epsilon>0$, $i=2,3$ to the past
of $T$. Similarly to the proof of the existence of an $\alpha$-limit
point, we have
\[
Z_{-1}'\leq c_{\epsilon}\No\Nt\Nth Z_{-1}.
\]
If there is an $S$ and a $\xi>0$ such that $q(\tau)\geq \xi>0$ for all
$\tau\leq S$, we get a contradiction as in the proof of Lemma
\ref{lemma:aexists}, since
$(\Nt\Nth)(t_{k})\rightarrow \infty$.
Thus there exists a sequence $\tau_{k}\rightarrow -\infty$ such that 
$q(\tau_{k})\rightarrow 0$. If $\Nt(\tau_{k})$ or $\Nth(\tau_{k})$
contains a bounded subsequence, we may refer to possibilities already
handled. By Lemma \ref{lemma:sppos}, we get $\Sp\geq 0$, a
contradiction. $\Box$

\section{Control over the density parameter}\label{section:ocon}

The idea behind the main argument is to use the existence of an 
$\alpha$-limit point on the Kasner circle to obtain a contradiction
to the assumption that the solution does not converge to the closure
of the set of vacuum type II points. The function 
\[
d=\O+\No\Nt+\Nt\Nth+\Nth\No
\]
is a measure of the distance from the attractor. We can consider
$d$ to be a function of $\tau$, if we evaluate it at a generic
Bianchi IX solution. If $\tau_{k}\rightarrow -\infty$ yields the
$\alpha$-limit point on the Kasner circle, then $d(\tau_{k})
\rightarrow 0$. If $d$ does not converge to zero, then it must grow 
from an arbitrarily small value up to some fixed number, say
$\delta>0$, as we go backward. In the contradiction argument, it is 
convenient to know that the growth occurs only in the sum of products 
of the $N_{i}$, and that during the growth one can assume $\O$ to 
be arbitrarily small. The following proposition achieves this goal,
assuming $\delta$ is small enough, which is not a restriction. The
proof is to be found at the end of this section.

\begin{figure}[hbt]
  \centerline{\psfig{figure=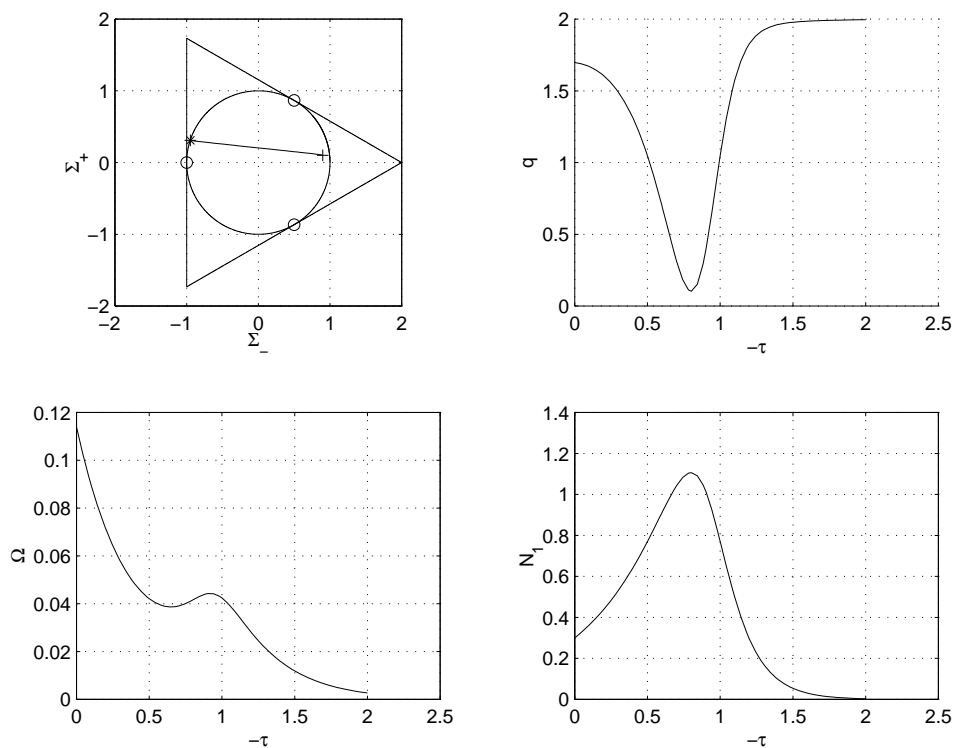,height=10cm}}
  \caption{Part of a type IX solution.}
\label{fig:ocon}
\end{figure}

\begin{prop}\label{prop:ocon}
Consider a Bianchi IX solution with $2/3<\g<2$.
There exists an $\epsilon>0$ such that if 
\begin{equation}\label{eq:sop}
\No\Nt+\Nt\Nth+\No\Nth\leq \epsilon
\end{equation}
in $[\tau_{1},\tau_{2}]$, then
\[
\O\leq c_{\g}\O(\tau_{2})
\]
in $[\tau_{1},\tau_{2}]$ if $\O(\tau_{2})\leq\epsilon$. Here 
$c_{\g}>0$ only depends on $\g$.
\end{prop}
The idea of the proof is the following. If the sum of product of 
the $N_{i}$ and $\O$ are small, the solution should behave in the
following way. If all the $N_{i}$ are small, then we are close to
the Kasner circle and $\O$ decays exponentially. One of the $N_{i}$ 
may become large alone, and then
$\O$ increases, but it can only be large for a short period of 
time. After that it must decay until some other $N_{i}$ becomes
large. But this process of the $N_{i}$ changing roles takes a long
time, and most of it occurs close to the Kasner circle, where $\O$
decays exponentially. Thus, $\O$ may increase by a certain factor, but 
after that it must decay by a larger factor until it can increase
again, hence the result. Figure \ref{fig:ocon} illustrates the 
behaviour. 

We divide the proof into lemmas, and begin by making the statement
that $\O$ decays exponentially close to the Kasner circle more
precise.
 
\begin{lemma}\label{lemma:dec}
Consider a Bianchi IX solution with $2/3<\g<2$.
If 
\[
\Sp^2+\Sm^2\geq \frac{1}{8}(3\g+2)
\]
in an interval $[s_{1},s_{2}]$, then
\[
\O(s)\leq \O(s_{2})e^{-\alpha_{\g}(s_{2}-s)}
\]
for $s\in[s_{1},s_{2}]$, where
\[
\alpha_{\g}=\frac{3}{2}(2-\g).
\]
\end{lemma}
\textit{Proof}. Observe that
\begin{equation}\label{eq:opr1}
\O'\geq 4[\Sp^2+\Sm^2-\frac{1}{4}(3\g-2)]\O,
\end{equation}
so that under the conditions of the lemma 
\[
\O'\geq\alpha_{\g}\O.
\]
The conclusion follows. $\Box$

Next, we prove that if the $N_{i}$ all stay sufficiently small under
a condition as in (\ref{eq:sop}) and $\O$ starts out small, then $\O$ 
will remain small.

\begin{lemma}\label{lemma:oinit}
Consider a Bianchi IX solution with $2/3<\g<2$.
There is an $\epsilon>0$ such that if
\begin{eqnarray}
&\frac{3}{4}N_{i}^2  \leq  \frac{1}{8}(6-3\g)\\
&\No\Nt+\Nt\Nth+\No\Nth  \leq  \epsilon \label{eq:nn}
\end{eqnarray}
in an interval $[s_{1},s_{2}]$, and $\O(s_{2})\leq \epsilon$,
then $\O(s)\leq\O(s_{2})$ for all $s\in[s_{1},s_{2}]$. 
\end{lemma}

\textit{Proof}. Let
\[
\mathcal{E}=\{\tau\in [s_{1},s_{2}] : t\in [\tau,s_{2}]
\Rightarrow \O(t)\leq \O(s_{2})\}.
\]
Let $\tau\in\mathcal{E}$, $\tau>s_{1}$. There must be two $N_{i}$, 
say $\Nt$ and 
$\Nth$, such that $\Nt\leq \epsilon^{1/2}$ and $\Nth\leq 
\epsilon^{1/2}$ in $\tau$,  by (\ref{eq:nn}). By the constraint
(\ref{eq:constraint}) and (\ref{eq:nn}), we have in $\tau$,
\[
\Sp^2+\Sm^2\geq 1-\frac{3}{4}\No^2-\O-h_{1}\geq
\frac{1}{8}(3\g+2)-4\epsilon,
\]
so that assuming $\epsilon$ small enough depending only on $\g$,
we have $\O'(\tau)>0$, cf. (\ref{eq:opr1}).
Thus there exists an $s<\tau$ such that $s\in\mathcal{E}$. In other
words, $\mathcal{E}$ is an open, closed, and non-empty subset of 
$[s_{1},s_{2}]$, so that $\mathcal{E}=[s_{1},s_{2}]$.
$\Box$

The next lemma describes the phase during which $\O$ may increase.

\begin{lemma}\label{lemma:oinc}
Consider a Bianchi IX solution with $2/3<\g<2$.
There is an $\epsilon>0$ such that if 
\begin{eqnarray}
&\frac{3}{4}N_{1}^2  \geq  \frac{1}{8}(6-3\g)\\
&\No\Nt+\Nt\Nth+\No\Nth  \leq  \epsilon \label{eq:nn2}
\end{eqnarray}
in $[s_{1},s_{2}]$, and $\O(s_{2})\leq \epsilon$, then
$s_{2}-s_{1}\leq c_{1,\g}$ and $\O(s)\leq c_{2,\g}\O(s_{2})$
for all $s\in [s_{1},s_{2}]$, where $c_{1,\g}$ and $c_{2,\g}$ are
positive constants depending on $\g$.
\end{lemma}

\textit{Proof}. Assume $\epsilon$ is small enough that 
\[
\frac{3}{4}\epsilon^{1/2}\leq\frac{1}{8}(6-3\g),
\]
so that $\No\geq \epsilon^{1/4}$ in $[s_{1},s_{2}]$. Assuming
$\epsilon<1$ we get $N_{i}\leq\epsilon^{1/2}$ in $[s_{1},s_{2}]$,
$i=2,3$. Use the constraint (\ref{eq:constraint}) 
to write
\begin{equation}\label{eq:omoss}
1-\O-\Sp^2-\Sm^2=\frac{3}{4}\No^2+h_{1}
\end{equation}
where $|h_{1}|\leq 3\epsilon$ by (\ref{eq:nn2}). Thus,
\[
1-\O-\Sp^2-\Sm^2\geq \frac{3}{4}\epsilon^{1/2}-3\epsilon,
\]
so that we may assume 
\begin{equation}\label{eq:osscon}
\O+\Sp^2+\Sm^2<1
\end{equation}
in $[s_{1},s_{2}]$.

We now compare the behaviour with a type II vacuum solution. By
(\ref{eq:sppr0}) and (\ref{eq:omoss}), we have
\begin{equation}\label{eq:sppr4}
\Sp'=-2(\frac{3}{4}\No^2+h_{1})(\Sp+1)-
\frac{3}{2}(2-\g)\O\Sp+\frac{9}{2}\No^2-
\end{equation}
\[
-\frac{9}{2}\No(\Nt+\Nth)
=\frac{3}{2}\No^2(2-\Sp)+h_{2}\O+h_{3},
\]
where $|h_{3}|\leq 17\epsilon$ and $|h_{2}|\leq 2$ in 
$[s_{1},s_{2}]$. Let $a_{\g}=(6-3\g)/4$. Then,
\[
\Sp(s_{2})-\Sp(s_{1})\geq a_{\g}(s_{2}-s_{1})+
\int_{s_{1}}^{s_{2}}(h_{2}\O+h_{3})dt.
\]
However,
\[
\O(s)\leq \O(s_{2})e^{-4(s-s_{2})}\leq \epsilon e^{-4(s-s_{2})}
\]
for all $s\in[s_{1},s_{2}]$, see (\ref{eq:whsu}). Thus, 
\[
|\int_{s_{1}}^{s_{2}}h_{2}\O ds|\leq \frac{1}{2}\O(s_{2})e^{4(s_{2}-
s_{1})}.
\]
We get
\[
\Sp(s_{2})-\Sp(s_{1})\geq a_{\g}(s_{2}-s_{1})-\frac{1}{2}\epsilon
e^{4(s_{2}-s_{1})}-17\epsilon (s_{2}-s_{1}).
\]
This inequality contradicts the statement that $s_{2}-s_{1}$ may
be taken equal to $4/a_{\g}$, by choosing $\epsilon$ small enough. 
We conclude that $s_{2}-s_{1}\leq 4/a_{\g}=c_{1,\g}$, and that we may
choose $c_{2,\g}=\exp(16/a_{\g})$. $\Box$

The following lemma deals with the decay in $\O$ that has to follow
an increase. The idea is that if $\No$ is on the boundary between big
and small, and its derivative is non-negative at a point, then it will
decrease as we go backward, and the solution will not move far from the
Kasner circle until one of the other $N_{i}$ has become large. That
takes a long time and $\O$ will decay.

\begin{lemma}\label{lemma:odec}
Consider a Bianchi IX solution such that $2/3<\g<2$.
There is an $\epsilon>0$ such that if
\begin{equation}\label{eq:nn3}
\No\Nt+\Nt\Nth+\Nth\No\leq \epsilon
\end{equation}
in $[s_{1},s_{2}]$,
\[
\frac{3}{4}\No^2(s_{2})=\frac{1}{8}(6-3\g),\ \No'(s_{2})\geq 0
\]
and $\O(s_{2})\leq c_{2,\g}\epsilon$, where $c_{2,\g}$ is the constant
appearing in Lemma \ref{lemma:oinc}, then $\O$ decays as we go
backward starting at $s_{2}$, until $s=s_{1}$, or we reach a point $s$
at which
\[
\O(s)\leq \frac{\O(s_{2})}{2c_{2,\g}}.
\]
\end{lemma}

\textit{Proof}. We begin by assuming that $\epsilon>0$ is a 
fixed number. As the proof progresses, we will restrict it
to be smaller than a certain constant depending on $\g$.
We could spell it out here, but prefer to add restrictions
successively. Let $\No\geq \epsilon^{1/4}$ in $[t_{1},s_{2}]$
and $\No(t_{1})=\epsilon^{1/4}$ or $t_{1}=s_{1}$, in case $\No$
does not attain $\epsilon^{1/4}$ in $[s_{1},s_{2}]$. As in the
proof of Lemma \ref{lemma:oinc}, we conclude that $N_{i}\leq
\epsilon^{1/2}$, $i=2,3$ in $[t_{1},s_{2}]$, and that we may assume
\begin{equation}\label{eq:osscon1}
\O+\Sp^2+\Sm^2<1.
\end{equation}

 The variables $(\O,\Sp,\Sm)$
have to belong to the interior of a paraboloid for $\No'$ to be 
negative. Since $\No'(s_{2})\geq 0$ we are on the boundary or outside 
the paraboloid. The boundary is given by $g=0$, where 
\[
g=\frac{1}{2}(3\g-2)\O+2\Sp^2+2\Sm^2-4\Sp.
\]
An outward pointing normal is given by $\nabla g$, where the 
derivatives are taken in the order: $\O$, $\Sp$ and $\Sm$. Let
\[
\mathcal{E}=\{\tau\in [t_{1},s_{2}] : t\in [\tau,s_{2}] \Rightarrow
\No'(t)\geq 0,\ \O(t)\leq c_{2,\g} \epsilon\}.
\]
Let $\tau\in \mathcal{E}$. By (\ref{eq:osscon1})
we get $q(\tau)<2$ and, as we are also outside the interior of the
paraboloid, $\Sp(\tau)\leq 1/2$.
For $\epsilon$, and thereby $\O$, small enough depending only on 
$\g$, we have
\[
\Sp'(\tau)\geq \epsilon^{1/2},
\]
cf. (\ref{eq:sppr4}). Using the above observations, we estimate
in $\tau$,
\[
\nabla g \cdot (\O',\Sp',\Sm')\leq C_{\g}\epsilon -\epsilon^{1/2},
\]
where $C_{\g}$ only depends on $\g$. For $\epsilon$ small enough, the
scalar product is negative. Thus, if $(\O(\tau),\Sp(\tau),\Sm(\tau))$
is on the surface of the paraboloid, the solution moves away from it 
as we go backward, so that $\No'\geq 0$ in $[s,\tau]$ for some 
$s<\tau$. If we are already outside the paraboloid, the existence of 
such an $s$ is guaranteed by less complicated arguments. As in the 
proof of Lemma \ref{lemma:oinit}, we get $\O'>0$
for $\epsilon$ small enough depending only on $\g$, so that 
$\mathcal{E}$ is open, closed and non-empty. Thus $\No$ decreases 
from $s_{2}$ to $t_{1}$ going backward. Now, 
\[
\Sp^2+\Sm^2\geq 1-\frac{3}{4}\No^2-\O-h_{1}\geq
\frac{1}{8}(3\g+2)-c_{2,\g}\epsilon-3\epsilon
\]
in $[t_{1},s_{2}]$, so that
\begin{equation}\label{eq:odec}
\O(t_{1})\leq \O(s_{2})e^{-(2-\g)(s_{2}-t_{1})},
\end{equation}
by an argument similar to Lemma \ref{lemma:dec}, if $\epsilon$ is
small enough. We can assume $\epsilon$ is small enough that the
time required for $\No$ to decrease to $\epsilon^{1/4}$ is great
enough that if $t_{1}\neq s_{1}$, then the conclusion of the lemma
follows by (\ref{eq:odec}). $\Box$

\textit{Proof of Proposition \ref{prop:ocon}}. Assume $\epsilon$
is small enough that all the conditions of Lemma 
\ref{lemma:oinit}-\ref{lemma:odec} are fulfilled. We divide the
interval $[\tau_{1},\tau_{2}]$ into suitable subintervals, such that
we may apply the above lemmas to them. If 
\begin{equation}\label{eq:nicon}
\frac{3}{4}N_{i}^2\leq \frac{1}{8}(6-3\g)
\end{equation}
in $\tau_{2}$ for $i=1,2,3$, then we let $t_{2}\in
[\tau_{1},\tau_{2}]$ be the smallest member of the interval such that
(\ref{eq:nicon}) holds in all of $[t_{2},\tau_{2}]$. Otherwise,
we chose $t_{2}=\tau_{2}$. Either $t_{2}=\tau_{1}$ or 
$3\No^2(t_{2})/4\geq(6-3\g)/8$, by a suitable permutation of the
variables. If $t_{2}\neq\tau_{1}$, let $t_{1}$ be the smallest
member of $[\tau_{1},t_{2}]$ such that $3\No^2/4\geq (6-3\g)/8$
in $[t_{1},t_{2}]$.

Because of Lemma \ref{lemma:oinit}, $\O$ decays in $[t_{2},\tau_{2}]$.
If $t_{2}=\tau_{1}$, we are done; let $c_{\g}=1$. Otherwise, we apply
Lemma \ref{lemma:oinc} to the interval $[t_{1},t_{2}]$ to conclude
that $\O(\tau)\leq c_{2,\g}\O(\tau_{2})$ in $[t_{1},\tau_{2}]$. If
$t_{1}=\tau_{1}$, we can choose $c_{\g}=c_{2,\g}$. Otherwise, we apply
Lemma \ref{lemma:odec} to $[\tau_{1},t_{1}]$. Either $\O$ decays until
we have reached $\tau_{1}$, or there is a 
point $s_{1}\in[\tau_{1},t_{1}]$ such that $\O(s_{1})\leq
\O(\tau_{2})/2$. By the proof of Lemma \ref{lemma:odec}, we can assume
that $\tau_{2}-s_{1}\geq 1$; some time has to elapse for the decay to
take place. 

Given an interval $[\tau_{1},\tau_{2}]$ as in the statement of the
proposition, there are thus two possibilities. Either $\O(\tau)\leq 
c_{2,\g}\O(\tau_{2})$ for all $\tau\in [\tau_{1},\tau_{2}]$ or
we can construct an $s_{1}\in [\tau_{1},\tau_{2}]$ such that
$\tau_{2}-s_{1}\geq 1$, $\O(s_{1})\leq
\O(\tau_{2})/2$, and $\O(\tau)\leq 
c_{2,\g}\O(\tau_{2})$ for all $\tau\in [s_{1},\tau_{2}]$. 
If the second possibility is the one that occurs, we can apply
the same argument to $[\tau_{1},s_{1}]$, and by repeated application,
the proposition follows. $\Box$

\begin{cor}\label{cor:otz}
Consider a Bianchi IX solution with $2/3<\g<2$. 
If 
\[
\lim_{\tau\rightarrow -\infty}(\No\Nt+\Nt\Nth+\No\Nth)= 0
\]
and there is a sequence $\tau_{k}\rightarrow -\infty$ such that 
$\O(\tau_{k})\rightarrow 0$, then
\[
\lim_{\tau\rightarrow -\infty}\O(\tau)=0.
\]
\end{cor}

\section{Generic attractor for Bianchi IX solutions}
\label{section:attractor}

\begin{figure}[hbt]
  \centerline{\psfig{figure=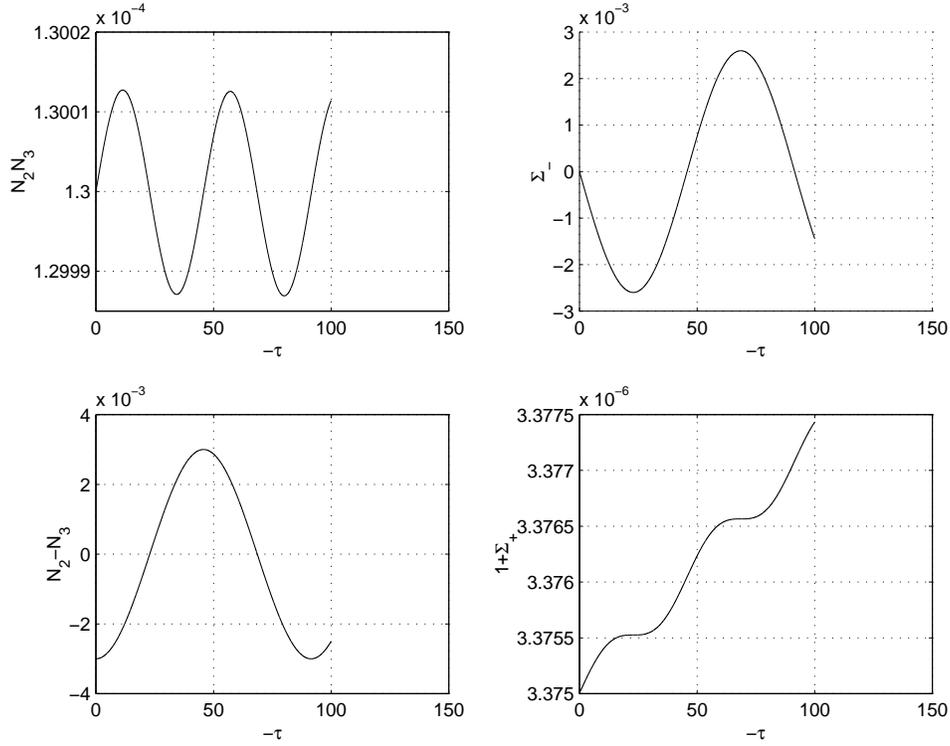,height=10cm}}
  \caption{Part of a Bianchi VI$\mathrm{I}_{0}$ vacuum solution.}
  \label{fig:osc}
\end{figure}

In this section, we prove that for a generic Bianchi IX solution, 
the closure of the set of type II vacuum points is an attractor, 
assuming $2/3<\g<2$. What we need to prove is that
\[
\lim_{\tau\rightarrow -\infty}(\O+\No\Nt+\Nt\Nth+\No\Nth)=0,
\]
since then we may for each $\epsilon>0$ choose a $T$ such that
at least two of the $N_{i}$ and $\O$ must be less than $\epsilon$
for $\tau\leq T$. The starting point is the existence of a limit 
point on the Kasner circle for a generic solution, given by 
Proposition \ref{prop:alke}. Since there is such a limit point, 
there is a sequence $\tau_{k} \rightarrow -\infty$ such that 
$N_{i}(\tau_{k})$ and $\O(\tau_{k})$ go to zero. If 
\begin{equation}\label{eq:hdef}
h=\No\Nt+\Nt\Nth+\No\Nth
\end{equation}
does not converge to zero, it must thus grow from an arbitrarily small
value up to some $\epsilon$. By choosing $\epsilon$ so that Proposition
\ref{prop:ocon} is applicable, we have control over $\O$. A few arguments
yield the conclusion that we may assume that it is the product 
$\Nt\Nth$ that grows, and that the growth occurs close to the special 
point $(\Sp,\Sm)=(-1,0)$. Close to this point, $\O$, $\No$ and 
$\No(\Nt+\Nth)$ decay exponentially, so as far as intuition goes, we 
may equate them with zero. We thus have a Bianchi VI$\mathrm{I}_{0}$ 
vacuum solution close to the special point $(-1,0)$. The 
behaviour of $\Nt\Nth$ will be oscillatory, and we may reduce the 
problem to one in which the product behaves essentially as a sine 
wave. However, by doing some technical estimates, one may see that 
one goes down going from top to top during the oscillation, and that 
that contradicts the assumed growth.
Figure \ref{fig:osc} illustrates the behaviour. It is a simulation of
part of a Bianchi VI$\mathrm{I}_{0}$ vacuum solution.

We begin by rewriting the solutions in a form that makes the 
oscillatory behaviour apparent. Consider a non Taub-NUT Bianchi IX 
solution in an interval such that $-1<\Sp<1$. Define the functions
\begin{eqnarray}
\tx & = & \frac{\Sm}{(1-\Sp^2)^{1/2}} \label{eq:tx}\\
\ty & = & \frac{\sqrt{3}}{2}\frac{\Nt-\Nth}{(1-\Sp^2)^{1/2}}
\label{eq:ty}.
\end{eqnarray}
The reason why these expressions are natural to consider is that, for
reasons mentioned above, $\No$, $\O$ and so forth may be considered to
be zero. In the situation we will need to consider $\Nt-\Nth$ and 
$\Sm$ will have much greater derivatives than $\Sp$, so that it is 
natural to consider $\tx$ and $\ty$ as sine and cosine, since the 
constraint essentially says $\tx^2+\ty^2=1$. Let
\begin{equation}\label{eq:gdef}
g= -3(\Nt+\Nth)-2(1+\Sp)\tx\ty=g_{1}+g_{2}.
\end{equation}
In our applications, $g_{1}$ will essentially be constant,
and $g_{2}$ will essentially be zero.

\begin{lemma}\label{lemma:oscillation}
The vector $\tbx=(\tx,\ty)^{t}$ satisfies
\[
\tbx'=A\tbx+\epsilon,
\]
where $A$ is defined as in (\ref{eq:Adef}), with $g$ as in
(\ref{eq:gdef}) and $\epsilon=(\epsilon_{x},\epsilon_{y})^{t}$,
where the components are given by (\ref{eq:exdef}) and
(\ref{eq:eydef}).
\end{lemma}

The error terms are
\begin{equation}\label{eq:exdef}
\epsilon_{x}=3\No\ty+(\frac{9}{2}\No(\No-\Nt-\Nth)-\frac{3}{2}
(2-\g)\O\Sp)\frac{\Sp\tx}{1-\Sp^2}-
\end{equation}
\[
-(\frac{3}{2}\No^2-3\No(\Nt+\Nth))\frac{\Sp\tx}{1-\Sp}-
\frac{3}{2}(2-\g)\O\tx-2(\frac{3}{4}\No^2-\frac{3}{2}\No(\Nt+\Nth))\tx
\]
and
\begin{equation}\label{eq:eydef}
\epsilon_{y}=[\frac{1}{2}(3\g-2)\O(1+\Sp)+
\frac{3}{2}(2-\g)\O+\frac{9}{2}\No(\No-\Nt-\Nth)]\frac{\ty\Sp}
{1-\Sp^2}
\end{equation}
\[
+\frac{1}{2}(3\g-2)\O\ty.
\]
It is clear that if we have a vacuum type VI$\mathrm{I}_{0}$ solution,
$\epsilon_{x}=\epsilon_{y}=0$, so that we may write 
$\tbx=(\sin (\xi(\tau)), \cos (\xi(\tau)))$, where $\xi$ is as in 
(\ref{eq:xi}). In our situation, there is an error term, but by the 
exponential decay mentioned above, it
only makes the technical details somewhat longer.

We begin by proving that we can assume that the growth occurs in
the product $\Nt\Nth$, and that $\O$ can be assumed to be negligible
during the growth. We also put bounds on $\Sp$. They constitute a
starting point for further restrictions.
The values of certain constants have been chosen for
future convenience.

The lemma below is formulated to handle more general situations than
the one above. One reason being the desire to prove uniform
convergence to the attractor. We will use the terminology that if
$x$ constitutes initial data for (\ref{eq:whsu})-(\ref{eq:constraint}),
then $\Sp(\tau,x)$ and so on will denote the solution of the equations
with initial value $x$ evaluated at $\tau$, assuming that $\tau$
belongs to the existence interval. We will use $\Phi(\tau,x)$ to 
summarize all the variables. The goal of this section is to
prove that the conditions of the lemma below are never met.

\begin{lemma}\label{lemma:setup1}
Let $2/3<\g<2$.
Consider a sequence $x_{l}$ of Bianchi IX initial data with all
$N_{i}>0$ and two sequences $s_{l}\leq\tau_{l}$ of real numbers, 
belonging to the existence interval corresponding to $x_{l}$, 
such that 
\begin{equation}\label{eq:dtz}
\lim_{l\rightarrow\infty}d(\tau_{l},x_{l})=0,
\end{equation}
where $d=\O+\No\Nt+\Nt\Nth+\No\Nth$, and
\begin{equation}\label{eq:hbig}
h(s_{l},x_{l})\geq \delta
\end{equation}
for some $\delta>0$ independent
of $l$. Then there is an $\epsilon>0$ and a $k_{0}$, such that for each 
$k\geq k_{0}$ there is an $l_{k}$, a symmetry operation on 
$\Phi(\cdot,x_{l_{k}})$, and an interval $[u_{k},v_{k}]$ belonging to
the existence interval of $\Phi(\cdot,x_{l_{k}})$, such that the 
transformed variables satisfy
\[
(\Nt\Nth)(u_{k},x_{l_{k}})=\epsilon,\ (\Nt\Nth)(v_{k},x_{l_{k}})
\leq\epsilon e^{-20k},\
\epsilon e^{-20k-1}\leq (\Nt\Nth)(\tau,x_{l_{k}})\leq\epsilon
\]
\begin{equation}\label{eq:n1n2n3}
\No(\tau,x_{l_{k}})\leq \epsilon \exp (-30k)\ and\
2\geq\Nt(\tau,x_{l_{k}}),\ \Nth(\tau,x_{l_{k}})\geq
\epsilon \exp (-25k)
\end{equation}
for $\tau\in [u_{k},v_{k}]$. Furthermore
\begin{equation}\label{eq:ospcon}
\O(\cdot,x_{l_{k}})\leq e^{-13k}\ \mathrm{and}\ -1<\Sp(\cdot,x_{l_{k}})\leq 0
\end{equation}
in $[u_{k},v_{k}]$. 
\end{lemma}

\textit{Remark}. Observe that for the main application of this lemma, 
the sequence $x_{l}$ will be independent of $l$.

\textit{Proof}. 
By (\ref{eq:dtz}) and (\ref{eq:hbig}), there is an $\epsilon>0$ 
such that for every $k$ there is a suitable 
$l_{k}$ and $u_{k}\leq v_{k}$ with $[u_{k},v_{k}]\subseteq
[s_{l_{k}},\tau_{l_{k}}]$ such that
\begin{equation}\label{eq:hcon1}
e^{-20k-1}\epsilon\leq h(\tau,x_{l_{k}})\leq 2\epsilon
\end{equation}
$h(u_{k},x_{l_{k}})=2\epsilon$, $h(v_{k},x_{l_{k}})=\exp (-20k-1)\epsilon$
where $\tau\in [u_{k},v_{k}]$.
We can also assume that 
\begin{equation}\label{eq:hcon2}
h(\tau,x_{l_{k}})\leq 2\epsilon
\end{equation}
for all $\tau\in [u_{k},\tau_{l_{k}}]$. Furthermore, we can
assume
\begin{equation}\label{eq:n123b}
(\No\Nt\Nth)(\cdot,x_{l_{k}})\leq \epsilon^2 \exp (-50k-1)/4
\end{equation}
in $[u_{k},\tau_{l_{k}}]$. The reason is that $d(\tau_{l},x_{l})$
converges to zero, so that $(\No\Nt\Nth)(\tau_{l},x_{l})$ also
converges to zero. Consequently, we can assume 
$(\No\Nt\Nth)(\tau_{l_{k}},x_{l_{k}})$ to be as small as we wish,
and thus we get (\ref{eq:n123b}) by the monotonicity of the product.
Since we may assume $\O(\tau_{l_{k}},x_{l_{k}})$ to be arbitrarily
small by (\ref{eq:dtz}), we may apply Proposition \ref{prop:ocon} in 
$[u_{k},\tau_{l_{k}}]$ by (\ref{eq:hcon2}), choosing $\epsilon$ small
enough. Thus we 
may assume $\O\leq \exp (-13k)$ in $[u_{k},v_{k}]$. From now on,
we consider the solution $\Phi(\cdot,x_{l_{k}})$ in the interval 
$[u_{k},\tau_{l_{k}}]$ and only use the observations above. 
To avoid cumbersome notation,
we will omit reference to the evaluation at
$x_{l_{k}}$. By (\ref{eq:hcon1}) and (\ref{eq:n123b}), we 
have in $[u_{k},v_{k}]$
\[
\epsilon e^{-20k-1}\leq h=\No\Nt\Nth(\frac{1}{\No}+
\frac{1}{\Nt}+\frac{1}{\Nth})\leq \frac{1}{4}\epsilon^2 e^{-50k-1}
(\frac{1}{\No}+\frac{1}{\Nt}+\frac{1}{\Nth}),
\]
so that 
\[
\frac{1}{\No}+\frac{1}{\Nt}+\frac{1}{\Nth}\geq \frac{4}{\epsilon}
e^{30k}.
\]
At a given $\tau\in [u_{k},v_{k}]$, one $N_{i}$, say $\No$, must be
smaller than $\epsilon \exp (-30k)$. If the second smallest is 
smaller than $\epsilon \exp (-25k)$, the largest cannot be bigger
than $2$, by Lemma \ref{lemma:B9N}, but that will contradict
$h\geq \epsilon \exp (-20k-1)$ if $k$ is great enough. Thus, if $\No$
is the smallest $N_{i}$ for one $\tau$, it is always the smallest.
We may thus assume 
\[
\No\leq \epsilon \exp (-30k)\ \mathrm{and}\ \Nt,\ \Nth\geq
\epsilon \exp (-25k)
\]
in $[u_{k},v_{k}]$. If $\epsilon$ is small enough, we can assume
$\Nt,\Nth\leq 2$ by Lemma \ref{lemma:B9N}. Thus,
\[
e^{-20k-1}\epsilon-4\epsilon e^{-30k}\leq \Nt\Nth \leq
2\epsilon+4\epsilon e^{-30k}.
\]
We may 
shift $u_{k}$ by adding a positive number to it so that 
\begin{equation}\label{eq:n23con}
(\Nt\Nth)(u_{k})=\epsilon\ \mathrm{and}\ (\Nt\Nth)(\tau)\leq \epsilon
\end{equation}
for $\tau\in [u_{k},v_{k}]$. We may also shift $v_{k}$ in the 
negative direction to achieve 
\[
(\Nt\Nth)(v_{k})\leq\epsilon e^{-20k},\ (\Nt\Nth)'(v_{k})<0\
\mathrm{and}\ (\Nt\Nth)(\tau)\geq \epsilon e^{-20k-1}
\]
for $\tau\in [u_{k},v_{k}]$. The condition on the derivative is 
there to get control on $\Sp$. 

We now establish rough control of $\Sp$. Since $(\Nt\Nth)'(v_{k})<0$,
$-1<\Sp(v_{k})<0$. Due to (\ref{eq:n1n2n3}), (\ref{eq:sppr0}) and
the constraint, $\Sp'<0$ if $\Sp=0$ or $\Sp=-1$. In other words,
$\Sp(w_{k})=0$ implies $\Sp\geq 0$ in $[u_{k},w_{k}]$. But if
$u_{k}<w_{k}$ then $\Sp(u_{k})>0$ so that 
$(\Nt\Nth)(u_{k})<(\Nt\Nth)(w_{k})$, contradicting
the construction as stated in (\ref{eq:n23con}). We thus have 
$\Sp\leq 0$ in $[u_{k},v_{k}]$. 
We also have $-1<\Sp$ in that interval. $\Box$

Below, we will omit reference to the evaluation at $x_{l_{k}}$ to 
avoid cumbersome notation, but it should be remembered that we
in general have a different solution for each $k$. Let
\[
r(\tau)=\int_{\tau}^{v_{k}}(q/2+\Sp)ds.
\]
Here we mean $q(s,x_{l_{k}})$ when we write $q$, and similarly for 
$\Sp$. Observe that $r$ depends on $k$, but that we omit reference
to this dependence.
All the information concerning the growth of $\Nt\Nth$ is contained 
in $r$, see (\ref{eq:whsu}), and this integral will be our main object
of study rather than the product $\Nt\Nth$. Let $[u_{k},v_{k}]$ be
an interval as in Lemma \ref{lemma:setup1}. Since 
\[
(\Nt\Nth)(v_{k})=e^{4r(u_{k})}(\Nt\Nth)(u_{k}),
\]
we have $r(u_{k})\leq -5k$. Let $u_{k}\leq\nu_{k}\leq \sigma_{k}
\leq \tau_{k}\leq r_{k}\leq v_{k}$. Starting at $u_{k}$, let
$\nu_{k}$ be the last point $r=-4k$, so that $r\geq -4k$ in 
$[\nu_{k},v_{k}]$. Furthermore, let $r\geq -k$ in $[r_{k},v_{k}]$ and 
finally, assume $r\leq -2k$ in $[\nu_{k},\tau_{k}]$. We also assume 
that $r$ evaluated at $r_{k}$, $\tau_{k}$, $\sigma_{k}$ and $\nu_{k}$
is $-k$, $-2k$, $-3k$ and $-4k$ respectively. See Table
\ref{table:setup}. Why? The interval we
will work with in the end is $[\sigma_{k},\tau_{k}]$, but the other
intervals are used to get control of the variables there. 
First of all, we want to get control of $\Sp$, and
the interval $[u_{k},\nu_{k}]$ together with the additional demand on
$\nu_{k}$ serves that purpose. The intervals at the other end, 
together with the associated demands, are there to yield us a 
quantitative statement of the intuitive idea that $\O$ and $\No$ are 
negligible relative to the other expressions of interest. Finally, we 
need to get quantitative bounds relating the different variables; as 
was mentioned earlier, the
main idea is to prove that $\Nt\Nth$ oscillates, but that it decreases
during a period. In order to prove the decrease, we need to have 
control over the relative sizes of different expressions, and 
$[\nu_{k},\sigma_{k}]$ is used to achieve the desired estimates.

From this point until the statement of Theorem \ref{thm:it},
we will assume that the conditions of Lemma \ref{lemma:setup1}
are fulfilled. We will use the consequences of this assumption,
as stated above, freely.

\begin{table}\label{table:setup}
\caption{Subdivision of the interval of growth.}
\begin{tabular}{@{}lc}
Interval & Bound on $r$  \\
$[\nu_{k},\sigma_{k}]$     & $-4k\leq r\leq -2k$ \\
$[\sigma_{k},\tau_{k}]$    & $-4k\leq r\leq -2k$  \\
$[\tau_{k},r_{k}]$         & $-4k\leq r$      \\
$[r_{k},v_{k}]$            & $-k\leq r$ \\
\end{tabular}
\end{table}

We improve the control of $\Sp$. Let us first give an
intuitive argument. Observe that under the
present circumstances, the solution is approximated by a 
Bianchi VI$\mathrm{I}_{0}$ vacuum solution. For such a 
solution, the function $Z_{-1}$, defined in (\ref{eq:zmodef}),
is monotone increasing going backwards. According to the Bianchi
VI$\mathrm{I}_{0}$ vacuum constraint, $Z_{-1}$ is proportional
to $(1-\Sp^2)/\Nt\Nth$. However, we know that $\Nt\Nth$ has to 
increase by a factor of $e^{20k}$ going from $v_{k}$ to $u_{k}$,
and consequently $1-\Sp^2$ has to increase by an even larger
factor. The only way this can occur, is if a large part of the
growth in $\Nt\Nth$ occurs when $\Sp$ is very close to $-1$.
Taking this into account, we see that the relevant variation in
$1-\Sp^2=(1-\Sp)(1+\Sp)$ occurs in the factor $1+\Sp$. 
Below, we will use the function $(1+\Sp)/\Nt\Nth$ instead of 
$Z_{-1}$. 
Let us begin by considering the vacuum case, in order 
to see the idea behind the argument, without the
technical difficulties associated with the non-vacuum case. 
We have
\begin{equation}\label{eq:spnn0}
\left( \frac{1+\Sp}{\Nt\Nth}\right)'<0
\end{equation}
in our situation, cf. Lemma \ref{lemma:dercom} and (\ref{eq:ospcon}). 
For $\tau\in [\nu_{k},v_{k}]$ we get
\[
0<1+\Sp(\tau)\leq (1+\Sp(u_{k}))\frac{(\Nt\Nth)(\tau)}{(\Nt\Nth)
(u_{k})}\leq e^{-4k}
\]
by our construction. 

Let us make some observations before we turn to the non-vacuum case.
First we analyze the derivative of $(1+\Sp)/\Nt\Nth$ in general. 
The estimates (\ref{eq:spnna}) and (\ref{eq:sppro}) will in fact
be important throughout this section.
\begin{lemma}\label{lemma:dercom}
Let $u_{k}$ and $v_{k}$ be as above. Then
\begin{equation}\label{eq:spnna}
\left( \frac{1+\Sp}{\Nt\Nth}\right)'\leq
\frac{-2[(1+\Sp)^2+\Sm^2](1+\Sp)+\frac{3}{2}(2-\g)\O}{\Nt\Nth}
\end{equation}
and 
\begin{equation}\label{eq:sppro}
\Sp'\leq\frac{3}{2}(2-\g)\O
\end{equation}
in the interval $[u_{k},v_{k}]$ for $k$ large enough.
\end{lemma}

\textit{Remark}. Observe that $1+\Sp>0$ in $[u_{k},v_{k}]$ by
(\ref{eq:ospcon}), so that the first term appearing in the
numerator of the right hand side of (\ref{eq:spnna}) has the
right sign.

\textit{Proof}. 
Using (\ref{eq:sppr0}), we have
\[
\left( \frac{1+\Sp}{\Nt\Nth}\right)'=
[-(2-2\O-2\Sp^2-2\Sm^2)(\Sp+1)-\frac{3}{2}(2-\g)\O\Sp+
\]
\[
+\frac{9}{2}
\No(\No-\Nt-\Nth)-(2q+4\Sp)(1+\Sp)](\Nt\Nth)^{-1}.
\]
Consider the numerator of the right hand side. The term involving
the $N_{i}$ has the right sign by (\ref{eq:n1n2n3}), and the terms
not involving $\O$ add up to the first term of the numerator of the
right hand side of (\ref{eq:spnna}). Let us consider the terms
involving $\O$. They are
\[
2\O(1+\Sp)-\frac{3}{2}(2-\g)\O(1+\Sp)+\frac{3}{2}(2-\g)\O-
(3\g-2)\O(1+\Sp)=
\]
\[
=-\frac{1}{2}(3\g-2)\O(1+\Sp)+\frac{3}{2}(2-\g)\O
\leq \frac{3}{2}(2-\g)\O
\]
proving (\ref{eq:spnna}). To prove (\ref{eq:sppro}), we observe that
by the constraint and the fact that $0<1+\Sp\leq 1$ in the interval
of interest, we have
\[
-(2-2\O-2\Sp^2-2\Sm^2)(\Sp+1)\leq 3\No(\Nt+\Nth)(1+\Sp)\leq
3\No(\Nt+\Nth).
\]
Inserting this inequality into (\ref{eq:sppr0}), we get 
\[
\Sp'\leq-\frac{3}{2}(2-\g)\O(1+\Sp)+\frac{3}{2}(2-\g)\O+
\frac{1}{2}\No(9\No-3\Nt-3\Nth)\leq\frac{3}{2}(2-\g)\O
\]
by (\ref{eq:n1n2n3}) and (\ref{eq:ospcon}) if $k$ is large enough,
proving (\ref{eq:sppro}). $\Box$

In the vacuum case, $\Sp$ is monotone in our situation,
see (\ref{eq:sppro}), but in the general case we have the
following weaker result.

\begin{lemma}\label{lemma:oed}
Consider an interval $[s,t]\subseteq [u_{k},v_{k}]$ such that
\[
\Sp^2\geq \frac{1}{8}(3\g+2).
\]
Then
\begin{equation}\label{eq:spainc}
(1+\Sp(t))-\O(t)\leq 1+\Sp(s)
\end{equation}
if $k$ is large enough.
\end{lemma}

\textit{Proof}. In $[s,t]$ we have
\[
\O'\geq\a_{\g}\O,
\]
where $\a_{\g}=3(2-\g)/2$, see the proof of Lemma \ref{lemma:dec}.
Thus,
\[
\O(u)\leq \O(t)\exp[\a_{\g}(u-t)]
\]
for all $u\in [s,t]$. Integrating (\ref{eq:sppro}) we get
(\ref{eq:spainc}). $\Box$

In connection with (\ref{eq:spnna}), the following lemma is
of interest.

\begin{lemma}\label{lemma:sp3o}
If $k$ is large enough and
\[
(1+\Sp(\tau))^3\geq e^{3k}\O(\tau)
\]
for some $\tau\in [u_{k},v_{k}]$, then
\[
(1+\Sp)^3\geq\frac{3}{4}(2-\g)\O
\]
in $[u_{k},\tau]$.
\end{lemma}

\textit{Proof}. If the solution is of vacuum type the lemma follows,
so assume $\O>0$. Let us first prove that $(1+\Sp(u))^3\geq
e^{k}\O(\tau)$ for $u\in [u_{k},\tau]$. Assume there is an
$s\in [u_{k},\tau]$ such that the reverse inequality holds.
Then there is a $t$ with $\tau\geq t\geq s$, such that $(1+\Sp)^3\leq
e^{3k}\O(\tau)$ in $[s,t]$, with equality at $t$. Because of 
(\ref{eq:ospcon}), Lemma \ref{lemma:oed} is applicable for $k$ 
large enough. Thus
\begin{equation}\label{eq:well}
e^{k}\O^{1/3}(\tau)-\O(t)\leq 1+\Sp(s)\leq e^{k/3}\O^{1/3}(\tau).
\end{equation}
However, by the proof of Lemma \ref{lemma:setup1}, Proposition
\ref{prop:ocon} is applicable in any subinterval of $[u_{k},v_{k}]$,
so that $\O(t)\leq c_{\g}\O(\tau)$. Substituting this into
(\ref{eq:well}), we get
\[
e^{k}\O^{1/3}(\tau)-c_{\g}\O(\tau)\leq e^{k/3}\O^{1/3}(\tau),
\]
which is impossible for $k$ large enough.

Thus we have, for $u\in [u_{k},\tau]$ and $k$ large enough,
\[
(1+\Sp(u))^{3}\geq e^{k}\O(\tau)\geq
e^{k}\frac{1}{\frac{3}{4}(2-\g)c_{\g}}\frac{3}{4}(2-\g)\O(u)\geq
\frac{3}{4}(2-\g)\O(u)
\]
where $c_{\g}$ is the constant appearing in the statement of 
Proposition \ref{prop:ocon}. The lemma follows. $\Box$

We now prove that we have control over $1+\Sp$ in $[\nu_{k},v_{k}]$.

\begin{lemma}\label{lemma:setup2}
Let $\nu_{k}$ and $v_{k}$ be as above. Then for $k$ large
enough,
\begin{equation}\label{eq:spcon}
0<1+\Sp<e^{-k}
\end{equation}
in $[\nu_{k},v_{k}]$.
\end{lemma}

\textit{Proof}. 
Assume $1+\Sp(\tau)\geq e^{-k}$ for some $\tau\in [\nu_{k},v_{k}]$. 
Because of (\ref{eq:ospcon}), we then conclude that Lemma
\ref{lemma:sp3o} is applicable, so that 
\[
\left( \frac{1+\Sp}{\Nt\Nth}\right)'\leq 0
\]
in $[u_{k},\tau]$ by (\ref{eq:spnna}). Thus 
\[
\frac{1+\Sp(u_{k})}{(\Nt\Nth)(u_{k})}\geq  
\frac{1+\Sp(\tau)}{(\Nt\Nth)(\tau)}\geq
\frac{e^{-k}}{(\Nt\Nth)(\tau)},
\]
but by our construction
\[
(\Nt\Nth)(\tau)=e^{4r(u_{k})-4r(\tau)}(\Nt\Nth)(u_{k})\leq
e^{-20k+16k}(\Nt\Nth)(u_{k}),
\]
so that
\[
e^{3k}\leq 1+\Sp(u_{k})\leq 1.
\]
The lemma follows. $\Box$

\begin{cor}\label{cor:control}
Let $\nu_{k}$ and $v_{k}$ be as above. For $k$ large enough,
\[
\O+\Sm^2+(1+\Sp)^2\leq 4e^{-k}
\]
in $[\nu_{k},v_{k}]$.
\end{cor}

\textit{Proof}. By (\ref{eq:n1n2n3}), we have
\[
\No(\Nt+\Nth)\leq 4\epsilon e^{-30k}
\]
in $[u_{k},v_{k}]$. This observation, the constraint, and
Lemma \ref{lemma:setup2} yield
\[
\O+\Sm^2\leq 1-\Sp^2+\frac{3}{2}\No(\Nt+\Nth)\leq 3e^{-k}
\]
in $[\nu_{k},v_{k}]$, for $k$ large enough. The corollary follows
using Lemma \ref{lemma:setup2}. $\Box$

The next thing to prove is that $\No$ and $\O$ are small
compared with $1+\Sp$. The fact that $r(r_{k})=-k$
will imply that the integral of $1+\Sp$ is large, but if $1+\Sp$ is
comparable with $\No$ or $\O$, it cannot be large since 
$\No$ and $\O$ decay exponentially. 

The reason $(1+\Sp)^{9}$ appears in the estimate (\ref{eq:onocon}) 
below is that the final argument will consist of an
estimate of an integral up to 'order of magnitude'. Expressions
of the form $(1+\Sp)^{n}$ and $(1+\Sp)^{m}/(\Nt+\Nth)^{l}$ will
will define what is 'big' and 'small', and here we see to it that
terms involving $\O$ and $\No$ are negligible in this order of
magnitude calculus. Finally, the factor $\exp (-3k)$ is there in
order for us to be able to ignore possible factors multiplying
expressions involving $\No$ and $\O$. We only turn up the number
$k$ and change $\exp (-3k)$ to $\exp (-2k)$ to eliminate constants
we do not want to think about; consider (\ref{eq:exdef}) and 
(\ref{eq:eydef}). 

\begin{lemma}\label{lemma:setup3}
Let $\nu_{k}$ and $\tau_{k}$ be as above. Then for $k$ large enough,
\begin{equation}\label{eq:onocon}
\O+\No+\No(\Nt+\Nth)\leq e^{-3k}e^{3b_{\g}(\tau-v_{k})}(1+\Sp)^{9}
\end{equation}
in $[\nu_{k},\tau_{k}]$ where $b_{\g}>0$. Furthermore,
\begin{equation}\label{eq:spnnder}
\left( \frac{1+\Sp}{\Nt\Nth}\right)'\leq
-2\Sm^2\frac{(1+\Sp)}{\Nt\Nth}
\end{equation}
in $[u_{k},\tau_{k}]$.
\end{lemma}

\textit{Proof}. Note that
\[
-\int_{r_{k}}^{v_{k}}(1+\Sp)d\tau\leq \int_{r_{k}}^{v_{k}}
(\Sp^2+\Sp)d\tau\leq \int_{r_{k}}^{v_{k}}(q/2+\Sp)d\tau=-k,
\]
so that
\begin{equation}\label{eq:spint}
k\leq \int_{r_{k}}^{v_{k}}(1+\Sp)d\tau.
\end{equation}
Let 
\[
\rho_{1}=\O+\No+\No(\Nt+\Nth).
\]
By the construction in Lemma \ref{lemma:setup1}, we may assume 
\[
\rho_{1}(v_{k})\leq e^{-12k}.
\]
Because of Corollary \ref{cor:control}, we have
\[
\rho_{1}(\tau)\leq e^{-12k}e^{4b_{\g}(\tau-v_{k})}
\]
for all $\tau\in [\nu_{k},v_{k}]$, where $b_{\g}>0$ is some constant
depending only on $\g$. Let 
\[
\rho_{2}(\tau)=e^{-9k}e^{b_{\g}(\tau-v_{k})}\geq
e^{3k}e^{-3b_{\g}(\tau-v_{k})}\rho_{1}(\tau).
\]
The assumption that $(1+\Sp)^{9}\leq \rho_{2}$ in $[r_{k},v_{k}]$
contradicts (\ref{eq:spint}). Thus there must be a 
$t_{0}\in[r_{k},v_{k}]$ such that $(1+\Sp(t_{0}))^{9}\geq 
\rho_{2}(t_{0})$. In the vacuum case, $1+\Sp$ increases as we go
backward, and $\rho_{2}$ obviously decreases, and thus we are in that case
able to conclude $(1+\Sp)^{9}\geq \rho_{2}$ in $[\nu_{k},r_{k}]$.
In the general case, we observe that $(1+\Sp(t_{0}))^3\geq
e^{3k}\O(t_{0})$ by the above constructions. We get 
\[
\left( \frac{1+\Sp}{\Nt\Nth}\right)'\leq -2\Sm^2\frac{(1+\Sp)}{\Nt\Nth}
\]
in $[u_{k},t_{0}]$, by combining Lemma \ref{lemma:sp3o} and
(\ref{eq:spnna}). Inequality (\ref{eq:spnnder}) follows. Thus, if
$\tau\in [\nu_{k},\tau_{k}]$, we have
\[
1+\Sp(\tau)\geq\frac{(\Nt\Nth)(\tau)}{(\Nt\Nth)(t_{0})}(1+\Sp(t_{0}))
\geq e^{4k}(1+\Sp(t_{0})).
\]
Consequently, we will have $(1+\Sp(\tau))^{9}\geq 
\rho_{2}(\tau)$, since $1+\Sp$ has increased from its value at
$t_{0}$ and $\rho_{2}$ has decreased. The lemma follows. $\Box$

Next we establish a relation between $1+\Sp$
and the product $\Nt\Nth$.  We prove that $(1+\Sp)/(\Nt\Nth)$ can be
chosen arbitrarily small in the interval $[\sigma_{k},\tau_{k}]$, by 
estimating it in $\nu_{k}$, and then comparing the
integral of $1+\Sp$ from $\nu_{k}$ to $\sigma_{k}$ with the integral
of $\Sm^2$ over the same interval. The following lemma is the 
starting point.

\begin{lemma}\label{lemma:setup4}
Let $\sigma_{k},\ \tau_{k}$ be as above. Then for $k$ large enough,
\begin{equation}\label{eq:spnest}
\frac{1+\Sp(\tau)}{(\Nt\Nth)(\tau)}\leq \frac{1}{\epsilon}
\exp(-2\int_{\nu_{k}}^{\sigma_{k}}\Sm^2 ds)
\end{equation}
if $\tau\in [\sigma_{k},\tau_{k}]$. Furthermore,
\[
\frac{1+\Sp(\tau)}{(\Nt\Nth)(\tau)}\leq\frac{1}{\epsilon}
\]
in $[u_{k},\tau_{k}]$. 
\end{lemma}

\textit{Proof}. The statement follows from (\ref{eq:spnnder}),
and the fact that 
\[
\frac{(1+\Sp)(u_{k})}{(\Nt\Nth)(u_{k})}\leq \frac{1}{\epsilon}.
\]
$\Box$

Considering the constraint, it is clear that $\Sm^2$ should be 
comparable with $1+\Sp$ when $\Nt-\Nth$ and $\Sm$ oscillate,
and thus the integral should be comparable with $k$,
cf. (\ref{eq:spint}). However, we have to work out the 
technical details.

We carry out the comparison between the integrals in three steps.
First, we estimate the error committed in viewing $\tx$ and
$\ty$ in (\ref{eq:tx}) and (\ref{eq:ty}) as sine and cosine. Then
we may, up to a small error, express the integral of $\Sm^2$ as the 
integral of $\sin^2(\eta/2)$, multiplied by some function  $f(\eta)$ by
changing variables. In order to make the comparison, we need to 
estimate the variation of $f$ during a period: the second step. The 
only expressions involved are $1+\Sp$ and $\Nt+\Nth$. The third step 
consists of making the comparison, using the information obtained in 
the earlier steps.

Let $\tx$, $\ty$, $g$, $g_{1}$
and $g_{2}$ be defined as in (\ref{eq:tx})-(\ref{eq:gdef}), and $\xi$,
$x$ and $y$ be defined as in the statement of Lemma \ref{lemma:oscest},
with $\tau_{0}$ replaced by $\tau_{k}$ and
$\phi_{0}$ by $\phi_{k}$. Observe that $x$, $y$ and $\xi$ in fact 
depend on $k$. We need to compare $x$ with $\tx$.

\begin{lemma}\label{lemma:setup5}
Let $\nu_{k}$ and $\tau_{k}$ be as above. Then for $k$ large enough,
\begin{equation}\label{eq:smosc}
|\Sm^2-(1-\Sp^2)x^2|\leq 12 e^{-2k}(1+\Sp)^{9}.
\end{equation}
in $[\nu_{k},\tau_{k}]$. Furthermore,
\begin{equation}\label{eq:txest}
|1-(\tx^{2}+\ty^{2})|\leq e^{-k}
\end{equation}
and
\begin{equation}\label{eq:tbxest}
\|\tbx-\bx\|\leq 3e^{-2k}(1+\Sp)^{8}
\end{equation}
in that interval.
\end{lemma}

\textit{Proof}. We
have
\[
|1-(\tx^2(\tau_{k})+\ty^2(\tau_{k}))^{1/2}|\leq |1-
\left(1+\frac{\frac{3}{2}\No(\Nt+\Nth)-\frac{3}{4}\No^2-\O}
{1-\Sp^2}\right)^{1/2}|\leq 
\]
\begin{equation}\label{eq:txcon}
\leq e^{-2k}(1+\Sp(\tau_{k}))^{8}
\end{equation}
by (\ref{eq:onocon}). Equation (\ref{eq:txest}) follows similarly.
By (\ref{eq:exdef}), (\ref{eq:eydef}),
(\ref{eq:onocon}) and (\ref{eq:txest}),  we have
\[
\|\epsilon(s)\|\leq 2b_{\g}e^{-2k}(1+\Sp(s))^{8}e^{3b_{\g}(s-v_{k})}
\]
for $k$ large enough.
Let us estimate how much $1+\Sp$ may decrease as we go backward in
time. By (\ref{eq:sppro}) and (\ref{eq:onocon}), we have
\[
(1+\Sp)'\leq \frac{3}{2}(2-\g)e^{-3k}e^{3b_{\g}(\tau-v_{k})}(1+\Sp)^9,
\]
so that if $[s,t]\subseteq [\nu_{k},\tau_{k}]$,
\begin{equation}\label{eq:psspinc}
1+\Sp(t)\leq \exp(\exp(-2k))(1+\Sp(s)),
\end{equation}
for $k$ large enough. Thus, for $\tau\leq \tau_{k}$, we get
\begin{equation}\label{eq:intest}
\int_{\tau}^{\tau_{k}}\|\epsilon(s)\|ds\leq
e^{-2k}(1+\Sp(\tau))^{8}.
\end{equation}
By (\ref{eq:errorest}), (\ref{eq:intest}), (\ref{eq:psspinc}) 
and (\ref{eq:txcon}), we thus have 
\[
\|\tbx-\bx\|\leq \frac{5}{2}e^{-2k}(1+\Sp)^{8}
\]
in $[\nu_{k},\tau_{k}]$, and (\ref{eq:tbxest}) follows. Since 
$|x|\leq 1$ and $|\tx|\leq 1.1$, 
cf. (\ref{eq:txest}), we have
\[
|\tx^2-x^2|\leq 6e^{-2k}(1+\Sp)^{8},
\]
so that
\[
|\Sm^2-(1-\Sp^2)x^2|\leq 12 e^{-2k}(1+\Sp)^{9}
\]
in the interval $[\nu_{k},\tau_{k}]$. $\Box$

Let us introduce
\begin{equation}\label{eq:etadef}
\eta(\tau)=2\xi(\tau)=2\int_{\tau_{k}}^{\tau}g(s)ds+2\phi_{k},
\end{equation}
where $g= -3(\Nt+\Nth)-2(1+\Sp)\tx\ty=g_{1}+g_{2}$. The reason we
study $\eta$ instead of $\xi$ is that the trigonometric expression 
we will be interested in is $\sin ^{2}(\xi)$, which has a period of 
length $\pi$, cf. Lemma \ref{lemma:setup5}. In the proof of Lemma
\ref{lemma:setup6}, it is shown that, in the interval
$[\nu_{k},\tau_{k}]$, the first term appearing in 
$g$ is much greater than the second. We can thus consider functions
of $\tau$ in the interval $[\nu_{k},\tau_{k}]$ to be functions of
$\eta$. We will mainly be interested in considering an interval
$[\eta_{0},\eta_{0}+2\pi]$ at a time, so that we will only need to
estimate the variation of the relevant expressions during one such
period.

\begin{lemma}\label{lemma:setup6}
Let $\eta_{1,k}=\eta(\sigma_{k})$ and $\eta_{2,k}=\eta(\nu_{k})$.
If $[\eta_{1},\eta_{1}+2\pi]\subseteq [\eta_{1,k},\eta_{2,k}]$ and
$\eta_{a},\eta_{b}\in[\eta_{1},\eta_{1}+2\pi]$, then for $k$ large
enough
\begin{equation}\label{eq:nrel}
e^{-6\pi/\epsilon}\leq\frac{(\Nt+\Nth)(\eta_{a})}{(\Nt+\Nth),
(\eta_{b})}\leq e^{6\pi/\epsilon},
\end{equation}
\begin{equation}\label{eq:sprel}
\frac{1}{2}\leq\frac{1+\Sp(\eta_{a})}{1+\Sp(\eta_{b})}\leq 2
\end{equation}
and
\begin{equation}\label{eq:gog}
|g_{1}|/2\leq |g|\leq 2|g_{1}|.
\end{equation}
\end{lemma}

\textit{Proof}. Because of Lemma \ref{lemma:setup4}, 
\begin{equation}\label{eq:spn}
\frac{1+\Sp}{\Nt+\Nth}\leq \frac{1+\Sp}{2(\Nt\Nth)^{1/2}}=
(\Nt\Nth)^{1/2}\frac{1+\Sp}{2\Nt\Nth}\leq
\end{equation}
\[
\leq\frac{1}{2\epsilon}\left(\frac{\Nt\Nth}{(\Nt\Nth)(u_{k})}\right)^
{1/2}(\Nt\Nth)^{1/2}(u_{k})\leq \frac{1}{2\epsilon^{1/2}}e^{-2k}
\]
in the interval $[\nu_{k},\tau_{k}]$.
By (\ref{eq:txest}) we may
assume $\tx^2+\ty^2\leq 2$ in $[\nu_{k},\tau_{k}]$. Combining this
fact with (\ref{eq:spn}) yields (\ref{eq:gog})
in $[\nu_{k},\tau_{k}]$. Thus, $d\eta/d\tau<0$ in that interval. 
We have
\[
|\frac{d(\Nt+\Nth)}{d\eta}|=|\frac{1}{2g}((q+2\Sp)(\Nt+\Nth)+
2\sqrt{3}\Sm (\Nt-\Nth))|\leq
\]
\[
\leq\frac{1}{2}(3\g-2)\O+|\Sp^2+(1-\Sp^2)\tx^2+\Sp|+
2\frac{|\tx\ty|}{|g|}(1-\Sp^2)
\leq 6(1+\Sp)+8\frac{1+\Sp}{\Nt+\Nth},
\]
so that
\[
|\frac{1}{\Nt+\Nth}\frac{d(\Nt+\Nth)}{d\eta}|\leq
6\frac{1+\Sp}{\Nt+\Nth}+8\frac{1+\Sp}{(\Nt+\Nth)^{2}}\leq
6\frac{1+\Sp}{\Nt+\Nth}+2\frac{1+\Sp}{\Nt\Nth}\leq\frac{3}{\epsilon}
\]
in $[\nu_{k},\sigma_{k}]$ for $k$ large, by Lemma
\ref{lemma:setup4} and (\ref{eq:spn}). If $\Nt+\Nth$ has a maximum
in $\eta_{\mathrm{max}}\in [\eta_{1},\eta_{1}+2\pi]$ and a minimum in
$\eta_{\mathrm{min}}$, we get
\[
\frac{(\Nt+\Nth)(\eta_{\mathrm{max}})}
{(\Nt+\Nth)(\eta_{\mathrm{min}})}\leq e^{6\pi/\epsilon},
\]
and (\ref{eq:nrel}) follows.
We also need to know how much $1+\Sp$ varies over one period. By
(\ref{eq:sppr0})
\[
(1+\Sp)'=(2\Sp^2+2\Sm^2-2)(1+\Sp)+f_{1},
\]
where $f_{1}$ is an expression that can be estimated as in
(\ref{eq:onocon}), so that we in $[\nu_{k},\tau_{k}]$ have 
\[
|\frac{(1+\Sp)'}{1+\Sp}|\leq 2(1-\Sp^2)(1+\tx^2)+(1+\Sp)\leq 13(1+\Sp),
\]
for $k$ large enough. Thus,
\begin{equation}\label{eq:dspdeta}
|\frac{1}{1+\Sp}\frac{d(1+\Sp)}{d\eta}|\leq \frac{10(1+\Sp)}{\Nt+\Nth},
\end{equation}
so that (\ref{eq:sprel}) holds
if $k$ is big enough and $|\eta_{a}-\eta_{b}|\leq 2\pi$ by 
(\ref{eq:spn}). $\Box$

\begin{lemma}\label{lemma:setup7}
Let $\sigma_{k}$ and $\tau_{k}$ be as above. Then if $k$ is 
large enough,
\[
\frac{1+\Sp}{\Nt\Nth}  \leq  \frac{1}{\epsilon}e^{-c_{\epsilon}k}
\]
in $[\sigma_{k},\tau_{k}]$ where $c_{\epsilon}>0$. 
\end{lemma}

\textit{Proof}. Observe that similarly to the proof of Lemma
\ref{lemma:setup3}, we have
\[
k\leq \int_{\nu_{k}}^{\sigma_{k}}(1+\Sp)d\tau=
\int_{\eta_{1,k}}^{\eta_{2,k}}\frac{(1+\Sp)}{-2g}d\eta.
\]
The contribution from one period in $\eta$ is negligible,
by (\ref{eq:spn}) and (\ref{eq:gog}). Compare this integral
with 
\[
\int_{\eta_{1,k}}^{\eta_{2,k}}\frac{\Sm^2}{-g}d\eta=
\int_{\eta_{1,k}}^{\eta_{2,k}}\frac{(1-\Sp^2)x^2}{-g}d\eta+
\int_{\eta_{1,k}}^{\eta_{2,k}}\frac{\Sm^2-(1-\Sp^2)x^2}{-g}d\eta=
I_{1,k}+I_{2,k}.
\]
Now,
\[
|I_{2,k}|\leq e^{-k}\int_{\eta_{1,k}}^{\eta_{2,k}}\frac{1+\Sp}{-g}
d\eta
\]
by (\ref{eq:smosc}). Consider an interval $[\eta_{1},\eta_{1}+2\pi]$.
Estimate, letting $\eta_{a}$ and $\eta_{b}$ be
the minimum and maximum of $\Sp$ respectively, and 
$\eta_{\mathrm{min}}$, 
$\eta_{\mathrm{max}}$ the min and max for $g_{1}$ in this interval,
\[
\int_{\eta_{1}}^{\eta_{1}+2\pi}\frac{(1-\Sp^2)x^2}{-g}d\eta\geq
\int_{\eta_{1}}^{\eta_{1}+2\pi}\frac{(1+\Sp)x^2}{-g}d\eta=
\int_{\eta_{1}}^{\eta_{1}+2\pi}\frac{(1+\Sp)\sin^2(\eta/2)}{-g}
d\eta\geq
\]
\[
\geq\pi\frac{1+\Sp(\eta_{a})}{2|g_{1}(\eta_{\mathrm{max}})|}\geq
\frac{\pi}{2}e^{-6\pi/\epsilon}\frac{1+\Sp(\eta_{a})}
{|g_{1}(\eta_{\mathrm{min}})|}
\geq
\frac{\pi}{4}e^{-6\pi/\epsilon}\frac{1+\Sp(\eta_{b})}{|g_{1}
(\eta_{\mathrm{min}})|}=
\]
\[
=\frac{1}{8}e^{-6\pi/\epsilon}
\int_{\eta_{1}}^{\eta_{1}+2\pi}\frac{1+\Sp(\eta_{b})}
{|g_{1}(\eta_{\mathrm{min}})|}d\eta\geq
\frac{1}{16}e^{-6\pi/\epsilon}
\int_{\eta_{1}}^{\eta_{1}+2\pi}\frac{1+\Sp(\eta)}
{-g(\eta)}d\eta,
\]
where we have used (\ref{eq:nrel}), (\ref{eq:sprel}) and
(\ref{eq:gog}). Assuming,
without loss of generality, that $\eta_{2,k}-\eta_{1,k}$ is an
integer multiple of $2\pi$, we get
\[
\int_{\nu_{k}}^{\sigma_{k}}2\Sm^2d\tau=
\int_{\eta_{1,k}}^{\eta_{2,k}}\frac{\Sm^2}{-g}d\eta=I_{1,k}+I_{2,k}
\geq
\left( \frac{1}{16}e^{-6\pi/\epsilon}-e^{-k}\right)
\int_{\eta_{1,k}}^{\eta_{2,k}}\frac{1+\Sp(\eta)}
{-g(\eta)}d\eta\geq
\]
\[
\geq\frac{1}{20}e^{-6\pi/\epsilon}\int_{\eta_{1,k}}^{\eta_{2,k}}
\frac{1+\Sp(\eta)}{-g(\eta)}d\eta=
\frac{1}{10}e^{-6\pi/\epsilon}
\int_{\nu_{k}}^{\sigma_{k}}(1+\Sp)d\tau
\geq\frac{k}{10}e^{-6\pi/\epsilon}=c_{\epsilon}k
\]
for $k$ large enough
and the lemma follows from (\ref{eq:spnest}). $\Box$

The following corollary summarizes the estimates that make the
order of magnitude calculus well defined.

\begin{cor}\label{cor:order}
Let $\sigma_{k}$ and $\tau_{k}$ be as above. Then
\begin{equation}\label{eq:spnn2}
\frac{1+\Sp}{(\Nt+\Nth)^{2}}\leq \frac{1}{\epsilon}e^{-c_{\epsilon}k},
\end{equation}
\begin{equation}\label{eq:spn2}
\frac{1+\Sp}{\Nt+\Nth}\leq e^{-2k}
\end{equation}
and
\begin{equation}\label{eq:gogrel}
1-e^{-2k}\leq \frac{g}{g_{1}}\leq 1+e^{-2k}
\end{equation}
in $[\sigma_{k},\tau_{k}]$ for $k$ large enough.
\end{cor}

\textit{Proof}. Observe that by Lemma \ref{lemma:setup7},
\[
\frac{1+\Sp}{(\Nt+\Nth)^2}\leq \frac{1+\Sp}{\Nt\Nth}\leq
\frac{1}{\epsilon}e^{-c_{\epsilon}k}
\]
and
\[
\frac{1+\Sp}{\Nt+\Nth}\leq \frac{1+\Sp}{2(\Nt\Nth)^{1/2}}
\leq \epsilon^{1/2}e^{-2k}\frac{1}{2\epsilon}e^{-c_{\epsilon}k}\leq,
e^{-2k}
\]
for $k$ large enough, cf. (\ref{eq:spn}). We have
\[
\frac{g}{g_{1}}=1+\frac{2(1+\Sp)\tx\ty}{3(\Nt+\Nth)}. 
\]
By (\ref{eq:txest})  and the above estimates, we get (\ref{eq:gogrel})
for $k$ large enough.
$\Box$

The interval we will work with from now on is $[\sigma_{k},\tau_{k}]$.
Let $\eta$ be defined as in (\ref{eq:etadef}), but define
$\eta_{1,k}=\eta(\tau_{k})$ and $\eta_{2,k}=\eta(\sigma_{k})$.
We need to improve the estimates of the variation of $1+\Sp$ and
$\Nt+\Nth$ during a period contained in $[\eta_{1,k},\eta_{2,k}]$. 

\begin{lemma}\label{lemma:variation}
Consider an interval $\mathcal{I}=[\eta_{1},\eta_{1}+2\pi]\subseteq
[\eta_{1,k},\eta_{2,k}]$, where $\eta_{1,k}=\eta(\tau_{k})$ and 
$\eta_{2,k}=\eta(\sigma_{k})$. Let $\eta_{a}$ and $\eta_{b}$
correspond to the max and min of $1+\Sp$ in $\mathcal{I}$, and let
$\eta_{\mathrm{max}}$ and $\eta_{\mathrm{min}}$ correspond to the max
and min of $\Nt+\Nth$ in the same interval. Then,
\begin{equation}\label{eq:spmsp}
|\Sp(\eta_{b})-\Sp(\eta_{a})|
\leq\frac{40\pi(1+\Sp(\eta_{b}))^2}{(\Nt+\Nth)(\eta_{\mathrm{max}})}
\end{equation}
and
\begin{equation}\label{eq:nqn}
\frac{(\Nt+\Nth)(\eta_{\mathrm{max}})}{(\Nt+\Nth)
(\eta_{\mathrm{min}})}
\leq \exp (\frac{20\pi}{\epsilon}\exp(-c_{\epsilon}k)).
\end{equation}
\end{lemma}

\textit{Proof}. The derivation of (\ref{eq:dspdeta}) is still valid,
so that
\[
|\frac{1}{1+\Sp}\frac{d(1+\Sp)}{d\eta}|\leq 
\frac{10(1+\Sp)}{\Nt+\Nth}.
\]
By (\ref{eq:spn2}) we conclude that 
$(1+\Sp(\eta_{a}))/(1+\Sp(\eta_{b}))$ can be chosen to be arbitrarily
close to one by choosing $k$ large enough. Now,
\[
\frac{1}{\Nt+\Nth}\frac{d(\Nt+\Nth)}{d\eta}=\frac{1}{\Nt+\Nth}
\frac{1}{2g}\frac{d(\Nt+\Nth)}{d\tau}=
\]
\[
=\frac{1}{\Nt+\Nth}\frac{1}{2g}((q+2\Sp)(\Nt+\Nth)+2\sqrt{3}
\Sm(\Nt-\Nth))=\frac{q+2\Sp}{2g}+\frac{4(1-\Sp^2)\tx\ty}{2(\Nt+\Nth)g},
\]
and consequently
\[
|\frac{1}{\Nt+\Nth}\frac{d(\Nt+\Nth)}{d\eta}|\leq \frac{10}{\epsilon}
e^{-c_{\epsilon}k}.
\]
Equation (\ref{eq:nqn}) follows, and the relative variation of
$\Nt+\Nth$ during one period can be chosen arbitrarily small.
Finally,
\[
|\Sp(\eta_{b})-\Sp(\eta_{a})|=(1+\Sp(\eta_{b}))
|\frac{1+\Sp(\eta_{a})}{1+\Sp(\eta_{b})}-1|\leq
\]
\[
\leq\frac{30\pi(1+\Sp(\eta_{b}))^2}{(\Nt+\Nth)(\eta_{\mathrm{min}})}
\]
by (\ref{eq:dspdeta}) and the above observations. We may also change 
$\eta_{\mathrm{min}}$ to $\eta_{\mathrm{max}}$ at the cost of
increasing the constant. $\Box$

As has been stated earlier, the goal of this section is to prove that
the conditions of Lemma \ref{lemma:setup1} are never met. We do this
by deducing a contradiction from the consequences of that lemma. On
the one hand, we have a rough picture of how the solution behaves in
$[\sigma_{k},\tau_{k}]$ by Lemma \ref{lemma:setup5}, Lemma 
\ref{lemma:variation} and Corollary \ref{cor:order}. On the other
hand, we know that, since $r(\sigma_{k})-r(\tau_{k})=-k$,
\begin{equation}\label{eq:nninc}
-k=\int_{\sigma_{k}}^{\tau_{k}}(\frac{1}{4}(3\g-2)\O+\Sp^2+\Sm^2+\Sp)
d\tau=
\alpha_{k}+\int_{\eta_{1,k}}^{\eta_{2,k}}\frac{\Sp^2+\Sm^2+\Sp}{-2g}
d\eta.
\end{equation}
We will use our knowledge of the behaviour of the solution in
$[\sigma_{k},\tau_{k}]$ to prove that (\ref{eq:nninc}) is false.
Observe that $\eta_{1,k}<\eta_{2,k}$, and that the contribution 
from one period is negligible, cf. Corollary \ref{cor:order}. Also, 
$\alpha_{k}\rightarrow 0$ as $k\rightarrow \infty$ so that we may
ignore it. We will prove that for $k$ great enough, the integral of
$(\Sp^2+\Sm^2+\Sp)/(-2g)$ over a suitably chosen period is positive. 
From here on, we consider an interval $[\eta_{1},\eta_{1}+2\pi]$ 
which, excepting intervals of length 
less than a period at each end of $[\eta_{1,k},\eta_{2,k}]$, we can 
assume to be of the form $[-\pi/2,3\pi/2]$. There is however one
thing that should be kept in mind; when translating the
$\eta$-variable by $2m\pi$ the $\xi$-variable is translated by $m\pi$.
In other words, there is a sign involved, and in order to keep track of
it we write out the details. By the above observations we have.

\begin{lemma}\label{lemma:division}
For each $k$ there are integers $m_{1,k}$ and $m_{2,k}$ such that
\begin{equation}\label{eq:nninc2}
-k=
\beta_{k}+\int_{-\pi/2+2m_{1,k}\pi}^{3\pi/2+2m_{2,k}\pi}
\frac{\Sp^2+\Sm^2+\Sp}{-2g}d\eta,
\end{equation}
where $\beta_{k}\rightarrow 0$ as $k\rightarrow \infty$, and
\[
\eta_{1,k}\leq -\pi/2+2m_{1,k}\pi\leq \eta_{1,k}+2\pi,\
\eta_{2,k}-2\pi\leq 3\pi/2+2m_{2,k}\pi\leq \eta_{2,k}.
\]
\end{lemma}

Consider now an interval 
\[[-\pi/2+2m\pi,3\pi/2+2m\pi]\subseteq
[-\pi/2+2m_{1,k},3\pi/2+2m_{2,k}\pi],
\]
where $m$ is an integer, and make the substitution 
\[
\te=\eta-2m\pi,\ \txi=\xi-m\pi
\]
in that interval. Compute
\[
\Sp^2+(1-\Sp^2)x^2+\Sp=(1+\Sp)(\Sp+(1-\Sp)\frac{1}{2}(1-\cos\eta))=
\]
\[
=(1+\Sp)(\frac{1}{2}(1+\Sp)-\frac{1}{2}(1-\Sp)\cos\te)=
\frac{1}{2}(1+\Sp)((1+\Sp)-(1-\Sp)\cos\te).
\]
This expression is the relevant part of the numerator of the integrand
in the right hand side of (\ref{eq:nninc2}). There is a drift term 
yielding a positive contribution to the integral, but the oscillatory 
term is arbitrarily much greater by Lemma \ref{lemma:setup2}. The interval
$[-\pi/2,3\pi/2]$ was not chosen at random. By considering the above
expression, one concludes that the oscillatory term is negative in 
$[-\pi/2,\pi/2]$ and positive in $[\pi/2,3\pi/2]$. As far as obtaining
a contradiction goes, the first interval is thus bad and the second
good. In order to estimate the integral over a period, the natural
thing to do is then to make a substitution in the interval  
$[\pi/2,3\pi/2]$, so that it becomes an integral over the interval
$[-\pi/2,\pi/2]$. It is then important to know how the different 
expressions vary with $\eta$. We will prove a lemma saying that
$\Sp$ roughly increases with $\eta$, and it will turn out to be useful
that $\Sp$ is greater in the good part than in the bad. Let
\begin{equation}\label{eq:jdef}
J=\int_{-\pi/2+2m\pi}^{3\pi/2+2m\pi}\frac{\Sp^2+\Sm^2+\Sp}{-2g}d\eta=
\frac{1}{2}\int_{-\pi/2}^{3\pi/2}\frac{(1+\Sp(\te+2m\pi))^2}
{-2g(\te+2m\pi)}d\te-
\end{equation}
\[
-\frac{1}{2}\int_{-\pi/2}^{3\pi/2}
\frac{(1-\Sp^2(\te+2m\pi))\cos\te}{-2g(\te+2m\pi)}d\te+
\]
\[
+\int_{-\pi/2}^{3\pi/2}\frac{\Sm^2(\te+2m\pi)-(1-\Sp(\te+2m\pi))^2
x^2(\te+2m\pi)}{-2g(\te+2m\pi)}d\te=
J_{1}+J_{2}+J_{3}.
\]
If we can prove that $J$ is positive regardless of $m$ we are done,
since $J$ positive contradicts (\ref{eq:nninc2}). The integral
$J_{1}$ is positive, and because the relative variation of the
integrand can be chosen arbitrarily small by choosing $k$ large
enough, $J_{1}$ is of the order of magnitude
\begin{equation}\label{eq:spn21}
\frac{(1+\Sp)^2}{\Nt+\Nth}.
\end{equation}
If negative terms in $J_{2}$ and $J_{3}$ of the orders of
magnitude
\begin{equation}\label{eq:spn32}
\frac{(1+\Sp)^3}{(\Nt+\Nth)^2}
\end{equation}
or
\begin{equation}\label{eq:spn33}
\frac{(1+\Sp)^3}{(\Nt+\Nth)^3}
\end{equation}
occur, we may ignore them by (\ref{eq:spn2}) and (\ref{eq:spnn2}).
By (\ref{eq:smosc}), $J_{3}$ may be ignored. Observe that the largest
integrand is the one appearing in $J_{2}$. However, it oscillates.
Considering (\ref{eq:jdef}), one can see that writing out
arguments such as $\te+2m\pi$ does not make things all that much
clearer. For that reason, we introduce the following convention.

\begin{convention}
By $\Sp(\te)$ and $\Sp(-\te+\pi)$, we will mean $\Sp(\te+2m\pi)$ 
and $\Sp(-\te+\pi+2m\pi)$ respectively, and similarly for
all expressions in the variables of Wainwright and Hsu. However,
trigonometric expressions should be read as stated.
Thus $\cos(\te/2)$ means just that and not $\cos(\te/2+m\pi)$.
\end{convention}

\begin{definition}
Consider an integral expression
\[
I=\int_{-\pi/2}^{3\pi/2}f(\te)d\te.
\]
Then we say that $I$ is less than or equal to zero \textit{up to 
order of magnitude}, if
\[
I\leq \int_{-\pi/2}^{3\pi/2}g(\te)d\te, 
\]
where $g$ satisfies a bound
\[
g\leq
C_{1}\frac{(1+\Sp)^3}{(\Nt+\Nth)^2}+
C_{2}\frac{(1+\Sp)^3}{(\Nt+\Nth)^3},
\]
for $k$ large enough, where $C_{1}$ and $C_{2}$ are positive constants
independent of $k$. We write $I\lesssim 0$.
The definition of $I\gtrsim 0$ is similar. We also define the concept 
similarly if the interval of integration is different.
\end{definition}

We will use the same terminology more generally in inequalities
between functions, if those inequalities, when inserted into
the proper integrals, yield inequalities in the sense of the 
definition above. We will write $\approx$ if the error is
of negligible order of magnitude.

\begin{lemma}\label{lemma:j2geq}
If $J_{2}$ as defined above satisfies $J_{2}\gtrsim 0$, then 
$J$ is non-negative for $k$ large enough.
\end{lemma}

\textit{Proof}. Under the assumptions of the lemma, we have
\[
J\geq 
\frac{1}{2}\int_{-\pi/2}^{3\pi/2}\frac{(1+\Sp)^2}{-2g}d\te-
\int_{-\pi/2}^{3\pi/2}(C_{1}\frac{(1+\Sp)^3}{(\Nt+\Nth)^2}+
C_{2}\frac{(1+\Sp)^3}{(\Nt+\Nth)^3})d\te+
\]
\[
+\int_{-\pi/2}^{3\pi/2}\frac{\Sm^2-(1-\Sp^2)x^2}{-2g}d\te.
\]
By Corollary \ref{cor:order}, Lemma \ref{lemma:variation} and
(\ref{eq:smosc}), we conclude that for $k$ large enough, $J$ is 
positive. $\Box$

The following lemma says that $\Sp$ almost increases with $\te$.

\begin{lemma}\label{lemma:spinc}
Let $-\pi/2\leq \te_{a}\leq\te_{b}\leq 3\pi/2$. Then
\[
\Sp(\te_{b})-\Sp(\te_{a})\geq -(1+\Sp(\te_{\mathrm{min}}))^{8},
\]
where $\te_{\mathrm{min}}$ corresponds to the minimum of
$1+\Sp$ in $[-\pi/2,3\pi/2]$.
\end{lemma}

\textit{Proof}. We have 
\[
\Sp'\leq \frac{3}{2}(2-\g)\O,
\]
so that 
\[
\frac{d\Sp}{d\te}\geq \frac{3}{2}(2-\g)\frac{\O}{2g}.
\]
Using (\ref{eq:onocon}), (\ref{eq:spn2}) and Lemma
\ref{lemma:variation}, we conclude that
\[
\frac{d\Sp}{d\te}\geq -\frac{1}{2\pi}
(1+\Sp(\te_{\mathrm{min}}))^{8}.
\]
The lemma follows. $\Box$

\begin{lemma}\label{lemma:ileq}
If
\[
I=\int_{-\pi/2}^{3\pi/2}\frac{1+\Sp}{-g}\cos\te d\te
\]
satisfies $I\lesssim 0$, then $J_{2}\gtrsim 0$.
\end{lemma}

\textit{Proof}. 
Consider
\[
-J_{2}=\int_{-\pi/2}^{3\pi/2}\frac{(1-\Sp^2)\cos\te}{-4g}d\te=
\int_{-\pi/2}^{3\pi/2}\frac{(\Sp(3\pi/2)-\Sp)(1+\Sp)}{-4g}
\cos\te d\te+
\]
\[
+(1-\Sp(3\pi/2))\int_{-\pi/2}^{3\pi/2}\frac{1+\Sp}{-4g}\cos\te d\te.
\]
The first integral is negligible by (\ref{eq:spmsp}). The lemma
follows. $\Box$

\begin{lemma}\label{lemma:i1tj2}
If
\[
I_{1}=\int_{-\pi/2}^{\pi/2}\frac{(1+\Sp(\te))(g_{1}(\te)-
g_{1}(-\te+\pi))}{g(\te)g(-\te+\pi)}\cos\te d\te
\]
satisfies $I_{1}\lesssim 0$, then $J_{2}\gtrsim 0$.
\end{lemma}

\textit{Proof}. We have
\[
I=\int_{-\pi/2}^{3\pi/2}\frac{1+\Sp}{-g}\cos\te d\te=
\int_{-\pi/2}^{\pi/2}\frac{1+\Sp}{-g}\cos\te d\te+
\int_{\pi/2}^{3\pi/2}\frac{1+\Sp}{-g}\cos\te d\te.
\]
Make the substitution $\chi=-\te+\pi$ in the second integral;
\[
\int_{\pi/2}^{-\pi/2}\frac{1+\Sp(-\chi+\pi)}{-g(-\chi+\pi)}
\cos(-\chi+\pi)
(-d\chi)=-\int_{-\pi/2}^{\pi/2}\frac{1+\Sp(-\chi+\pi)}{-g(-\chi+\pi)}
\cos(\chi)d\chi.
\]
Thus,
\[
I=\int_{-\pi/2}^{\pi/2}\left(\frac{1+\Sp(\te)}{-g(\te)}-
\frac{1+\Sp(-\te+\pi)}{-g(-\te+\pi)}\right) \cos\te d\te=
\]
\[
=\int_{-\pi/2}^{\pi/2}\frac{(1+\Sp(-\te+\pi))g(\te)-
(1+\Sp(\te))g(-\te+\pi)}{g(\te)g(-\te+\pi)}\cos\te d\te.
\]
But
\[
(1+\Sp(-\te+\pi))g(\te)\lesssim (1+\Sp(\te))g(\te),
\]
by Lemma \ref{lemma:spinc}, so that
\begin{equation}\label{eq:ileq}
I\lesssim \int_{-\pi/2}^{\pi/2}\frac{(1+\Sp(\te))(g(\te)-
g(-\te+\pi))}{g(\te)g(-\te+\pi)}\cos\te d\te.
\end{equation}
Now,
\[
g(\te)-g(-\te+\pi)=g_{1}(\te)-g_{1}(-\te+\pi)+
g_{2}(\te)-g_{2}(-\te+\pi),
\]
but since $2xy=\sin\te$ and the error committed in replacing
$\tx$ with $x$ and $\ty$ with $y$ is negligible by (\ref{eq:tbxest}), 
we have
\[
g_{2}(\te)-g_{2}(-\te+\pi)\approx -(1+\Sp(\te))\sin\te+
(1+\Sp(-\te+\pi))\sin(-\te+\pi)=
\]
\[
=(\Sp(-\te+\pi))-\Sp(\te))\sin\te.
\]
The corresponding contribution to the integral may consequently be
neglected; the error in the integral will be of type (\ref{eq:spn33})
by (\ref{eq:spmsp}). Consequently, if 
\[
I_{1}=\int_{-\pi/2}^{\pi/2}\frac{(1+\Sp(\te))(g_{1}(\te)-
g_{1}(-\te+\pi))}{g(\te)g(-\te+\pi)}\cos\te d\te
\]
satisfies $I_{1}\lesssim 0$, then $I\lesssim 0$  by 
(\ref{eq:ileq}), so that the lemma
follows by Lemma \ref{lemma:ileq}. $\Box$

Let 
\[
h_{1}(\te)=g_{1}(\te)-g_{1}(-\te+\pi).
\]
We estimate $h_{1}$ by estimating the derivative. We have 
$h_{1}(\pi/2)=0$. 

\begin{lemma}
Let $h_{1}$ be as above. In the interval $[-\pi/2,\pi/2]$, we have
\begin{equation}\label{eq:dh1de}
\frac{dh_{1}}{d\te}\gtrsim 3\left(\frac{1-\Sp^2(\te)}{-g(\te)}+
\frac{1-\Sp^2(-\te+\pi)}{-g(-\te+\pi)}\right)\sin\te.
\end{equation}
\end{lemma}

\textit{Proof}. Compute
\[
\frac{dh_{1}}{d\te}(\te)=\frac{dg_{1}}{d\te}(\te)+
\frac{dg_{1}}{d\te}(-\te+\pi).
\]
But
\[
\frac{dg_{1}}{d\te}=-\frac{3}{2g}((q+2\Sp)(\Nt+\Nth)+2\sqrt{3}\Sm
(\Nt-\Nth))=
\]
\[
=\frac{1}{2}(q+2\Sp)\frac{g-g_{2}}{g}-3\sqrt{3}
\frac{\Sm(\Nt-\Nth)}{g}.
\]
Observe that $x$ and $y$ are trigonometric expressions, and that
\[
2x(\te+2m\pi)y(\te+2m\pi)=2\sin(\te/2+m\pi)\cos(\te/2+m\pi)=
\sin\te.
\]
We have
\[
\sqrt{3}\Sm(\Nt-\Nth)\approx 2(1-\Sp^2)xy=(1-\Sp^2)\sin\te,
\]
so that
\[
\frac{dg_{1}}{d\te}\approx (\frac{1}{4}(3\g-2)\O+\Sp^2+\Sm^2+\Sp)-
\frac{g_{2}}{g}(\frac{1}{4}(3\g-2)\O+\Sp^2+\Sm^2+\Sp)-
\]
\[
-\frac{3(1-\Sp^2)\sin\te}{g}.
\]
The middle term and all terms involving $\O$ may be ignored. 
Estimate
\[
\Sp^2(\te)+\Sm^2(\te)+\Sp(\te)+\Sp^2(-\te+\pi)+
\Sm^2(-\te+\pi)+\Sp(-\te+\pi)\approx
\]
\[
\approx\Sp^2(\te)+(1-\Sp^2(\te))
(\sin^2(\te/2+m\pi)-1/2)+\frac{1}{2}(1-\Sp^2(\te))+\Sp(\te)+
\]
\[
+\Sp^2(-\te+\pi)+(1-\Sp^2(-\te+\pi))
(\cos^2(\te/2+m\pi)-1/2)+\frac{1}{2}(1-\Sp^2(-\te+\pi))+
\]
\[
+\Sp(-\te+\pi)=
\frac{1}{2}(1+\Sp(\te))^2+\frac{1}{2}(1+\Sp(-\te+\pi))^2+
\]
\[
+(1-\Sp^2(\te))(\sin^2(\te/2)-1/2)
+(1-\Sp^2(-\te+\pi))(\cos^2(\te/2)-1/2).
\]
The first equality is a consequence of (\ref{eq:smosc}). Due to the
fact that $\te\in[-\pi/2,\pi/2]$, we have $\cos^2(\te/2)-1/2\geq 0$.
Since $-\te+\pi\geq \te$ and $\Sp$ increases with $\te$ up to 
order of magnitude according to Lemma \ref{lemma:spinc}, we have
\[
1-\Sp^2(-\te+\pi)\gtrsim 1-\Sp^2(\te).
\]
Consequently,
\[
\frac{1}{2}(1+\Sp(\te))^2+\frac{1}{2}(1+\Sp(-\te+\pi))^2
+(1-\Sp^2(\te))(\sin^2(\te/2)-1/2)+
\]
\[
+(1-\Sp^2(-\te+\pi))(\cos^2(\te/2)-1/2)\gtrsim
\frac{1}{2}(1+\Sp(\te))^2+\frac{1}{2}(1+\Sp(-\te+\pi))^2+
\]
\[
+(1-\Sp^2(\te))(\sin^2(\te/2)-1/2)
+(1-\Sp^2(\te))(\cos^2(\te/2)-1/2)\geq 0.
\]
In other words, we have (\ref{eq:dh1de}). Here the importance of the
fact that $\Sp$ is greater in the good part than in the bad
becomes apparent. $\Box$

\begin{lemma}\label{lemma:i1leq}
Let $I_{1}$ be defined as above. Then $I_{1}\lesssim 0$.
\end{lemma}

\textit{Proof}.
Let $\te_{\mathrm{max}}$ and $\te_{\mathrm{min}}$ correspond to
the max and min of $-g$ in the interval $[-\pi/2,3\pi/2]$, and
let $\te_{a}$ and $\te_{b}$ correspond to the max and min
of $\Sp$, in the same interval. Observe that for $\te\in
[-\pi/2,3\pi/2]$, we have
\[
1-\Sp^2(\te_{a})\geq 1-\Sp^2(\te)\geq 1-\Sp^2(\te_{b}).
\]
In order not to obtain too complicated expressions, let us introduce
the following terminology:
\[
a_{1}=6\frac{1-\Sp^2(\te_{b})}{-g(\te_{\mathrm{max}})}\leq
6\frac{1-\Sp^2(\te)}{-g(\te)}\leq
6\frac{1-\Sp^2(\te_{a})}{-g(\te_{\mathrm{min}})}=a_{2}\ \mathrm{and}
\]
\[
b_{1}=\frac{1+\Sp(\te_{b})}{g^2(\te_{\mathrm{max}})}\leq
\frac{1+\Sp(\te)}{g(\te)g(-\te+\pi)}\leq
\frac{1+\Sp(\te_{a})}{g^2(\te_{\mathrm{min}})}=b_{2},
\]
where $\te\in [-\pi/2,3\pi/2]$. 
Observe that
\begin{equation}\label{eq:ablim}
\lim_{k\rightarrow \infty}\frac{a_{1}}{a_{2}}=
\lim_{k\rightarrow \infty}\frac{b_{1}}{b_{2}}=1,
\end{equation}
by Corollary \ref{cor:order} and Lemma \ref{lemma:variation}.
Consider the interval $[0,\pi/2]$. By (\ref{eq:dh1de}), we have 
\begin{equation}\label{eq:dhde}
\frac{dh_{1}}{d\te}\gtrsim
a_{1}\sin\te,
\end{equation}
so that
\[
h_{1}(\te)=h_{1}(\pi/2)-\int_{\te}^{\pi/2}\frac{dh_{1}}{d\te}
d\te\lesssim -a_{1}\cos\te
\]
in the interval $[0,\pi/2]$. Now consider the interval $[-\pi/2,0]$. 
We have 
\[
\frac{dh_{1}}{d\te}\gtrsim
a_{2}\sin\te.
\]
Consequently,
\[
h_{1}(\te)=h_{1}(0)-\int_{\te}^{0}\frac{dh_{1}}{d\te}
d\te\lesssim -a_{1}+a_{2}(1-\cos\te)
\]
in the interval $[-\pi/2,0]$. Estimate
\[
\int_{0}^{\pi/2}\frac{(1+\Sp(\te))(g_{1}(\te)-
g_{1}(-\te+\pi))}{g(\te)g(-\te+\pi)}\cos\te d\te=
\]
\[
=\int_{0}^{\pi/2}\frac{(1+\Sp(\te))h_{1}(\te)}{g(\te)g(-\te+\pi)}
\cos\te d\te\lesssim
\int_{0}^{\pi/2}\frac{(1+\Sp(\te))}{g(\te)g(-\te+\pi)}(-a_{1}\cos^2\te)
d\te\leq
\]
\[
\leq -a_{1}b_{1}\int_{0}^{\pi/2}\cos^2\te
 d\te=-\frac{\pi a_{1}b_{1}}{4}.
\]
We also estimate
\[
\int_{-\pi/2}^{0}\frac{(1+\Sp(\te))(g_{1}(\te)-
g_{1}(-\te+\pi))}{g(\te)g(-\te+\pi)}\cos\te d\te=
\]
\[
=\int_{-\pi/2}^{0}\frac{(1+\Sp(\te))h_{1}(\te)}{g(\te)g(-\te+\pi)}
\cos\te d\te\lesssim
 -a_{1}
\int_{-\pi/2}^{0}\frac{(1+\Sp(\te))}{g(\te)g(-\te+\pi)}
\cos\te d\te+
\]
\[
+a_{2}\int_{-\pi/2}^{0}\frac{(1+\Sp(\te))}{g(\te)g(-\te+\pi)}
(1-\cos\te)\cos\te d\te\leq
-a_{1}b_{1}\int_{-\pi/2}^{0}\cos\te d\te+
\]
\[
+a_{2}b_{2}
\int_{-\pi/2}^{0}(1-\cos\te)\cos\te d\te\leq
-a_{1}b_{1}+
(1-\frac{\pi}{4})a_{2}b_{2}.
\]
Adding up, we conclude that
\[
I_{1}\lesssim
-(1+\pi/4)a_{1}b_{1}
+(1-\pi/4)a_{2}b_{2}=[-(1+\pi/4)\frac{a_{1}b_{1}}{a_{2}b_{2}}+
(1-\pi/4)]a_{2}b_{2},
\]
which is negative for $k$ large enough by (\ref{eq:ablim}). 
Thus $I_{1}\lesssim 0$. $\Box$

\begin{thm}\label{thm:it}
The conditions of Lemma \ref{lemma:setup1} are never met.
\end{thm}

\textit{Proof}. If the conditions are met, then Lemma
\ref{lemma:division} follows,
and also that it is false, by Lemmas \ref{lemma:i1leq}, 
\ref{lemma:i1tj2}, \ref{lemma:j2geq} and (\ref{eq:jdef}). $\Box$

\begin{cor}\label{cor:nsf}
Let $2/3<\g<2$. For every $\epsilon>0$ there is a $\delta>0$ such 
that if $x$ constitutes Bianchi IX initial data for 
(\ref{eq:whsu})-(\ref{eq:constraint}) and
\[
\inf_{y\in\mathcal{A}}\|x-y\|\leq \delta
\]
then
\[
\inf_{y\in\mathcal{A}}\|\Phi(\tau,x)-y\|\leq \epsilon
\]
for all $\tau\leq 0$, where $\Phi$ is the flow of 
(\ref{eq:whsu})-(\ref{eq:constraint}).
\end{cor}

\textit{Proof}. Assuming the contrary, there is an $\epsilon>0$ and 
a sequence $x_{l}\rightarrow \mathcal{A}$ such that 
\[
\inf_{y\in\mathcal{A}}\|\Phi(s_{l},x_{l})-y\|\geq \epsilon
\]
for some $s_{l}\leq 0$. Let $\tau_{l}=0$. Since
$d(\tau_{l},x_{l})\rightarrow 0$ and we can assume $\epsilon$
is small enough that Proposition \ref{prop:ocon} is applicable,
there must be an $\eta>0$ such that $h(s_{l},x_{l})>\eta$
for $l$ large enough, contradicting Theorem \ref{thm:it}.
$\Box$

\begin{cor}\label{cor:it}
Consider a generic Bianchi IX solution with $2/3<\g<2$. Then
\[
\lim_{\tau\rightarrow -\infty}(\O+\No\Nt+\Nt\Nth+\No\Nth)=0.
\]
\end{cor}

\textit{Proof}. If $h$ does not converge to zero, then the conditions
of Lemma \ref{lemma:setup1} are met, since there for a generic solution
is an $\a$-limit point on the Kasner circle by Proposition 
\ref{prop:alke}. Corollary \ref{cor:otz} then yields the desired
conclusion. $\Box$

Let $\mathcal{A}$ be the set of vacuum type I and II points as in
Definition \ref{def:attractor}. By
Corollary \ref{cor:it}, a generic type IX solution with $2/3<\g<2$ 
converges to $\mathcal{A}$.

\begin{cor}\label{cor:ffaIX}
Let $2/3<\g<2$.
The closure of $\mathcal{F}_{\mathrm{IX}}$ and the closure of
$\mathcal{P}_{\mathrm{IX}}$ do not intersect $\mathcal{A}$. Furthermore,
the set of generic Bianchi IX points is open in the set of
Bianchi IX points.
\end{cor}

\textit{Remark}. The closure of the Taub type IX points does intersect
$\mathcal{A}$.

\textit{Proof}. Assume there is a sequence 
$x_{l}\in\mathcal{F}_{\mathrm{IX}}$ such that $x_{l}\rightarrow x
\in\mathcal{A}$. Let $\tau_{l}=0$. Observe that then
$d(x_{l},\tau_{l})\rightarrow 0$. By Theorem \ref{thm:it},
there is for each $\epsilon>0$ and for each $L$ an
$l\geq L$ such that $h(\tau,x_{l})\leq \epsilon$ for
$\tau\leq\tau_{l}=0$. By choosing $L$ large enough, we can
assume $\O(\tau_{l},x_{l})$ to be arbitrarily small and by
choosing $\epsilon$ small enough, we can assume that Proposition
\ref{prop:ocon} is applicable. Consequently, we can assume
$\O(\tau,x_{l})$ to be as small as we wish for $\tau\in
(-\infty,\tau_{l}]$, contradicting the fact that $\O(\tau,x_{l}) 
\rightarrow 1$ as $\tau\rightarrow -\infty$. 
The argument for $\mathcal{P}_{\mathrm{IX}}$
is similar, since the $\O$-coordinate of $P_{i}^{+}(II)$ is
positive.

Consider now a generic point $x$ in the set of Bianchi IX points.
There is a neighbourhood of $x$ that does not intersect the 
Taub points. Let us prove the similar statement for 
$\mathcal{F}_{\mathrm{IX}}$ and $\mathcal{P}_{\mathrm{IX}}$.
Assume there is a sequence 
$x_{l}\in\mathcal{F}_{\mathrm{IX}}$ such that $x_{l}\rightarrow x$.
For each $\epsilon>0$ there is a $T\leq 0$ such that 
$d(T,\Phi(T,x))\leq \epsilon/2$, by Corollary \ref{cor:it}. 
By continuity of the flow and the
function $d$, we conclude that for $l$ large enough we have
$d(T,\Phi(T,x_{l}))\leq \epsilon$. Since $\Phi(T,x_{l})\in
\mathcal{F}_{\mathrm{IX}}$, we get a contradiction to the first part of
the lemma. Thus, there is an open neighbourhood of $x$ that does not
intersect $\mathcal{F}_{\mathrm{IX}}$. The argument for 
$\mathcal{P}_{\mathrm{IX}}$ is similar. $\Box$

\begin{cor}\label{cor:ffaVII0}
Let $2/3<\g<2$.
The closure of $\mathcal{F}_{\mathrm{VII}_{0}}$ and the closure of
$\mathcal{P}_{\mathrm{VII}_{0}}$ do not intersect $\mathcal{A}$.
Furthermore, the generic Bianchi $\mathrm{VII}_{0}$ points are
open in the set of Bianchi $\mathrm{VII}_{0}$ points.
\end{cor}

\textit{Proof}. The argument proving the first part is as in 
the Bianchi IX case, once one has checked that analogues of 
Proposition \ref{prop:ocon} and Theorem \ref{thm:it} hold 
in the Bianchi VI$\mathrm{I}_{0}$ case. The second part
then follows as in the Bianchi IX case, using Proposition
\ref{prop:typeVII0}.
$\Box$

\section{Regularity of the set of non-generic points}
\label{section:regular}

Observe that the constraint (\ref{eq:constraint}) together with the 
additional assumption $\O\geq 0$ defines a 5-dimensional submanifold of
$\mathbb{R}^{6}$ which has a 4-dimensional boundary given
by the vacuum points. We have the following.

\begin{thm}\label{thm:regular}
Let $2/3<\g<2$.
The sets $ \mathcal{F}_{\mathrm{II}},
\mathcal{F}_{\mathrm{VII}_{0}}$, $\mathcal{F}_{\mathrm{IX}}$,
$\mathcal{P}_{\mathrm{VII}_{0}}$ and
$\mathcal{P}_{\mathrm{IX}}$ are $C^{1}$ submanifolds of $\mathbb{R}^{6}$
of dimensions $1$, $2$, $3$, $1$ and $2$ respectively.
\end{thm}

We prove this theorem at the end of this section.
The idea is as follows. The only obstruction to 
e. g. $\mathcal{F}_{\mathrm{II}}$ being a $C^{1}$ submanifold,
is if there is an open set $O$ containing $F$  and a sequence
$x_{k}\in \mathcal{F}_{\mathrm{II}}$ such that $x_{k}\rightarrow F$,
but each $x_{k}$ has to leave $O$ before it can converge to $F$.
If there is such a sequence, we produce a sequence
$y_{k}\in \mathcal{F}_{\mathrm{II}}$ such that the distance from
$y_{k}$ to $\mathcal{A}$ converges to zero, contradicting
Lemma \ref{lemma:ffaII}. The argument is similar in the other cases.

We will need some results from \cite{hart}. 
The theorem stated below is a special case of Theorem 6.2, p. 243. 

\begin{thm}\label{thm:hart}
In the differential equation 
\begin{equation}\label{eq:de}
\xi'=E\xi+G(\xi)
\end{equation}
let $G$ be of class $C^{1}$ and $G(0)=0,\ \partial_{\xi}G(0)=0$.
Let $E$ have $e>0$ eigenvalues with positive real parts,
$d>0$ eigenvalues with negative real parts and no eigenvalues with
zero real part. Let
$\xi_{t}=\xi(t,\xi_{0})$ be the solution of  (\ref{eq:de})
satisfying $\xi(0,\xi_{0})=\xi_{0}$ and $T^{t}$ the corresponding
map $T^{t}(\xi_{0})=\xi(t,\xi_{0})$. Then there exists a map
$R$ of a neighbourhood of $\xi=0$ in $\xi$-space onto a neighbourhood
of the origin in Euclidean $(u,v)$-space, where $\mathrm{dim}(u)=d$
and  $\mathrm{dim}(v)=e$, such that $R$ is $C^{1}$ with non-vanishing
Jacobian and $RT^{t}R^{-1}$ has the form
\begin{equation}\label{eq:flow}
\left(\begin{array}{c}
    u_{t}\\
    v_{t}
      \end{array}
\right)=
\left(\begin{array}{c}
    e^{tP}u_{0}+U(t,u_{0},v_{0})\\
    e^{tQ}v_{0}+V(t,u_{0},v_{0})
      \end{array}
\right).
\end{equation}
$U,\ V$ and their partial derivatives with respect to $u_{0},\ v_{0}$
vanish at $(u_{0},v_{0})=0$. Furthermore $V=0$ if $v_{0}=0$ and
$U=0$ if $u_{0}=0$. Finally $\|e^{P}\|<1$ and $\|e^{-Q}\|<1$.
\end{thm}

Let us begin by considering the local behaviour close to the 
fixed points.

\begin{lemma}\label{lemma:local}
Consider the critical point $F$. There is an open neighbourhood
$O$ of $F$ in $\mathbb{R}^6$,
and a 1-dimensional $C^{1}$ submanifold
$M_{\mathrm{II}}\subseteq \mathcal{F}_{\mathrm{II}}$ of 
$O\cap \mathcal{I}_{\mathrm{II}}$, such that for
each $x\in O\cap \mathcal{I}_{\mathrm{II}}$, either $x\in 
M_{\mathrm{II}}$, or $x$ will leave $O$ as the flow of
(\ref{eq:whsu})-(\ref{eq:constraint}) is applied 
to $x$ in the negative time direction. Similarly, we get a
2-dimensional $C^{1}$ submanifold $M_{\mathrm{VII}_{0}}$
of $O\cap \mathcal{I}_{\mathrm{VII}_{0}}$, and a 3-dimensional
$C^{1}$ submanifold $M_{\mathrm{IX}}$ of 
$O\cap \mathcal{I}_{\mathrm{IX}}$ with the same properties.
Consider the critical point $P_{1}^{+}(II)$. We then have a
similar situation. Give the neighbourhood corresponding to $O$ the
name $P$, and use the letter $N$ instead of the letter $M$ to denote
the relevant submanifolds. Then $N_{\mathrm{VII}_{0}}$ has dimension
1 and $N_{\mathrm{IX}}$ has dimension 2.
\end{lemma}

\textit{Proof}. 
Observe that when $\O>0$, we can consider 
(\ref{eq:whsu})-(\ref{eq:constraint}) to be an unconstrained
system of equations in five variables. Using the constraint
(\ref{eq:constraint}) to express $\O$ in terms of the other
variables, we can ignore $\O$ and consider the first five
equations of (\ref{eq:whsu}) as a set of equations on an
open submanifold of $\mathbb{R}^{5}$, defined by the condition
$\O>0$ (considering $\O$ as a function of the other variables). 
In the Bianchi VI$\mathrm{I}_{0}$ case, we can consider the 
system to be unconstrained in four variables.

Let us first deal with the Bianchi VI$\mathrm{I}_{0}$ case. 
Consider the fixed point $P_{1}^{+}(II)$. Considering the
Bianchi $\mathrm{VII_{0}}$ points with $\No,\ \Nt>0$ and
$\Nth=0$, the linearization has one eigenvalue with positive 
real part and three with negative real part, cf. \cite{whsu}. 
By a suitable translation of the variables, reversal of time, and
a suitable definition of $G$ and $E$ in (\ref{eq:de}), we
can consider a solution to (\ref{eq:whsu})-(\ref{eq:constraint})
converging to $P_{1}^{+}(II)$ as $\tau\rightarrow -\infty$ as a
solution $\xi$ to (\ref{eq:de}) converging to $0$ as $t\rightarrow
\infty$. $E$ has one eigenvalue with negative real part and 
three with positive real part, so that Theorem \ref{thm:hart}
yields a $C^{1}$ map $R$ of a neighbourhood of $0$ with 
non-vanishing Jacobian to a 
neighbourhood of the origin in $\mathbb{R}^{4}$, such that 
the flow takes the form (\ref{eq:flow}) where $u\in\mathbb{R}$
and $v\in\mathbb{R}^{3}$.

Observe that since $\xi=0$ is a fixed point, there is a
neighbourhood of that point such that the flow is defined for 
$|t|\leq 1$. There is also an open bounded ball $B$ centered at the origin
in $(u_{0},v_{0})$-space such that $U$ and $V$ are defined in
a neighbourhood $N$ of $[-1,1]\times B$. Let $a=\|e^{P}\|$
and $1/c=\|e^{-Q}\|$. For any $\epsilon>0$,
we can choose $B$ and then $N$ small enough that the norms
of $U,\ V$ and their partial derivatives with respect to $u$ and
$v$ are smaller than
$\epsilon$ in $N$. Assume $B$ and $N$ are such for some
$\epsilon$ satisfying
\begin{equation}\label{eq:ed}
\epsilon<\mathrm{min}\{\frac{c-1}{2},\frac{1-a}{2}\}.
\end{equation}
Consider a solution $\xi$ to (\ref{eq:de})
such that  $R\circ\xi(t)\in B$ for all 
$t\geq T$. Let $(u_{t},v_{t})=R(\xi(t))$ for $t\geq T$. 
We wish to prove that $v_{t}=0$, and assume therefore that 
$v_{t_{0}}\neq 0 $ for some $t_{0}\geq T$. We have 
\[
\|v_{t_{0}+n}\|\geq \|e^{Q}v_{t_{0}+n-1}+V(1,v_{t_{0}+n-1},
u_{t_{0}+n-1})\|\geq 
\]
\[
\geq c\|v_{t_{0}+n-1}\|-
\epsilon\|v_{t_{0}+n-1}\|\geq\frac{1+c}{2}\|v_{t_{0}+n-1}\|,
\]
where we have used (\ref{eq:ed}), the fact that $V$ is zero 
when $v_{0}=0$, and the fact that $(u_{t},v_{t})$ remain in
$B$ for $t\geq T$. Thus,
\[
\|v_{t_{0}+n}\|\geq\left(\frac{1+c}{2}\right)^{n}\|v_{t_{0}}\|,
\]
which is irreconcilable with the fact that $v_{t}$ remains bounded.

If $(u_{t_{0}},v_{t_{0}})\in B$ and $v_{t_{0}}=0$, 
(\ref{eq:flow}) yields $v_{t_{0}+1}=0$ and
\[
\|u_{t_{0}+1}\|\leq (a+\frac{1-a}{2})\| u_{t_{0}}\|=
\frac{1+a}{2}\| u_{t_{0}}\|.
\]
Consequently, all points $(u,v)\in B$ with $v=0$ converge to
$(0,0)$ as one applies the flow.

We are now in a position to go backwards in order to obtain the 
conclusions of the lemma. The set $R^{-1}(B)$ will, after 
suitable operations, including non-unique extensions, 
turn into the set $P$ and $R^{-1}(\{ v=0\}\cap B)$ turns
into $N_{\mathrm{VII}_{0}}$. One can carry out a similar
construction in the Bianchi IX case. Observe that one might
then get a different $P$, but by taking the intersection
we can assume them to be the same. The dimension of $N_{\mathrm{IX}}$ 
follows from a computation of the eigenvalues. 

The argument concerning the fixed point $F$ is similar. $\Box$

\textit{Proof of Theorem \ref{thm:regular}}. 
Let $O$, $M_{\mathrm{II}}$ and so on be as in the statement of 
Lemma \ref{lemma:local}. Observe that if there
is a neighbourhood $\tilde{O}\subseteq O$ of $F$ such that 
$\mathcal{F}_{\mathrm{II}}\cap \tilde{O}=M_{\mathrm{II}}\cap
\tilde{O}$, then $\mathcal{F}_{\mathrm{II}}$ is a $C^{1}$
submanifold. The reason is that given any $x\in 
\mathcal{F}_{\mathrm{II}}$, there is a $T$ such that 
$\Phi(\tau,x)\in\tilde{O}$ for all $\tau\leq T$. By 
Lemma \ref{lemma:local}, we conclude that $\Phi(T,x)
\in M_{\mathrm{II}}$. Then there is a neighbourhood $O'\subseteq
\tilde{O}$ of $\Phi(T,x)$ such that $O'\cap
\mathcal{F}_{\mathrm{II}}=O'\cap M_{\mathrm{II}}$. We thus get, for
$O'$ suitably chosen, a $C^{1}$ map $\psi:O'\rightarrow \mathbb{R}^6$
with $C^{1}$ inverse, sending $\mathcal{F}_{\mathrm{II}}\cap O'$ to 
a one dimensional hyperplane. If $O'$ is small enough, we can apply
$\Phi(-T,\cdot)$ to it obtaining a neighbourhood of $x$. By the
invariance of $\mathcal{F}_{\mathrm{II}}$, we have
\[
\Phi(-T,O')\cap\mathcal{F}_{\mathrm{II}}=
\Phi(-T,O'\cap\mathcal{F}_{\mathrm{II}}).
\]
In other words, $\Phi(T,\psi(\cdot))$ defines coordinates on
$\Phi(-T,O')$ straightening out $\mathcal{F}_{\mathrm{II}}$. 
The arguments for the other cases are similar. 

Let us now assume, in order to reach a contradiction, that there
is a sequence $x_{k}\in\mathcal{F}_{\mathrm{II}}\cap O$ such
that $x_{k}\rightarrow F$ but $x_{k}\notin M_{\mathrm{II}}$
for all $k$. If we let $O'\subseteq O$ be a small enough ball
containing $F$, we can assume that $|N_{i}|'\geq 0$ for $i=1,2,3$
in $O'$, cf. the proof of Lemma \ref{lemma:falp}. 
For $k$ large enough, $x_{k}\in O'$ and applying the flow
to them we obtain points $y_{k}\in\mathcal{F}_{\mathrm{II}}\cap
\partial O'$. By choosing a suitable subsequence, we can assume
that $y_{k}$ converges to a type I point $y$ which is not $F$.
Given $\epsilon>0$, there is a $T$ such that $\Phi(-T,y)$ is at 
distance less than $\epsilon/2$ from $\mathcal{A}$. For $k$ large 
enough, $\Phi(-T,y_{k})\in \mathcal{F}_{\mathrm{II}}$ will then 
be at distance less than $\epsilon$ from $\mathcal{A}$. We get
a contradiction to Lemma \ref{lemma:ffaII}. The arguments for
$\mathcal{F}_{\mathrm{VII}_{0}}$ and $\mathcal{F}_{\mathrm{IX}}$
are similar, due to Corollaries \ref{cor:ffaVII0} and
\ref{cor:ffaIX}. 

For $\mathcal{P}_{\mathrm{VII}_{0}}$ and $\mathcal{P}_{\mathrm{IX}}$,
we need to modify the argument. Assume there is a sequence
$x_{k}\in\mathcal{P}_{\mathrm{VII}_{0}}\cap P$ such that 
$x_{k}\rightarrow P_{1}^{+}(II)$, but $x_{k}\notin
N_{\mathrm{VII}_{0}}$ for all $k$. By choosing $P'\subseteq P$
as a small enough ball, we can assume that $|N_{i}|'\geq 0$ in
$P'$ for $i=2,3$, cf. the proof of Lemma \ref{lemma:palp}. 
For $k$ large enough, $x_{k}\in P'$, and applying the flow
to them we obtain points $y_{k}\in\mathcal{P}_{\mathrm{VII}_{0}}\cap
\partial P'$. By choosing a suitable subsequence, we can assume
that $y_{k}$ converges to a type II point $y$ which is not
$P_{1}^{+}(II)$. If $y\notin\mathcal{F}_{\mathrm{II}}$, we can
apply the same kind of reasoning as before, using Proposition
\ref{prop:typeII} to get a contradiction to the consequences
of Corollary \ref{cor:ffaVII0}. If $y\in\mathcal{F}_{\mathrm{II}}$ 
we get, by applying the flow to the points $y_{k}$, a sequence
$z_{k}\in \mathcal{P}_{\mathrm{VII}_{0}}$ converging to $F$.
Applying the flow again, as before, we get a contradiction.
The Bianchi IX case is similar using Corollary \ref{cor:ffaIX}.
$\Box$

\section{Uniform convergence to the attractor}\label{section:uniform}
If $x$ constitutes initial data to
(\ref{eq:whsu})-(\ref{eq:constraint}) at $\tau=0$, then we denote the
corresponding solution $\Sp(\tau,x)$ and so on.

\begin{prop}
Let $2/3< \g\leq 2$ and let $K$ be a compact set of Bianchi IX 
initial data. Then $\No\Nt\Nth$
converges uniformly to zero on $K$. That is, for all $\epsilon>0$ 
there is a $T$ such that
\[
(\No\Nt\Nth)(\tau,x)\leq \epsilon
\]
for all $\tau\leq T$ and all $x\in K$.
\end{prop}

\textit{Proof}. Assume that $\No\Nt\Nth$ does not converge to zero
uniformly. Then there is an $\epsilon>0$, a sequence 
$\tau_{k}\rightarrow -\infty$ and $x_{k}\in K$ such that 
\[
(\No\Nt\Nth)(\tau_{k},x_{k})\geq \epsilon.
\]
We may assume, by choosing a convergent subsequence, that 
$x_{k}\rightarrow x_{*}$ as $k\rightarrow \infty$. Because of the
monotonicity of $(\No\Nt\Nth)(\cdot,x_{k})$, we conclude that 
\[
(\No\Nt\Nth)(\tau,x_{k})\geq \epsilon.
\]
for $\tau\in [\tau_{k},0]$. Thus
\[
(\No\Nt\Nth)(\tau,x_{*})=\lim_{k\rightarrow \infty}
(\No\Nt\Nth)(\tau,x_{k})\geq \epsilon
\]
for all $\tau\leq 0$. We have a contradiction. $\Box$

\begin{cor}\label{cor:uniparabcon}
Let $2/3<\g\leq 2$ and let $K$ be a compact set of Bianchi IX 
initial data. Then for every
$\epsilon>0$, there is a $T$ such that 
\[
\O+\Sp^2+\Sm^2\leq 1+\epsilon
\]
for all $x\in K$ and $\tau\leq T$.
\end{cor}

\textit{Proof}. As before. $\Box$

Consider
\[
d=\O+\No\Nt+\Nt\Nth+\Nth\No.
\]
\begin{prop}
Let $K$ be a compact set of generic Bianchi IX initial data with
$2/3<\g<2$. Then $d$ converges uniformly to zero on $K$.
\end{prop}

\textit{Proof}. Assume that $d$ does not converge to zero uniformly.
Then there is an $\eta>0$, a sequence $\tau_{k}\rightarrow -\infty$ 
and a sequence $x_{k}\in K$ such that 
\begin{equation}\label{eq:strange}
d(\tau_{k},x_{k})\geq \eta.
\end{equation}
We now prove that there is no sequence $s_{k_{n}}$ such that
$\tau_{k_{n}}\leq s_{k_{n}}\leq 0$ and
\[
d(s_{k_{n}},x_{k_{n}})\rightarrow 0.
\]
Assume there is. By Theorem \ref{thm:it}, there is no 
$\delta>0$ such that maximum of $h(\cdot,x_{k_{n}})$ in 
$[\tau_{k_{n}},s_{k_{n}}]$ exceeds $\delta$ for all $n$.
For $\delta$ small enough, we can apply Proposition 
\ref{prop:ocon} to the interval $[\tau_{k_{n}},s_{k_{n}}]$
to conclude that for some $n$, $\O$ cannot grow in very much in that 
interval either. We obtain a contradiction to (\ref{eq:strange})
for $\delta$ small enough and $n$ big enough.

Thus there is an $\epsilon>0$ such that 
\[
d(\tau,x_{k})\geq \epsilon
\]
for all $\tau\in[\tau_{k},0]$ and all $k$. Assume $x_{k}\rightarrow
x_{*}$. Then
\[
d(\tau,x_{*})=\lim_{k\rightarrow \infty}d(\tau,x_{k})\geq \epsilon>0
\]
for all $\tau\leq 0$.
But $x_{*}$ constitutes generic initial data. $\Box$

\section{Existence of non-special $\a$-limit points on the Kasner
circle}\label{section:nsale}

We know that there is an $\a$-limit point on the Kasner circle,
but in order to prove curvature blow up we wish to prove the 
existence of a non-special $\a$-limit point on the Kasner circle.

\begin{lemma}\label{lemma:iale}
Consider a generic Bianchi IX solution with $2/3<\g<2$. If 
it has a special point on the Kasner circle as an $\a$-limit point
then it has an infinite number of $\a$-limit points on the Kasner
circle.
\end{lemma}

\textit{Proof}. By applying the symmetries, we can assume that there 
is an $\a$-limit point on the Kasner circle with $(\Sp,\Sm)=(-1,0)$. 
Since the solution is not of Taub type, $(\Sp,\Sm)$ cannot converge
to $(-1,0)$ by Proposition \ref{prop:limchar}. Thus there is an
$1>\epsilon>0$ such that for each $T$ there is a $\tau\leq T$ such 
that $1+\Sp(\tau)\geq \epsilon$. Let $\tau_{k}\rightarrow -\infty$
be such that $\Sp(\tau_{k})\rightarrow -1$. 

Let $\eta>0$ satisfy
$\eta< \epsilon$. We wish to prove that there is a non-special 
$\a$-limit point on the Kasner circle with $1+\Sp\leq \eta$.
There is a sequence $t_{k}\leq \tau_{k}$ such
that $1+\Sp(t_{k})=\eta$ and $\Sp'(t_{k})\leq 0$ assuming $k$ is 
large enough. The condition on the derivative is possible to 
impose due to the fact that $1+\Sp$ eventually has to become greater 
than $\epsilon$. Choosing a suitable subsequence of $\{t_{k}\}$, we get
an $\a$-limit point which has to be a vacuum type I or II point
by Corollary \ref{cor:it}. If it is of type I, we get an $\a$-limit
point on the Kasner circle with $1+\Sp=\eta$ and we are done. The
$\a$-limit point cannot have $\No>0$, because of the condition on
the derivative, cf. the proof of Proposition \ref{prop:b2}. If it
is of type II with $\Nt$ or $\Nth$ greater than zero, we can apply 
the flow to get a type II solution, call it $x$, of $\a$-limit 
points to the original solution. Since a type II solution with 
$\Nt$ or $\Nth$ greater than zero satisfies $\Sp'<0$, the $\o$-limit
point $y$ of $x$ must have $1+\Sp<\eta$. By Proposition \ref{prop:b2}
$y\in \mathcal{K}_{2}\cup \mathcal{K}_{3}$, so that it is non-special.

Let $0<\eta_{1}<\epsilon$. As above, we can then construct a
non-special $\a$-limit point $x_{1}$ on the Kasner circle 
with $\Sp$ coordinate $\Sigma_{+,1}$
such that $1+\Sigma_{+,1}\leq \eta_{1}$. Assume we have constructed
non-special $\a$-limit points $x_{i}$ on the Kasner circle, 
$i=1,...,m$ with $\Sp$
coordinates $\Sigma_{+,i}$ satisfying $\Sigma_{+,i}<\Sigma_{+,i-1}$.
Let $0<\eta_{m+1}<1+\Sigma_{+,m}$. Then by the above we can construct
a non-special $\a$-limit point $x_{m+1}$ on the Kasner circle 
with $\Sp$ coordinate 
$\Sigma_{+,m+1}$, satisfying $\Sigma_{+,m+1}<\Sigma_{+,m}$. Thus the
solution has an infinite number of $\a$-limit points on the Kasner
circle. $\Box$

\begin{cor}\label{cor:nsale}
A generic Bianchi IX solution with $2/3<\g<2$ has at least three
non-special $\a$-limit points on the Kasner circle. Furthermore,
no $N_{i}$ converges to zero. 
\end{cor}

\textit{Proof}. Assume first that the solution has a special 
$\a$-limit point on the Kasner circle. By Lemma \ref{lemma:iale},
the first part of the lemma follows.
By the proof of Lemma \ref{lemma:iale}, there is a non-special
$\a$-limit point on the Kasner circle with $\Sp$ coordinate 
arbitrarily close to $-1$, say that it belongs to $\mathcal{K}_{2}$.
Repeated application of Proposition \ref{prop:BKL} then gives
$\a$-limit points first in $\mathcal{K}_{3}$, and after enough 
iterates, either an $\a$-limit point in $\mathcal{K}_{1}$, or 
a special $\a$-limit point on the Kasner circle with $\Sp=1/2$.
If the latter case occurs, a similar argument to the proof of 
Lemma \ref{lemma:iale} yields an $\a$-limit point on
$\mathcal{K}_{1}$. By Proposition \ref{prop:BKL}, we conclude
that there are $\a$-limit points with $\No>0$, with $\Nt>0$ and
with $\Nth>0$. 

Assume that there is no special $\a$-limit point on the Kasner circle.
Repeated application of the Kasner map yields $\a$-limit points in
$\mathcal{K}_{i}$, $i=1,2,3$, and the conclusions of the lemma follow 
as in the previous situation. $\Box$

\section{Conclusions}\label{section:conclusions}

Let us first state the conclusions concerning the asymptotics
of solutions to the equations of Wainwright and Hsu. We begin
with the stiff fluid case.

\begin{thm}\label{thm:stiff}
Consider a solution to (\ref{eq:whsu})-(\ref{eq:constraint})
with $\g=2$ and $\O>0$. Then the solution converges to a type I
point with $\Sp^2+\Sm^2<1$. For the Bianchi types other than I, 
we have the following additional restrictions. 
\begin{enumerate}
\item If the solution is of type II with $\No>0$, then 
  $\Sp<1/2$. 
\item For a type V$I_{0}$ or VI$I_{0}$ with
  $\Nt$ and $\Nth$ non-zero, then $\Sp\pm\sqrt{3}\Sm>-1$.
\item If the solution is of type VIII or IX, then
  $\Sp\pm\sqrt{3}\Sm>-1$ and $\Sp<1/2$.
\end{enumerate}
\end{thm}

\begin{figure}[hbt]
  \centerline{\psfig{figure=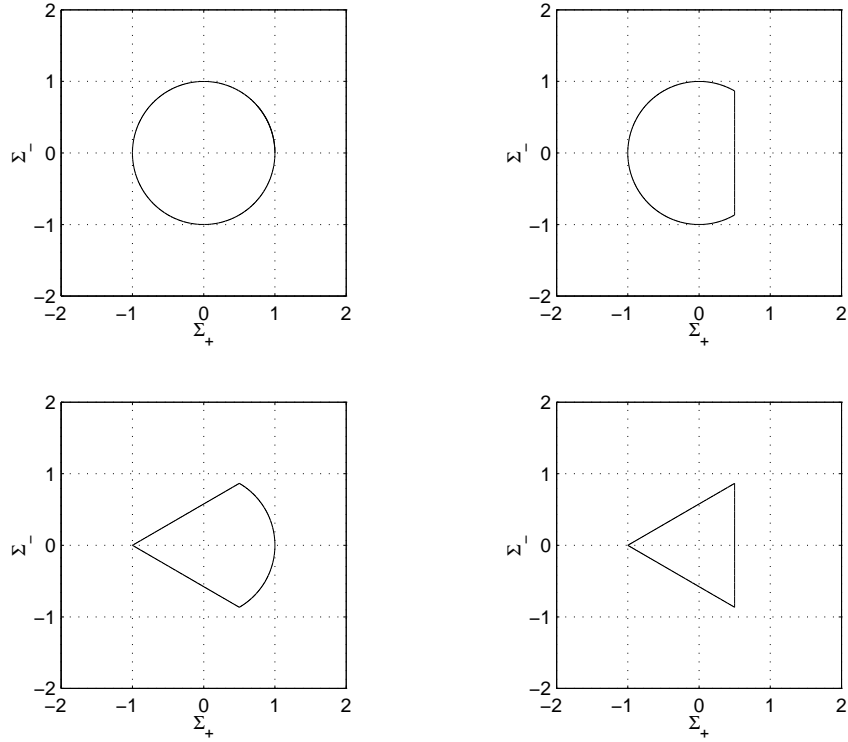,height=10cm}}
  \caption{The points to which the shear variables may converge
    for a stiff fluid.}
  \label{fig:stifftypes}
\end{figure}

\textit{Remark}. Figure \ref{fig:stifftypes} 
illustrates the restriction on the shear variables.
The types depicted are I, II, V$\mathrm{I}_{0}$ and
VI$\mathrm{I}_{0}$, and VIII and IX, counting from top left
to bottom right.

\textit{Proof}. The theorem follows from Propositions \ref{prop:notmi}
and \ref{prop:notli}. $\Box$

Consider now the case $2/3<\g<2$. Let $\mathcal{A}$ be the closure
of the type II vacuum points.

\begin{thm}
Consider a generic Bianchi IX solution $x$ with $2/3<\g<2$. 
Then it converges to the closure of the set of vacuum type II
points, that is
\[
\lim_{\tau\rightarrow -\infty}\inf_{y\in\mathcal{A}} \|x(\tau)-y\|=0
\]
where $\|\cdot\|$ is the Euclidean norm on $\mathbb{R}^{6}$. 
Furthermore, there are at least three non-special $\a$-limit points
on the Kasner circle.
\end{thm}

\textit{Remark}. One can start out arbitrarily close to this set
without converging to it, cf. Proposition \ref{prop:taub9}.

\textit{Proof}. The first part follows from Corollary \ref{cor:it}
and the second part follows from Corollary \ref{cor:nsale}. $\Box$

\textit{Proof of Theorems \ref{thm:classification} and 
\ref{thm:main}}. Let $(M,g)$ be the 
Lorentz manifold obtained in Lemma \ref{lemma:utveckling} with
topology $I\times G$. It is globally hyperbolic by Lemma 
\ref{lemma:hyperbolic}.

If the initial data satisfy
$\mathrm{tr}_{g}k=0$ for a development not of type IX, then 
it is causally geodesically complete and satisfies $\mu=0$ for the
entire development, by Lemma \ref{lemma:orient}
and Lemma \ref{lemma:completeness}. The first
part of Theorem \ref{thm:classification} follows. 

Consider initial data of type I, II, V$\mathrm{I}_{0}$,
VI$\mathrm{I}_{0}$ or VIII such that 
$\mathrm{tr}_{g}k\neq 0$. By Lemma \ref{lemma:orient} and 
Lemma \ref{lemma:completeness}, we may then time orient the development
so that it is future causally geodesically complete and past 
causally geodesically incomplete, and the second part of 
Theorem \ref{thm:classification} follows. The third part follows
from Lemma \ref{lemma:completeness}.

Consider an inextendible future directed causal geodesic 
in the above development. Since each hypersurface $\{v\}\times G$ is
a Cauchy hypersurface by Lemma \ref{lemma:hyperbolic}, the causal 
curve exhausts the interval $I$. 

1. If the solution is not of type IX, then the solution to 
(\ref{eq:dndt})-(\ref{eq:dmdt}), which is used in constructing the
class A development, corresponds to a solution to 
(\ref{eq:whsu})-(\ref{eq:constraint}), because of Lemma
\ref{lemma:orient}. Furthermore, $t\rightarrow t_{-}$ corresponds
to $\tau\rightarrow -\infty$, because of Lemma \ref{lemma:bn9corr}.  

a. In all the stiff fluid cases, the solution to 
(\ref{eq:whsu})-(\ref{eq:constraint}) converges to a non-vacuum
type I point by Theorem \ref{thm:stiff}, so that Lemma
\ref{lemma:rlp} and Lemma \ref{lemma:ricci2} yield the desired
conclusions in that case.

b. Type I, II and VI$\mathrm{I}_{0}$ with $1\leq\g<2$. That the 
Kretschmann scalar is unbounded in the cases stated in
Theorem \ref{thm:main} follows from
Proposition \ref{prop:typeI}, Proposition \ref{prop:typeII},
Proposition \ref{prop:typeVII0}, Lemma \ref{lemma:rlp} and 
and Lemma \ref{lemma:taublp}.

c. Non-vacuum solutions which are not of type IX. Then 
$\bar{R}_{\a\beta}\bar{R}^{\a\beta}$ is unbounded using
Lemma \ref{lemma:ricci2}. 

2. If the solution is of type IX, then half of a solution
to (\ref{eq:dndt})-(\ref{eq:dmdt}) corresponds to a
Bianchi IX solution to (\ref{eq:whsu})-(\ref{eq:constraint}),
because of Lemma \ref{lemma:recollapse}. By Lemma 
\ref{lemma:b9corr}, $t\rightarrow t_{\pm}$ corresponds
to $\tau\rightarrow -\infty$. 

a. In the stiff fluid case, we get the desired statement as
before.

b. If $1\leq\g<2$, we get the desired conclusions, concerning
blow up of the Kretschmann scalar, from
Corollary \ref{cor:nsale}, Proposition \ref{prop:taub9},
Lemma \ref{lemma:rlp} and Lemma \ref{lemma:taublp}.

c. Non-vacuum solutions. Then 
$\bar{R}_{\a\beta}\bar{R}^{\a\beta}$ is unbounded using
Lemma \ref{lemma:ricci2}.

Let us now prove that the development  is inextendible in the relevant
cases. Assume there is a connected Lorentz manifold 
$(\hat{M},\hat{g})$ of the same dimension, and a map 
$i:M\rightarrow \hat{M}$ which is 
an isometry onto its image,  with $i(M)\neq \hat{M}$. Then there
is a $p\in \hat{M}-i(M)$ and a timelike geodesic
$\gamma:[a,b]\rightarrow \hat{M}$ such that $\gamma([a,b))\subseteq
i(M)$ and $\gamma(b)=p$. Since $\gamma |_{[a,b)}$ can be considered
to be a future or past inextendible timelike geodesic in $M$, either it
has infinite length or a curvature invariant blows up along it, by
the above arguments. Both possibilities lead to a contradiction. 
Theorem \ref{thm:main} follows. $\Box$

\section{Asymptotically velocity term dominated behaviour near the
  singularity}\label{section:avtds}
In this section, we consider the asymptotic behaviour of Bianchi
VIII and IX stiff fluid solutions from another point of view.
We wish to compare our results with \cite{fuchs}, a paper which deals 
with analytic solutions of Einstein's equations
coupled to a scalar field or a stiff fluid. In \cite{fuchs}, Andersson 
and Rendall prove that given a certain kind of solution to the so 
called velocity dominated system, there is a unique solution of Einstein's
equations coupled to a stiff fluid approaching the velocity dominated 
solution asymptotically. We will be more specific concerning the
details below. The question which arises is to what extent it is
natural to assume that a solution has the asymptotic behaviour
they prescribe. We show here that all Bianchi VIII and IX stiff fluid
solutions exhibit such asymptotic behaviour. 

In order to speak about
velocity term dominance, we need to have a foliation. In our case, there
is a natural foliation given by the spatial hypersurfaces of
homogeneity. Relative to this foliation, we can express the metric as
in (\ref{eq:metric}) according to Lemma \ref{lemma:utveckling}.
In what follows, we will use the frame $e_{i}'$ appearing in Lemma
\ref{lemma:utveckling}, and Latin indices will refer to this frame.
Let $g$ be the Riemannian metric, and $k$ the second fundamental form
of the spatial hypersurfaces of homogeneity, so that
\begin{equation}\label{eq:gij}
g_{ij}=\bar{g}(e_{i}',e_{j}')=a_{i}^{-2}\delta_{ij},
\end{equation}
where $\bar{g}$ is as in (\ref{eq:metric}). The constraint equations
in our situation are
\begin{eqnarray}
R-k_{ij}k^{ij}+(\mathrm{tr}k)^2 & = & 2\mu \label{eq:eincon1} \\
\nabla^{i}k_{ij}-\nabla_{j}(\mathrm{tr}k) & = & 0\label{eq:eincon2},
\end{eqnarray}
which are the same as (\ref{eq:constraint1}) and (\ref{eq:commute2})
respectively. The evolution equations are
\begin{eqnarray}
\partial_{t}g_{ij} & = & -2k_{ij} \label{eq:ev1} \\
\partial_{t}k^{i}_{\ j} & = & R^{i}_{\ j}+(\mathrm{tr}k)k^{i}_{\ j}.
\label{eq:ev2}
\end{eqnarray}
The evolution equation for the matter is
\begin{equation}\label{eq:evmat}
\partial_{t}\mu=2(\mathrm{tr}k)\mu.
\end{equation}
We wish to compare solutions to these equations with solutions
to the so called velocity dominated system. This system also consists
of constraints and evolution equations, and we will denote the velocity
dominated solution with a left superscript zero. The constraints are
\begin{eqnarray}
-{}^{0}k_{ij}{}^{0}k^{ij}+(\mathrm{tr}{}^{0}k)^2 & = & 2{}^{0}\mu
 \label{eq:vtdcon1}
 \\
{}^{0}\nabla^{i}({}^{0}k_{ij})-{}^{0}\nabla_{j}(\mathrm{tr}{}^{0}k) 
& = & 0. \label{eq:vtdcon2}
\end{eqnarray}
The evolution equations are 
\begin{eqnarray}
\partial_{t}{}^{0}g_{ij} & = & -2{}^{0}k_{ij} \label{eq:vtdev1} \\
\partial_{t}{}^{0}k^{i}_{\ j}&=&(\mathrm{tr}{}^{0}k){}^{0}k^{i}_{\ j},
\label{eq:vtdev2}
\end{eqnarray}
and the matter equation is
\begin{equation}\label{eq:vtdmat}
\partial_{t}{}^{0}\mu=2(\mathrm{tr}{}^{0}k){}^{0}\mu.
\end{equation}
We raise and lower indices of the velocity dominated system with
the velocity dominated metric.
In \cite{fuchs}, Andersson and Rendall prove that given an analytic 
solution to (\ref{eq:vtdcon1})-(\ref{eq:vtdmat}) on $S\times
(0,\infty)$ such that $t\mathrm{tr}{}^{0}k=-1$, and such that 
the eigenvalues of $-t{}^{0}k^{i}_{\ j}$ are positive, there is
a unique analytic solution to (\ref{eq:eincon1})-(\ref{eq:evmat})
asymptotic, in a suitable sense, to the solution of the velocity
dominated system. In fact, they prove this statement in a more 
general setting than the one given above. We have specialized to 
our situation. Observe the condition on the eigenvalues of 
$-t{}^{0}k^{i}_{\ j}$. Our goal is to prove that this is a 
natural condition in the Bianchi VIII and IX cases.

\begin{thm}
Consider a Bianchi VIII or IX stiff fluid development as in
Lemma \ref{lemma:utveckling} with $\mu_{0}>0$. 
Choose time coordinate so that $t_{-}=0$. Then there is a solution to 
(\ref{eq:vtdcon1})-(\ref{eq:vtdmat}) such that 
$t\mathrm{tr}{}^{0}k=-1$, the eigenvalues of $-t{}^{0}k^{i}_{\ j}$ are
positive, and the following estimates hold
\begin{enumerate}
\item
${}^{0}g^{il}g_{lj}=\delta^{i}_{\ j}+o(t^{\a^{i}_{\ j}})$
\item
$k^{i}_{\ j}={}^{0}k^{i}_{\ j}+o(t^{-1+\a^{i}_{\ j}})$
\item
$\mu={}^{0}\mu+o(t^{-2+\beta_{1}})$,
\end{enumerate}
where $\a^{i}_{\ j}$ and $\beta_{1}$ are positive real numbers.
\end{thm}

\textit{Remark}. In \cite{fuchs} two more estimates occur. They are
not included here as they are replaced by equalities in our situation.
Observe that the difficulties encountered in \cite{fuchs} concerning
the non-diagonal terms of $k^{i}_{\ j}$ disappear in the present 
situation.

\textit{Proof}. Below we will use the results of Lemma 
\ref{lemma:utveckling} and its proof implicitly.
When we speak of $\theta_{ij}$, $\sigma_{ij}$, $\theta$, $n_{ij}$
and $\mu$, we will refer to the solution of
(\ref{eq:dndt})-(\ref{eq:dmdt}) and the indices of these objects
should not be understood in terms of evaluation on a frame.
Since $\theta_{ij}$ and so on are all diagonal, we will sometimes
write $\theta_{i}$ etc instead, denoting diagonal component $i$.
There are two relevant frames: $e_{i}'$ and $e_{i}=a_{i}e_{i}'$.
The latter frame yields $n_{ij}$ through (\ref{eq:ndef}).
When we speak of $k^{i}_{\ j}$, $R_{ij}$ and so on, we will always
refer to the frame $e_{i}'$.
We have 
\[
k^{i}_{\ j}=-\theta_{i}\delta^{i}_{\ j}
\]
(no summation on $i$). The metric is given by (\ref{eq:gij}) above.
Let us choose
\begin{equation}\label{eq:nollkdef}
{}^{0}k^{i}_{\ j}=-{}^{0}\theta_{i}\delta^{i}_{\ j},
\end{equation}
let
${}^{0}\theta={}^{0}\theta_{1}+{}^{0}\theta_{2}+{}^{0}\theta_{3}$
and
\[
{}^{0}g_{ij}={}^{0}a_{i}^{-2}\delta_{ij}
\]
(no summation on $i$). Because of (\ref{eq:nollkdef}), equation
(\ref{eq:vtdcon2}) will be satisfied since it is a statement 
concerning the commutation of ${}^{0}k^{i}_{\ j}$ and $n_{ij}$.
The existence interval for the solution to 
Einstein's equations is $(0,t_{+})$ by our conventions, and 
since we wish to have $t\mathrm{tr}{}^{0}k=-1$ we need to define
${}^{0}\theta(t)=1/t$. Observe that 
${}^{0}\theta_{i}/{}^{0}\theta$ is constant in time,
and that $\theta_{i}/\theta$ converges to a positive value as
$t\rightarrow 0$; this is a consequence of Theorem \ref{thm:stiff}
and the definition (\ref{eq:spsmdef}) of the variables $\Sp$ and
$\Sm$. Choose ${}^{0}\theta_{i}$ so that ${}^{0}\theta_{i}/{}^{0}\theta$
coincides with the limit of $\theta_{i}/\theta$. Similarly
${}^{0}\mu/{}^{0}\theta^2$ is constant, $\mu/\theta^2$ converges
to a positive value, and we choose ${}^{0}\mu/{}^{0}\theta^2$ to be
the limit. Since $R/\theta^2$ is a polynomial in the
$N_{i}$ and the $N_{i}$ converge to zero by Theorem \ref{thm:stiff},
equation (\ref{eq:vtdcon1}) will be fulfilled. By our choices, 
(\ref{eq:vtdev2}) and (\ref{eq:vtdmat}) will also be fulfilled.
We will specify the initial value of ${}^{0}a_{i}$ later on, and
then define ${}^{0}a_{i}$ by demanding that (\ref{eq:vtdev1}) 
holds.

It will be of interest to estimate terms of the form 
$R^{i}_{\ j}/\theta^2$. These terms are quadratic polynomials
in the $N_{i}$. By abuse of notation, we will write $N_{i}(\tau)$
when we wish to evaluate $N_{i}$ in the Wainwright-Hsu time
(\ref{eq:dtdtau}) and $N_{i}(t)$ when we wish to evaluate in the
time used in this theorem. By Theorem \ref{thm:stiff}, there is an 
$\epsilon>0$ and a $\tau_{0}$ such that 
\[
|N_{i}(\tau)|\leq \exp(\epsilon\tau)
\]
for all $\tau\leq \tau_{0}$. We wish to rewrite this estimate in terms
of $t$. Let us begin with (\ref{eq:raychaudhuri2}). Since we can
assume that $q\leq 3$ for $\tau\leq \tau_{0}$ we get
\[
\theta(\tau)\leq \exp[-4(\tau-\tau_{0})]\theta(\tau_{0}),
\]
so that for $\tau_{1},\tau\leq \tau_{0}$ we get, using
(\ref{eq:dtdtau}),
\[
t(\tau)-t(\tau_{1})=\int_{\tau_{1}}^{\tau}\frac{3}{\theta}
ds\geq\frac{3}{4\theta(\tau_{0})}(\exp[4(\tau-\tau_{0})]-
\exp[4(\tau_{1}-\tau_{0})]).
\]
Letting $\tau_{1}$ go to $-\infty$ and observing that 
$t(-\infty)=0$, cf. Lemma \ref{lemma:bn9corr} and Lemma
\ref{lemma:b9corr}, we get for some constant $c$
\[
e^{4\tau}\leq c t(\tau),
\]
so that
\[
N_{i}(t)\leq \exp(\epsilon\tau(t))\leq C t^{\eta}
\]
for some positive number $\eta$. Consequently expressions
such as $R^{i}_{\ j}/\theta^2$ and $R/\theta^2$ satisfy
similar bounds.

Let us now prove the estimates formulated in the statement of the
theorem. Observe that for $t$ small enough, we have
\[
-\theta=\mathrm{tr}k(t)=-(\int_{0}^{t}[\frac{R}{\theta^2}+1]ds)^{-1},
\]
since the singularity is at $t=0$ and $\mathrm{tr}k$ must become
unbounded at the singularity, cf. Lemma \ref{lemma:bn9corr},
\ref{lemma:b9corr} and (\ref{eq:raychaudhuri2}). Thus we get
\begin{equation}\label{eq:trkest}
\theta-{}^{0}\theta=-\int_{0}^{t}\frac{R}{\theta^2}ds
\{t\int_{0}^{t}[\frac{R}{\theta^2}+1]ds\}^{-1}=o(t^{-1+\eta_{1}})
\end{equation}
for some $\eta_{1}>0$. In order to make the estimates concerning
$k^{i}_{\ j}$, we need only consider $\theta_{i}$ and
${}^{0}\theta_{i}$. We have
\[
\partial_{t}(\frac{\theta_{i}}{\theta}-\frac{{}^{0}\theta_{i}}
{{}^{0}\theta})=\partial_{t}\frac{\theta_{i}}{\theta}=
\frac{\theta_{i}R-R^{i}_{\ i}\theta}{\theta^2}
\]
with no summation on the $i$ in $R^{i}_{\ i}$. This computation, 
together with the estimates above and the fact that
$\theta_{i}/\theta-{}^{0}\theta_{i}/{}^{0}\theta$ converges to 
zero, yields the estimate 
\begin{equation}\label{eq:relest}
\frac{\theta_{i}}{\theta}-\frac{{}^{0}\theta_{i}}
{{}^{0}\theta}=o(t^{\eta_{2}}),
\end{equation}
for some $\eta_{2}>0$. However,
\begin{equation}\label{eq:junk}
\frac{\theta_{i}}{\theta}-\frac{{}^{0}\theta_{i}}
{{}^{0}\theta}=
\frac{\theta_{i}-{}^{0}\theta_{i}}{\theta}+
\frac{{}^{0}\theta_{i}}{{}^{0}\theta}
\frac{{}^{0}\theta-\theta}{\theta}.
\end{equation}
Combining (\ref{eq:trkest}), (\ref{eq:relest}) and (\ref{eq:junk}),
we get estimate $2$ of the theorem. Similarly, we have
\[
\partial_{t}(\frac{\mu}{\theta^2}-\frac{{}^{0}\mu}{{}^{0}\theta^2})=
\partial_{t}\frac{\mu}{\theta^2}=
\frac{2\mu R}{\theta^3}.
\]
Integrating, using the fact that $\mu/\theta^2$ converges
to ${}^{0}\mu/{}^{0}\theta^2$, we get
\begin{equation}\label{eq:junk2}
\frac{\mu}{\theta^2}-\frac{{}^{0}\mu}{{}^{0}\theta^2}=o(t^{\eta_{3}})
\end{equation}
where $\eta_{3}>0$. Using
\[
\frac{\mu}{\theta^2}-\frac{{}^{0}\mu}{{}^{0}\theta^2}=
\frac{\mu-{}^{0}\mu}{\theta^2}+
\frac{{}^{0}\mu}{{}^{0}\theta^2}
\frac{{}^{0}\theta^2-\theta^2}{\theta^2},
\]
(\ref{eq:trkest}) and (\ref{eq:junk2}), we get estimate $3$ of the 
theorem. Finally, we need to specify the initial value of 
${}^{0}a_{i}$ and prove estimate $1$. Since 
\[
\partial_{t}a_{i}=-\theta_{i}a_{i},
\]
(no summation on $i$) and similarly for ${}^{0}a_{i}$, we get
\[
\partial_{t}\frac{a_{i}}{{}^{0}a_{i}}=
\frac{a_{i}}{{}^{0}a_{i}}({}^{0}\theta_{i}-\theta_{i}).
\]
By our estimates on ${}^{0}\theta_{i}-\theta_{i}$, we see that this
implies that $a_{i}/{}^{0}a_{i}$ converges as $t\rightarrow 0$.
Choose the value of ${}^{0}a_{i}$ at one point in time  so that this
limit is $1$. We thus get, using estimate $2$ of the theorem,
\[
\frac{a_{i}}{{}^{0}a_{i}}-1=o(t^{\a^{i}_{\ j}}).
\]
Estimate $1$ of the theorem now follows from this estimate and the
fact that
\[
{}^{0}g^{il}g_{lj}=\left(\frac{{}^{0}a_{i}}{a_{i}}\right)^{2}
\delta^{i}_{\ j}.
\]
The theorem follows. $\Box$

\section{Appendix}

The goal of this appendix is to relate the asymptotic behaviour
of solutions to the ODE (\ref{eq:whsu})-(\ref{eq:constraint})
to the behaviour of the spacetime in the incomplete directions
of inextendible causal curves. We proceed as follows.
\begin{enumerate}
\item First, we formulate Einstein's equations as an ODE, assuming that
the spacetime has a given structure (\ref{eq:structure}). The
first formulation is due to Ellis and MacCallum. We also
relate this formulation to the one by Wainwight and Hsu. 

\item Given initial data as in Definition \ref{def:data}, we then 
  show how to construct a Lorentz manifold as in (\ref{eq:structure}),
  satisfying Einstein's equations and with initial data as specified,
  using the equations of Ellis and MacCallum. We also prove some
  properties of this development such as Global hyperbolicity
  and answer some questions concerning causal geodesic completeness.

\item Finally, we relate the asymptotic behaviour of solutions to 
  (\ref{eq:whsu})-(\ref{eq:constraint}) to the question of curvature
  blow up in the development obtained by the above procedure.
\end{enumerate}

We consider a special class of spatially homogeneous
four dimensional spacetimes of the form
\begin{equation}\label{eq:structure}
(\bar{M},\bar{g})=(I\times G,-d t^2+\chi_{ij}(t)\xi^{i}\otimes 
\xi^{j}),
\end{equation}
where $I$ is an open interval, $G$ is a Lie group of class A, 
$\chi_{ij}$ is a smooth positive definite matrix and the $\xi^{i}$ 
are the duals of a left invariant basis on $G$.
The stress energy tensor is assumed to be given by
\begin{equation}\label{eq:set}
T=\mu dt^{2}+p(\bar{g}+dt^{2}),
\end{equation}
where $p=(\g-1)\mu$. 
Below, Latin indices will be raised and lowered by $\delta_{ij}$.

Consider a four dimensional $(\bar{M},\bar{g})$ as in 
(\ref{eq:structure}) with $G$ of class A. 
In order to define the different variables, we specify a
suitable orthonormal basis. Let 
$e_{0}=\partial_{t}$ and $e_{i}=a_{i}^{\ j}Z_{j}$, i=1,2,3, be an 
orthonormal basis, where  $a$ is a $C^{\infty}$ matrix valued function
of $t$  and the $Z_{i}$ are the duals of $\xi^{i}$. 

By the following argument, we can assume that
$<\bar{\nabla}_{e_{0}}e_{i},e_{j}>=0$. Let
the matrix valued function $A$ satisfy $e_{0}(A)+AB=0$,
$A(0)=\mathrm{Id}$ where $B_{ij}=
<\bar{\nabla}_{e_{0}}e_{i},e_{j}>$ and $\mathrm{Id}$ is the 
$3\times 3$ identity matrix. Then $A$ is smooth and $SO(3)$ valued 
and if $e_{i}'=A_{i}^{\ j}e_{j}$, then  
$<\bar{\nabla}_{e_{0}}e_{i}',e_{j}'>=0$.

Let
\begin{equation}
\theta(X,Y)=<\bar{\nabla}_{X}e_{0},Y>,
\end{equation}
$\theta_{\alpha \beta}=\theta(e_{\alpha},e_{\beta})$ and
$[e_{\beta},e_{\gamma}]=\gamma_{\beta \gamma}^{\alpha}e_{\alpha}$  
where Greek indices run from $0$ to $3$. The objects $\theta_{\alpha 
\beta}$ and $\gamma_{\beta \gamma}^{\alpha}$ will be viewed as
smooth functions from $I$ to some suitable $\mathbb{R}^{k}$, and
our variables will be defined in terms of them. 

Observe that 
$[Z_{i},e_{0}]=0$. The $e_{i}$ span the tangent space of $G$, and 
$<[e_{0},e_{i}],e_{0}>=0$. 
We get $\theta_{00}=\theta_{0i}=0$ and $\theta_{\alpha\beta}$
symmetric. We also have $\gamma_{ij}^{0}=\gamma_{0i}^{0}=0$ and
$\gamma_{0j}^{i}=-\theta_{ij}$. We let $n$ be defined as in 
(\ref{eq:ndef}) and
\[
\sigma_{ij}=\theta_{ij}-\frac{1}{3}\theta \delta_{ij},
\]
where we by abuse of notation have written $\mathrm{tr}(\theta)$ as 
$\theta$. 

We express Einstein's equations in terms of $n$, $\sigma$ 
and $\theta$. The Jacobi identities for $e_{\alpha}$ yield
\begin{equation}\label{eq:dndt}
e_{0}(n_{ij})-2n_{k(i}^{}\sigma_{j)}^{\ k}+\frac{1}{3}\theta n_{ij}=0.
\end{equation}
The $0i$-components of the Einstein equations are equivalent to
\begin{equation}\label{eq:commute2}
\sigma_{i}^{\ k}n_{kj}-n_{i}^{\ k}\sigma_{kj}=0.
\end{equation}
Letting  
$b_{ij}=2n_{i}^{\ k}n_{kj}-\mathrm{tr}(n) n_{ij}$ and
$s_{ij}=b_{ij}-\frac{1}{3}\mathrm{tr}(b)\delta_{ij}$,
the trace free part of the $ij$ equations are
\begin{equation}\label{eq:dsdt}
e_{0}(\sigma_{ij})+\theta\sigma_{ij}+s_{ij}=0.
\end{equation}
The $00$-component yields the Raychaudhuri equation
\begin{equation}\label{eq:raychaudhuri}
e_{0}(\theta)+\theta_{ij}\theta^{ij}+\frac{1}{2}(3\g-2)\mu=0,
\end{equation}
and using this together with the trace of the $ij$-equations yields a
constraint
\begin{equation}\label{eq:constraint1}
\sigma_{ij}\sigma^{ij}+(n_{ij}n^{ij}-\frac{1}{2}\mathrm{tr}(n)^2)+
2\mu=\frac{2}{3}\theta^2.
\end{equation}
Equations (\ref{eq:dndt})-(\ref{eq:constraint1}) are special cases
of equations given in Ellis and MacCallum \cite{emac}. 
At a point $t_{0}$, we may diagonalize $n$ and $\sigma$ simultaneously
since they commute (\ref{eq:commute2}). Rotating $e_{\alpha}$ by the 
corresponding
element of $SO(3)$ yields upon going through the definitions that the
new $n$ and $\sigma$ are diagonal at $t_{0}$. Collect the 
off-diagonal terms of $n$ and $\sigma$ in one vector $v$. By 
(\ref{eq:dndt}) and (\ref{eq:dsdt}), there is a time dependent matrix
$C$ such that $\dot{v}=Cv$ so that $v(t)=0$ for all $t$, since 
$v(t_{0})=0$. Since the rotation was time independent, 
$<\nabla_{e_{0}}e_{i},e_{j}>=0$ holds in the new basis. 

The fact that $T$ is divergence free yields
\begin{equation}\label{eq:dmdt}
e_{0}(\mu)+\g\theta\mu=0.
\end{equation}

Introduce, as in 
Wainwright and Hsu \cite{whsu},
\begin{eqnarray*}
\Sigma_{ij}=\sigma_{ij}/\theta \\
N_{ij}=n_{ij}/\theta \\
\O=3\mu/\theta^2
\end{eqnarray*}
and define a new time coordinate $\tau$, independent of time
orientation, satisfying
\begin{equation}\label{eq:dtdtau}
\frac{d t}{d\tau}=\frac{3}{\theta}.
\end{equation}
For Bianchi IX developments,
we only consider the part of spacetime where 
$\theta$ is strictly positive or strictly negative. Let
\begin{equation}\label{eq:spsmdef}
\Sp=\frac{3}{2}(\Sigma_{22}+\Sigma_{33})\ \mathrm{and}\ 
\Sm=\frac{\sqrt{3}}{2}(\Sigma_{22}-\Sigma_{33}). 
\end{equation}
If we let $N_{i}$ be the 
diagonal elements of $N_{ij}$, equations (\ref{eq:dndt}) and 
(\ref{eq:dsdt}) turn into (\ref{eq:whsu}) with definitions as in
(\ref{eq:whsudef}), except for the expression for $\O'$. It can 
however be derived from (\ref{eq:dmdt}). The constraint
(\ref{eq:constraint1}) turns into (\ref{eq:constraint}).
The Raychaudhuri equation (\ref{eq:raychaudhuri}) takes the form
\begin{equation}\label{eq:raychaudhuri2}
\theta'=-(1+q)\theta.
\end{equation}
Before using the equations of Ellis and MacCallum to construct 
a development, it is convenient to know that one can make some 
simplifying assumptions concerning the choice of basis. The next
lemma fulfills this objective, and also proves the classification
of the class A Lie algebras mentioned in the introduction.

\begin{lemma}\label{lemma:liealg}
Table \ref{table:bianchiA} constitutes a classification of the class
A Lie algebras. Consider an arbitrary basis $\{e_{i}\}$ of the Lie 
algebra. Then by applying an orthogonal matrix to it, we can 
construct a basis $\{e_{i}'\}$ such that the corresponding $n'$ 
defined by (\ref{eq:ndef}) is diagonal, with diagonal elements of one 
of the types given in Table \ref{table:bianchiA}.
\end{lemma}
\textit{Proof}. Let $e_{i}$ be a 
basis for the Lie algebra and $n$ be defined as in (\ref{eq:ndef}).
If we change the basis according to 
$e_{i}'=(A^{-1})_{i}^{\ j}e_{j}$, then $n$ transforms to
\begin{equation}\label{eq:transform}
n'=(\det A)^{-1}A^{t}nA
\end{equation}
Since $n$ is symmetric, we assume from here on that the basis is such 
that it is diagonal. The matrix $A=\mathrm{diag}(1\ 1\ -1)$ changes 
the sign of $n$. A suitable orthogonal matrix performs even 
permutations of the diagonal. The number of non-zero elements on the 
diagonal is invariant under transformations (\ref{eq:transform})
taking one diagonal matrix to another. If $A=(a_{ij})$ and the 
diagonal matrix $n'$ is constructed as in (\ref{eq:transform}), we have
$n'_{kk}=(\det A)^{-1}\sum_{i=1}^{3}a_{ik}^{2}n_{ii}$, so that if all 
the diagonal elements of $n$ have the same sign, the same is true
for $n'$. The statements of the lemma follow. $\Box$

We now prove that if we begin with initial data as in Definition
\ref{def:data}, we get a development as in Definition
\ref{def:development} of the form (\ref{eq:structure}), with certain
properties.

\begin{lemma}\label{lemma:utveckling}
Fix $2/3<\g\leq 2$.
Let $G,\ g,\ k$ and $\mu_{0}$ be initial data as in Definition
\ref{def:data}. Then there is an orthonormal basis $e_{i}'$ $i=1,2,3$ 
of the Lie algebra such that $n_{ij}'$ defined by (\ref{eq:ndef})
and $k_{ij}=k(e_{i}',e_{j}')$ are diagonal and $n_{ij}'$ is of one 
of the forms given in Table \ref{table:bianchiA}. Let 
\[
\theta(0)=-\mathrm{tr}_{g} k,\
\sigma_{ij}(0)=-k(e_{i}',e_{j}')+\frac{1}{3}\theta(0)\delta_{ij},\
n_{ij}(0)=n_{ij}'\ and\ \mu(0)=\mu_{0}.
\]
Solve (\ref{eq:dndt}), (\ref{eq:dsdt}), (\ref{eq:raychaudhuri}) and
(\ref{eq:dmdt}) with these conditions as initial
data to obtain $n,\ \sigma,\ \theta$ and $\mu$, and let $I$ be the
corresponding existence interval. 
Then there are smooth functions $a_{i}:I\rightarrow (0,\infty)$
$i=1,2,3$, with $a_{i}(0)=1$, such that 
\begin{equation}\label{eq:metric}
\bar{g}=-dt^{2}+\sum_{i=1}^{3}a_{i}^{-2}(t)\xi^{i}\otimes \xi^{i},
\end{equation}
where $\xi^{i}$ is the dual of $e_{i}'$, satisfies  Einstein's
equations (\ref{eq:einstein}) on $\bar{M}=I\times G$, with 
$T$ as in (\ref{eq:tdef}) with $u=e_{0}$, $\mu$ as above and
$p=(\g-1)\mu$. Furthermore,
\[
<\bar{\nabla}_{e_{i}}e_{0},e_{j}>=\sigma_{ij}+\frac{1}{3}\theta
\delta_{ij},
\]
where $\bar{\nabla}$ is the Levi-Civita connection of $\bar{g}$
and $e_{i}=a_{i}e_{i}'$, 
if we consider the left hand side to be a function of
$t$. Consequently, the induced metric and second fundamental form on 
$\{0\}\times G$ are $g$ and $k$, and we have a development satisfying 
the conditions of Definition \ref{def:development}.
\end{lemma} 

\textit{Proof}. Let $e_{i}'$, $i=1,2,3$ 
be a left invariant orthonormal basis. We can assume the 
corresponding $n'$ to be of one of the forms given in Table 
\ref{table:bianchiA} by Lemma \ref{lemma:liealg}. 
The content of (\ref{eq:con2}) is that $k_{ij}=k(e_{i}',e_{j}')$ and 
$n'$ are to commute. We may thus also assume $k_{ij}$ to 
be diagonal without changing the earlier conditions of the
construction. If we let $n(0)=n'$, $\theta(0)=
-\mathrm{tr}_{g}k$, $\sigma_{ij}(0)=-k_{ij}+\theta\delta_{ij}/3$ 
and $\mu(0)=\mu_{0}$,
then (\ref{eq:con1}) is the same as (\ref{eq:constraint1}). Let $n$,
$\sigma$, $\theta$ and $\mu$ satisfy (\ref{eq:dndt}), (\ref{eq:dsdt}),
(\ref{eq:raychaudhuri}) and (\ref{eq:dmdt}) with initial values as 
specified above. 
Since (\ref{eq:constraint1}) is satisfied at $0$, it is satisfied 
for all times. For reasons given in connection with
(\ref{eq:constraint1}), $n$ and $\sigma$ will 
remain diagonal so that (\ref{eq:commute2}) will always hold. 
Let $n_{i}$ and $\sigma_{i}$ denote the diagonal elements of $n$ and
$\sigma$ respectively. 

How are we to define the $a_{i}$ in the statement of the lemma?
The $\tilde{n}$ obtained from $e_{i}$ by (\ref{eq:ndef}) should 
coincide with $n$. This leads us to the following definitions.
Let $f_{i}(0)=1$ and
$\dot{f_{i}}/f_{i}=2\sigma_{i}-\theta/3$.
Let $a_{i}=(\Pi_{j\neq i} f_{j})^{1/2}$ and
define $e_{i}=a_{i}e_{i}'$. Then $\tilde{n}$ associated to $e_{i}$ 
equals $n$. We complete the basis by letting $e_{0}=\partial_{t}$. 
Define a metric $<\cdot,\cdot>$ on $\bar{M}$ by demanding 
$e_{\alpha}$ to be orthonormal with $e_{0}$ timelike and $e_{i}$ 
spacelike, and let $\bar{\nabla}$ be the associated Levi-Civita
connection. Compute $<\bar{\nabla}_{e_{0}}e_{i},e_{j}>=0$.
If $\tilde{\theta}(X,Y)=<\bar{\nabla}_{X}e_{0},Y>$ and
$\tilde{\theta}_{\mu\nu}=\tilde{\theta}(e_{\mu},e_{\nu})$, then 
$\tilde{\theta}_{00}=\tilde{\theta}_{i0}=
\tilde{\theta}_{0i}=0$. Furthermore,
\[
\frac{1}{a_{j}}e_{0}(a_{j})\delta_{ij}=-\tilde{\theta}_{ij}
\]
(no summation over $j$) so that $\tilde{\theta}_{ij}$ is diagonal and
$\mathrm{tr}\tilde{\theta}=\theta$. Finally,
\[
-\tilde{\sigma}_{ii}=-\tilde{\theta}_{ii}+\frac{1}{3}\theta=
-\sigma_{i}.
\]
The lemma follows by considering the derivation of the equations of 
Ellis and MacCallum. $\Box$

\begin{definition}\label{def:utveckling}
A development as in Lemma \ref{lemma:utveckling} will be called
a \textit{class A development}. We will also assign a type to
such a development according to the type of the initial data.
\end{definition}

The next thing to prove is that each $\bar{M}_{v}=\{ v\}\times G$ is
a Cauchy surface, but first we need a lemma.

\begin{lemma}\label{lemma:complete}
Let $\rho$ be a left invariant Riemannian metric on a Lie group $G$.
Then $\rho$ is geodesically complete.
\end{lemma}
\textit{Proof}. 
Assume $\g:(t_{-},t_{+})\rightarrow G$ is a geodesic satisfying
$\rho(\g',\g')=1$, with $t_{+}<\infty$. There is a $\delta>0$ such 
that every geodesic $\lambda$ satisfying $\lambda(0)=e$, the identity 
element of $G$, and $\lambda'(0)=v$ with $\rho(v,v)\leq 1$ is defined 
on $(-\delta,\delta)$. If $L_{h}:G\rightarrow G$ is defined by
$L_{h}(h_{1})=hh_{1}$, then $L_{h}$ is by definition an isometry.
Let $t_{0}\in (t_{-},t_{+})$ satisfy  $t_{+}-t_{0}\leq \delta/2$. 
Let $v\in T_{e}G$ be the vector corresponding to 
$\g'(t_{0})$ under the isometry $L_{\g(t_{0})}$. Let $\lambda$ be 
a geodesic with $\lambda(0)=e$ and $\lambda'(0)=v$. Then
$L_{\g(t_{0})}\circ \lambda$ is a geodesic extending $\g$. $\Box$

Let us be precise concerning the concept Cauchy surface. 

\begin{definition}
Consider a time oriented  Lorentz manifold $(M,g)$.
Let $I$ be an interval in $\mathbb{R}$ and $\g:I\rightarrow M$ be  a
continuous map which is smooth except for a finite
number of points. We say that $\g$ is a \textit{future directed 
causal}, \textit{timelike} or \textit{null}
curve if at each $t\in I$ where $\g$ is differentiable, $\g'(t)$ is
a future oriented causal, timelike or null vector respectively.
We define past directed curves similarly. A \textit{causal} curve
is a curve which is either a future directed causal curve or a 
past directed causal curve and similarly for timelike and null 
curves. If there is a curve $\lambda:I_{1}\rightarrow M$ such that
$\g(I)$ is properly contained in $\lambda(I_{1})$, then $\g$ is
said to be \textit{extendible}, otherwise it is called
\textit{inextendible}. A subset $S\subset M$
is called a \textit{Cauchy surface} if it is intersected exactly
once by every inextendible causal curve. A Lorentz manifold as above
which admits a Cauchy surface is said to be \textit{Globally
  hyperbolic}.
\end{definition}

\begin{lemma}\label{lemma:hyperbolic}
For a class A development, each
$\bar{M}_{v}=\{ v\}\times G$ is a Cauchy surface.
\end{lemma}

\textit{Proof}. The metric is given by (\ref{eq:metric}). A causal 
curve cannot intersect $\bar{M}_{v}$ twice since the $t$-component 
of such a curve must
be strictly monotone. Assume that $\gamma:(s_{-},s_{+})\rightarrow M$ 
is an inextendible causal curve that never intersects $\bar{M}_{v}$. 
Let $\tt:\bar{M}\rightarrow I$ be defined by $\tt[(s,h)]=s$. Let
$s_{0}\in(s_{-},s_{+})$ and assume that  
$\tt(\gamma(s_{0}))=t_{1}<v$ and that $<\gamma',\partial_{t}><0$ 
where it is defined. Thus $\tt(\gamma(s))$ increases with $s$ and 
$\tt(\gamma([s_{0},s_{+})))\subseteq [t_{1},v]$. Since we have
uniform bounds on $a_{i}$ from below and above on $[t_{1},v]$ and 
the curve is causal, we get 
\begin{equation}\label{eq:causal}
(\sum_{i=1}^{3}\xi^{i}(\gamma')^{2})^{1/2}\leq -C<\gamma',e_{0}>
\end{equation}
on that interval, with $C>0$. Since 
\begin{equation}\label{eq:length}
\int_{s_{0}}^{s_{+}}-<\gamma',e_{0}>d s=\int_{s_{0}}^{s_{+}}
\frac{d \tt\circ \gamma}{d s}d s\leq v-t_{1},
\end{equation}
the curve $\gamma|_{[s_{0},s_{+})}$, projected to $G$, will have 
finite length in the
metric $\rho$ on $G$ defined by making $e_{i}'$ an orthonormal basis.
Since $\rho$ is a left invariant metric on a Lie group, it is complete 
by Lemma \ref{lemma:complete}, and sets closed and bounded in the
corresponding topological metric must be compact. Adding the 
above observations, we conclude that $\gamma([s_{0},s_{+}))$ is 
contained in a compact set, and thus there is a sequence $s_{k}\in 
[s_{0},s_{+})$ with $s_{k}\rightarrow s_{+}$ such that $\g(s_{k})$ 
converges. Since $\tt(\gamma(s))$ is
monotone and bounded it converges. Using (\ref{eq:causal}) and
an analogue of (\ref{eq:length}), we conclude that $\g$ has to 
converge as $s\rightarrow s_{+}$. Consequently, $\g$ is extendible
contradicting our assumption. By this and similar arguments
covering the other cases, we conclude that $\bar{M}_{v}$
is a Cauchy surface for each $v\in (t_{-},t_{+})$. $\Box$

Before we turn to the questions concerning causal geodesic
completeness, let us consider the evolution of $\theta$ for solutions
to the equations of Ellis and MacCallum. This is relevant also for 
the definition of the variables of Wainwright and Hsu, since there
one divides by $\theta$. We first consider developments as in 
Lemma \ref{lemma:utveckling} which are not of type IX. 

\begin{lemma}\label{lemma:orient}
Consider class A developments which are 
not of type IX. Let the existence interval be $I=(t_{-},t_{+})$.
Then there are two possibilities.
\begin{enumerate}
\item  $\theta\neq 0$ for the entire development. We then time orient
  the manifold so that $\theta>0$. With this time orientation,
  $t_{+}=\infty$.
\item $\theta=0$, $\sigma_{ij}=0$ and $\mu=0$ for the entire
  development. Furthermore, $n_{ij}$ is constant and diagonal and
  two of the diagonal components are equal and the third is zero.
  The only Bianchi types which admit this possibility are thus
  type I and type VI$I_{0}$. Furthermore $I=(-\infty,\infty)$. 
\end{enumerate}
\end{lemma}

\textit{Proof}. 
Since $n_{ij}$ is diagonal, see the proof of Lemma
\ref{lemma:utveckling}, we can formulate the constraint
(\ref{eq:constraint1}) as
\[
\sigma_{ij}\sigma^{ij}+\frac{1}{2}[n_{1}^{2}+(n_{2}-n_{3})^{2}
-2n_{1}(n_{2}+n_{3})]+2\mu=\frac{2}{3}\theta^2,
\]
where the $n_{i}$ are the diagonal components of $n_{ij}$. Considering
Table \ref{table:bianchiA}, we see that, excepting type IX, the
expression in the $n_{i}$ is always non-negative. Thus we deduce the
inequality
\begin{equation}\label{eq:smucon}
\sigma_{ij}\sigma^{ij}+2\mu\leq\frac{2}{3}\theta^2.
\end{equation}
Combining it with (\ref{eq:raychaudhuri}), we get 
$|e_{0}(\theta)|\leq \theta^{2}$, using the fact that $2/3<\g\leq 2$. 
Consequently, if $\theta$ is zero once, it is always zero. 
Time orient the developments with $\theta\neq 0$ so that $\theta>0$.

Consider the possibility $\theta=0$. Equation (\ref{eq:raychaudhuri})
then implies $\sigma_{ij}=0$ and $\mu=0$, since $\g>2/3$. Equations
(\ref{eq:constraint1}) and (\ref{eq:dsdt}) then imply $b_{ij}=0$,
and (\ref{eq:dndt}) implies $n_{ij}$ constant. All the statements 
except the the fact that $t_{+}=\infty$ in the $\theta>0$ case follow
from the above.

Observe that $\theta$
decreases in magnitude with time, so that it is bounded to the future.
By the (\ref{eq:smucon}), the same is true of
$\sigma_{ij}$ and $\mu$. Using (\ref{eq:dndt}), we get control of 
$n_{ij}$ and conclude that the solution may not blow up in finite 
time. We must thus have $t_{+}=\infty$. $\Box$

By a theorem of Lin and Wald \cite{law}, Bianchi IX developments
recollapse.

\begin{lemma}\label{lemma:recollapse}
Consider a Bianchi IX class A development
with $1\leq \g\leq 2$ and $I=(t_{-},t_{+})$. Then there is a 
$t_{0}\in I$ such that $\theta>0$ in $(t_{-},t_{0})$ and 
$\theta<0$ in $(t_{0},t_{+})$.
\end{lemma}

\textit{Proof}. Let us begin by proving that $\theta$ can be zero
at most once. If $\theta(t_{i})=0$, $i=1,2$ and $t_{1}<t_{2}$, then
$\theta=0$ in $(t_{1},t_{2})$ since it is monotone 
by (\ref{eq:raychaudhuri}). Thus (\ref{eq:raychaudhuri}) implies
$\sigma_{ij}=0=\mu$ in $(t_{1},t_{2})$ as well. Combining this fact
with (\ref{eq:constraint1}) and (\ref{eq:dsdt}), we get $b_{ij}=0$,
which is impossible for a Bianchi IX solution.
Assume $\theta$ is never zero. By a suitable choice of time 
orientation, we can assume that $\theta>0$ on $I$. Let us prove that 
$t_{+}=\infty$. Since $\theta$ is decreasing
on $I_{1}=[0,t_{+})$ and non-negative on $I$ it is bounded on $I_{1}$.
By (\ref{eq:dndt}), $n_{1}n_{2}n_{3}$ decreases so that it is bounded
on $I_{1}$. By an argument similar to the proof of Lemma
\ref{lemma:B9N}, one can combine this bound with (\ref{eq:constraint1})
to conclude that $\sigma_{ij}$ and $\mu$ are bounded on $I_{1}$.
By (\ref{eq:dndt}), we conclude that $n_{ij}$ cannot grow faster than
exponentially. Consequently, the future existence interval must be
infinite, that is $t_{+}=\infty$, since $I$ was the maximal existence
interval and solutions cannot blow up in finite time. In order to
use the arguments of Lin and Wald, we define
\[
\beta_{i}(t)=\int_{0}^{t}\sigma_{i}(s)ds+\beta_{i}^{0},\
\a(t)=\int_{0}^{t}\frac{1}{3}\theta(s)ds+\a_{0},
\]
where $2\beta_{i}^{0}-\a_{0}=\ln (n_{i}(0))$ and $\sum_{i=1}^{3}
\beta_{i}^{0}=0$. Then
\[
n_{i}=\exp(2\beta_{i}-\a).
\]
Let $\rho=\mu/8\pi$ and $P_{i}=p/8\pi=(\g-1)\mu/8\pi$, $i=1,2,3$.
Equations (\ref{eq:constraint1}) and (\ref{eq:raychaudhuri}) then
imply equations (1.4) and (1.5) of \cite{law}, and equations 
(1.6) and (1.7) of \cite{law} follow from (\ref{eq:dsdt}). 
We have thus constructed a solution to (1.4)-(1.7) of \cite{law}
on an interval $[0,\infty)$ with $d\a/dt>0$. Lin and Wald prove
in their paper \cite{law} that this assumption leads to a 
contradiction, if one assumes that $|P_{i}|\leq \rho$ and
$P_{1}+P_{2}+P_{3}\geq 0$. However, these conditions are fulfilled
in our situation, assuming $1\leq \g\leq 2$. In other words, there
is a zero and since $\theta$ is decreasing it must be positive
before the zero and negative after it. The lemma follows. $\Box$

The lemma concerning causal geodesic completeness will build on the
following estimate. 

\begin{lemma}\label{lemma:geoest}
Consider a class A development. Let 
$\gamma:(s_{-},s_{+})\rightarrow \bar{M}$ be a 
future directed inextendible causal geodesic, and
\begin{equation}\label{eq:fdef}
f_{\nu}(s)=<\gamma'(s),e_{\nu}|_{\gamma(s)}>.
\end{equation}
If $\theta=0$ for the entire development, then $f_{0}$ is constant.
Otherwise,
\begin{equation}\label{eq:geoest}
\frac{d}{d s}(f_{0}\theta)\geq \frac{2-\sqrt{2}}{3}\theta^{2}
f_{0}^{2}.
\end{equation}
\end{lemma}

\textit{Remark}. We consider functions of $t$ as functions of $s$ by 
evaluating them at $\tt(\gamma(s))$, where $\tt$ is the function 
defined in Lemma \ref{lemma:hyperbolic}.

\textit{Proof}.
Compute, using the proof of Lemma \ref{lemma:utveckling},
\[
\frac{d f_{0}}{d s}=<\gamma'(s),\nabla_{\gamma'(s)}e_{0}>=
\sum_{k=1}^{3}\theta_{k}f_{k}^{2},
\]
where $\theta_{k}$ are the diagonal elements of $\theta_{ij}$.
If $\theta=0$ for the entire development, then $\theta_{k}=0$ for
the entire development by Lemma \ref{lemma:orient} and Lemma
\ref{lemma:recollapse}, so that $f_{0}$ is constant. Compute,
using Raychaudhuri's equation (\ref{eq:raychaudhuri}),
\[
\frac{d}{d s}(f_{0}\theta)=
\frac{1}{3}\theta^{2}\sum_{k=1}^{3}f_{k}^{2}+\sum_{k=1}^{3}\theta
\sigma_{k}f_{k}^{2}+f_{0}^{2}\sum_{k=1}^{3}\sigma_{k}^{2}+
\frac{1}{3}\theta^{2}f_{0}^{2}+\frac{1}{2}(3\g-2)\mu f_{0}^{2}
\]
where $\sigma_{k}$ are the diagonal elements of $\sigma_{ij}$.
Estimate
\[
|\sum_{k=1}^{3}\sigma_{k}f_{k}^{2}|\leq \left(\frac{2}{3}\right)^{1/2}
\left(\sum_{k=1}^{3}\sigma_{k}^{2}\right)^{1/2}\sum_{k=1}^{3}f_{k}^{2},
\]
using the tracelessness of $\sigma_{ij}$. By making a division into
the three cases $\sum_{k=1}^{3}\sigma_{k}^{2}\leq \theta^{2}/3$,
$\theta^{2}/3\leq \sum_{k=1}^{3}\sigma_{k}^{2}\leq 2\theta^{2}/3$
and $2\theta^{2}/3\leq \sum_{k=1}^{3}\sigma_{k}^{2}$, and using the
causality of $\gamma$ we deduce (\ref{eq:geoest}). $\Box$

\begin{lemma}\label{lemma:completeness}
Consider a class A development with 
existence interval $I=(t_{-},t_{+})$. There are
three possibilities.
\begin{enumerate}
\item $\theta=0$ for the entire development, in which case the
  development is causally geodesically complete. 
\item The development is not of type IX and $\theta> 0$. Then all
  inextendible causal geodesics are future complete and past
  incomplete. Furthermore, $t_{-}>-\infty$ and $t_{+}=\infty$.
\item If the development is of type IX with $1\leq \g\leq 2$, 
  then all inextendible causal geodesics are past and future 
  incomplete. We also have $t_{-}>-\infty$ and $t_{+}<\infty$.
\end{enumerate}
\end{lemma}

\textit{Proof}. 
Let $\gamma:(s_{-},s_{+})\rightarrow \bar{M}$ be a future directed 
inextendible causal geodesic and $f_{\nu}$ be defined as in
(\ref{eq:fdef}). Let furthermore $I=(t_{-},t_{+})$ be the existence 
interval mentioned in Lemma \ref{lemma:utveckling}. Since every 
$\bar{M}_{v}$, $v\in I$ is a Cauchy surface by Lemma
\ref{lemma:hyperbolic}, $\tt(\gamma(s))$ must cover the interval
$I$ as $s$ runs through $(s_{-},s_{+})$. Furthermore, $\tt(\gamma(s))$
is monotone increasing so that
\begin{equation}\label{eq:stt}
\tt(\g(s))\rightarrow t_{\pm} \ \mathrm{as}\ s\rightarrow s_{\pm}.
\end{equation}
Let $s_{0}\in (s_{-},s_{+})$ and compute
\begin{equation}\label{eq:intf}
\int_{s_{0}}^{s}-f_{0}(u)d u=\tt(\gamma(s))-\tt(\gamma(s_{0})).
\end{equation}

Consider the case $\theta=0$ for the entire development. By Lemma
\ref{lemma:geoest}, $f_{0}$ is then constant, and  $I=(-\infty,\infty)$ 
by Lemma \ref{lemma:orient}. Equations (\ref{eq:intf}) and
(\ref{eq:stt}) then prove that we must have
$(s_{-},s_{+})=(-\infty,\infty)$. Thus, all inextendible causal
geodesics must be complete.

Assume that the development is not of type IX and that $\theta>0$.
Since $f_{0}\theta$ is negative on $[s_{0},s_{+})$, its absolute
value is bounded on that interval by (\ref{eq:geoest}). If $s_{+}$ 
were finite, $\theta$ would be bounded from below by a positive
constant on $[s_{0},s_{+})$, since
\[
|\frac{d \theta}{d s}|\leq -f_{0}\theta^{2}\leq C\theta
\]
on that interval for some $C>0$, cf. (\ref{eq:smucon}) and the
observations following that equation. Since $f_{0}\theta$ is
bounded, we then deduce that $f_{0}$ is bounded on $[s_{0},s_{+})$.
But then (\ref{eq:stt}) and (\ref{eq:intf}) cannot both hold,
since $t_{+}=\infty$ by Lemma \ref{lemma:orient}.
Thus, $s_{+}=\infty$ and all inextendible causal geodesics must 
future complete. Since $f_{0}\theta$ is negative on $(s_{-},s_{+})$,
(\ref{eq:geoest}) proves that this expression must blow up in finite 
$s$-time going backward, so that $s_{-}>-\infty$. Since the curve 
$\g(s)=(s,e)$ is an inextendible timelike geodesic, we conclude that
$t_{-}>-\infty$. 

Consider the Bianchi IX case. By Lemma \ref{lemma:hyperbolic} and
\ref{lemma:recollapse}, we conclude the existence of an $s_{0}\in
(s_{-},s_{+})$ such that $f_{0}\theta$ is negative on $(s_{-},s_{0})$
and positive on $(s_{0},s_{+})$. By (\ref{eq:geoest}), $f_{0}\theta$
must blow up a finite $s$-time before $s_{0}$, and a finite $s$-time
after $s_{0}$. Every inextendible causal geodesic is thus future and
past incomplete. We conclude $t_{-}>-\infty$ and
$t_{+}<\infty$. $\Box$

\section{Appendix}

In this appendix, we consider the curvature expressions. According
to \cite{wald}, p. 40, the Weyl tensor $\bar{C}_{\a\beta\g\delta}$ 
is defined by
\[
\bar{R}_{\a\beta\g\delta}=\bar{C}_{\a\beta\g\delta}+
(\bar{g}_{\a[\g}\bar{R}_{\delta]\beta}-
\bar{g}_{\beta[\g}\bar{R}_{\delta]\a})-
\frac{1}{3}\bar{R}\bar{g}_{\a[\g}\bar{g}_{\delta]\beta},
\]
where the bar in $\bar{g}_{\a\beta}$ and so on indicates that
we are dealing with spacetime objects as opposed to objects on
a spatial hypersurface. Using this relation and the fact that
our spacetime satisfies (\ref{eq:einstein}), where $T$ is given
by (\ref{eq:tdef}) and (\ref{eq:eqofst}), one can derive the 
following expression for the Kretschmann scalar
\begin{equation}\label{eq:kexpr}
\kappa=\bar{R}_{\a\beta\g\delta}\bar{R}^{\a\beta\g\delta}
=\bar{C}_{\a\beta\g\delta}\bar{C}^{\a\beta\g\delta}+
2\bar{R}_{\a\beta}\bar{R}^{\a\beta}-\frac{1}{3}\bar{R}^2=
\end{equation}
\[
=\bar{C}_{\a\beta\g\delta}\bar{C}^{\a\beta\g\delta}+
\frac{1}{3}[4+(3\g-2)^2]\mu^2.
\]
However, according to \cite{dyn}, p. 19, we have
\begin{equation}\label{eq:weyleh}
\bar{C}_{\a\beta\g\delta}\bar{C}^{\a\beta\g\delta}=
8(E_{\a\beta}E^{\a\beta}-H_{\a\beta}H^{\a\beta}),
\end{equation}
where, relative to the frame $e_{\a}$ appearing in Lemma
\ref{lemma:utveckling}, all components of $E$ and $H$ involving
$e_{0}$ are zero, and the $ij$ components are given by
\begin{eqnarray*}
E_{ij}&=&\frac{1}{3}\theta\sigma_{ij}-(\sigma_{i}^{\ k}\sigma_{kj}-
\frac{1}{3}\sigma_{kl}\sigma^{kl}\delta_{ij})+s_{ij}\\
H_{ij}&=&-3\sigma^{k}_{\ (i}n_{j)k}+n_{kl}\sigma^{kl}\delta_{ij}+
\frac{1}{2}n^{k}_{\ k}\sigma_{ij},
\end{eqnarray*}
where $s_{ij}$ is the same expression that appears in
(\ref{eq:dsdt}), see p. 40 of \cite{dyn}. Observe that in our
situation, $E$ and $H$ are diagonal, since we are interested in
the developments obtained in Lemma \ref{lemma:utveckling}.  
It is natural to normalize $\tilde{E}_{ij}=E_{ij}/\theta^2$ and
similarly for $H$. We will denote the diagonal components of
$\tilde{E}_{ij}$ by $\tilde{E}_{i}$. We want to have expressions
in $\Sp$, $\Sm$ and so on, and therefore we compute
\begin{eqnarray*}
\tilde{H}_{1}&=&\No\Sp+\frac{1}{\sqrt{3}}(\Nt-\Nth)\Sm  \\
\tilde{H}_{2}&=&-\frac{1}{2}\Nt(\Sp+\sqrt{3}\Sm)+\frac{1}{2}(\Nth-
\No)(\Sp-\frac{1}{\sqrt{3}}\Sm) \\
\tilde{E}_{2}-\tilde{E}_{3}&=&\frac{2}{3\sqrt{3}}\Sm(1-2\Sp)+(\Nt-\Nth)(\Nt
+\Nth-\No) \\
\tilde{E}_{2}+\tilde{E}_{3}&=&\frac{2}{9}\Sp(1+\Sp)-\frac{2}{9}\Sm^2
-\frac{2}{3}\No^2+\frac{1}{3}(\Nt-\Nth)^2+\frac{1}{3}\No(\Nt+\Nth).
\end{eqnarray*}
Observe that all other components of $\tilde{E}_{i}$ and
$\tilde{H}_{i}$ can be computed from this, as $E_{ij}$ and 
$H_{ij}$ are both traceless.

It is convenient to define the normalized Kretschmann scalar
\begin{equation}\label{eq:nk}
\tilde{\kappa}=R_{\a\beta\g\delta}R^{\a\beta\g\delta}/\theta^{4}.
\end{equation}
The latter object can be expressed as a polynomial in
the variables of Wainwright and Hsu. By the above observations and
the fact that $\O=3\mu/\theta^2$, we have
\[
\tilde{\kappa}=8[\frac{3}{2}(\tilde{E}_{2}+\tilde{E}_{3})^2+
\frac{1}{2}(\tilde{E}_{2}-\tilde{E}_{3})^2-2\tilde{H}_{1}^2-
2\tilde{H}_{2}^2-2\tilde{H}_{1}\tilde{H}_{2}]+
\frac{1}{27}[4+(3\g-2)^2]\O^2.
\]

We will associate a
$\kappa$ and a $\bar{R}_{\a\beta}\bar{R}^{\a\beta}$ to a solution to 
(\ref{eq:whsu})-(\ref{eq:constraint}) in the following way. Since 
$\kappa/\theta^4$ can be expressed in terms of the variables of 
Wainwright and Hsu, it is natural to define $\kappa$ by this expression
multiplied by $\theta^4$, where $\theta$ obeys (\ref{eq:raychaudhuri2}).
There is of course an ambiguity as to the initial value of $\theta$,
but we are only interested in the asymptotics, and any non-zero
value will yield the same conclusion. We associate 
$\bar{R}_{\a\beta}\bar{R}^{\a\beta}$ to a solution similarly.

\begin{lemma}\label{lemma:rlp}
The normalized Kretschmann scalar (\ref{eq:nk}) is non-zero at the
fixed points $F,\ P_{i}^{+}(II)$, at the non-special points on the
Kasner circle, and at the type I stiff fluid points with $\O>0$.
Consequently
\begin{equation}\label{eq:ktau}
\limsup_{\tau\rightarrow -\infty}|\kappa(\tau)|=\infty
\end{equation}
for all solutions to (\ref{eq:whsu})-(\ref{eq:constraint})
which have one such point as an $\a$-limit point.
\end{lemma}

\textit{Proof}. The statement concerning the normalized Kretschmann
scalar is a computation. Equation (\ref{eq:ktau}) is a consequence
of this computation, the fact that $\kappa=\tilde{\kappa}\theta^4$
and the fact that $\theta\rightarrow \infty$ as $\tau\rightarrow 
-\infty$, cf. (\ref{eq:raychaudhuri2}). $\Box$ 

For some non-vacuum Taub type solutions with $2/3<\g<2$,
the following lemma is needed.

\begin{lemma}\label{lemma:taublp}
Consider a solution to (\ref{eq:whsu})-(\ref{eq:constraint})
with $\O>0$ and $2/3<\g<2$ such that
\begin{equation}\label{eq:splim}
\lim_{\tau\rightarrow -\infty}(\Sp,\Sm)=(-1,0).
\end{equation}
Then
\[
\lim_{\tau\rightarrow -\infty}\kappa(\tau)=\infty.
\]
\end{lemma}

\textit{Proof}. By Proposition \ref{prop:limchar}, the solution
must satisfy $\Sm=0$ and $\Nt=\Nth$. Observe that because of
(\ref{eq:splim}), we have $\O\rightarrow 0$, since $\O$ decays
exponentially for $\Sp^2$ large, cf. the proof of Lemma
\ref{lemma:dec}. Consequently, $q\rightarrow 2$.
One can then prove that for any $\epsilon>0$, there is a $T$ such 
that
\begin{eqnarray}
&\exp[(a_{\g}+\epsilon)\tau]\leq\O(\tau)\leq\exp[(a_{\g}-\epsilon)\tau]
\label{eq:asoest}\\
&\exp[(6+\epsilon)\tau]\leq \No(\tau)\leq\exp[(6-\epsilon)\tau]\\
&\exp[(6+\epsilon)\tau]\leq [\No(\Nt+\Nth)](\tau)
\leq\exp[(6-\epsilon)\tau]\label{eq:nottest}\\
&\exp[(-6+\epsilon)\tau]\leq\theta^{2}(\tau)\leq \exp[(-6-\epsilon)\tau]
\label{eq:thetaest}
\end{eqnarray}
for all $\tau\leq T$, where $a_{\g}=3(2-\g)$. However, the constraint
can be written
\[
(1-\Sp)(1+\Sp)=\O+\frac{3}{4}\No^2-\frac{3}{2}\No(\Nt+\Nth).
\]
By (\ref{eq:asoest})-(\ref{eq:nottest}), $\O$ will dominate the right
hand side, since it is non-zero. Since $1-\Sp$ converges to $2$,
$1+\Sp$ will consequently have to be positive and of the order of 
magnitude $\O$. In particular, for every $\epsilon>0$ there is a $T$ 
such that 
\begin{equation}\label{eq:aspest}
\exp[(a_{\g}+\epsilon)\tau]\leq(1+\Sp)(\tau)\leq\exp[(a_{\g}-\epsilon)\tau]
\end{equation}
Observe that since $a_{\g}<4$, $\O\theta^2$ and $(1+\Sp)\theta^2$ both
diverge to infinity as $\tau\rightarrow -\infty$, by (\ref{eq:asoest}),
(\ref{eq:thetaest}) and (\ref{eq:aspest}). Other expressions
of interest are $\No\theta^2$ and $\No(\Nt+\Nth)\theta^2$. The
estimates (\ref{eq:asoest})-(\ref{eq:thetaest}) do not yield any
conclusions concerning whether they are bounded or not. However, using
(\ref{eq:raychaudhuri2}), we have
\[
\No(\tau)\theta^2(\tau)=\No(0)\theta^2(0)\exp[\int_{\tau}^{0}
(2+q+4\Sp)ds]=
\]
\[
=\No(0)\theta^2(0)\exp[\int_{\tau}^{0}
(2(1+\Sp)+\frac{1}{2}(3\g-2)\O+2\Sp(1+\Sp))ds],
\]
which is bounded since all the terms appearing in the integral are 
integrable by (\ref{eq:asoest}) and (\ref{eq:aspest}). A similar argument
yields the same conclusion concerning $\No(\Nt+\Nth)\theta^2$.

Since the solution is of Taub type, we have $\tilde{H}_{1}=\No\Sp$ and
$\tilde{H}_{2}=\tilde{H}_{3}=-\tilde{H}_{1}/2$. We also have
$\tilde{E}_{2}=\tilde{E}_{3}$ and
\[
2\tilde{E}_{2}=\frac{2}{9}\Sp(1+\Sp)
-\frac{2}{3}\No^2+\frac{1}{3}\No(\Nt+\Nth).
\]
Consequently the $E$ field blows up and the $H$ field remains
bounded, and the lemma follows. $\Box$

Finally, we observe that 
$\bar{R}_{\a\beta}\bar{R}^{\a\beta}$ becomes unbounded
in the matter case.

\begin{lemma}\label{lemma:ricci2}
Consider a solution to (\ref{eq:whsu})-(\ref{eq:constraint})
with $\O>0$. Then
\[
\lim_{\tau\rightarrow -\infty}
\bar{R}_{\a\beta}\bar{R}^{\a\beta}=\infty.
\]
\end{lemma}
\textit{Remark}. How to associate $\bar{R}_{\a\beta}\bar{R}^{\a\beta}$
to a solution of (\ref{eq:whsu})-(\ref{eq:constraint}) is clarified in
the remarks preceding the statement of Lemma \ref{lemma:rlp}.

\textit{Proof}. We have
\[
\bar{R}_{\a\beta}\bar{R}^{\a\beta}=\mu^2+3p^2=
[1+3(\g-1)^2]\mu^2=\frac{1}{9}[1+3(\g-1)^2]\O^2\theta^4.
\]
But by (\ref{eq:whsu}) and (\ref{eq:raychaudhuri2}), we have
\[
\O^2(\tau)\theta^4(\tau)=\O^2(0)\theta^4(0)\exp(\int_{\tau}^{0}
(-4q+2(3\g-2)+4+4q)ds)=
\]
\[
=\O^2(0)\theta^4(0)\exp(-3\g\tau),
\]
and the lemma follows. $\Box$

\begin{lemma}\label{lemma:bn9corr}
Consider a class A development, not of type IX, with 
$I=(t_{-},t_{+})$ and $\theta>0$. Then the corresponding solution
to the equations of Wainwright and Hsu has existence interval
$\mathbb{R}$, and $t\rightarrow t_{\pm}$ corresponds to
$\tau\rightarrow \pm\infty$.
\end{lemma}

\textit{Proof}. 
The function $\theta$ has to converge to 
infinity as $t\rightarrow t_{-}$ for the following reason.
Assume it does not. As $\theta$ is monotone decreasing, we can 
assume it to be bounded on $(t_{-},0]$. By the constraint
(\ref{eq:constraint1}), $\sigma_{ij}$ and $\mu$ are then bounded
on $(t_{-},0]$, so that the same will be true of $n_{ij}$ by
(\ref{eq:dndt}) and the fact that $t_{-}>-\infty$. But then one can
extend the solution beyond $t_{-}$, contradicting
the fact that $I$ is the maximal existence interval. 
By (\ref{eq:raychaudhuri}), $\theta\rightarrow 0$ as
$t\rightarrow \infty=t_{+}$. Equation (\ref{eq:dtdtau}) defines
a diffeomorphism $\tilde{\tau}:(t_{-},t_{+})\rightarrow
(\tau_{-},\tau_{+})$, and we get a solution to the equations of
Wainwright and Hsu on $(\tau_{-},\tau_{+})$. By
(\ref{eq:raychaudhuri2}), we conclude that the statement of the lemma
holds. $\Box$

\begin{lemma}\label{lemma:b9corr}
Consider a Bianchi IX class A development with $I=(t_{-},t_{+})$
and $1\leq \g\leq 2$. 
According to Lemma \ref{lemma:recollapse}, there is a $t_{0}\in I$ such
that $\theta>0$ in $I_{-}=(t_{-},t_{0})$ and $\theta<0$ in
$I_{+}=(t_{0},t_{+})$. The solution to the equations of Wainwright
and Hsu corresponding to the interval $I_{-}$ has existence interval
$(-\infty,\tau_{-})$, and $t\rightarrow t_{-}$ corresponds to 
$\tau\rightarrow -\infty$. Similarly, $I_{+}$ corresponds to 
$(-\infty,\tau_{+})$ with $t\rightarrow t_{+}$ corresponding to 
$\tau\rightarrow -\infty$.
\end{lemma}

\textit{Proof}. Let us relate the different time coordinates on
$I_{-}$. According to equation (\ref{eq:dtdtau}), $\tau$ has to satisfy
$d t/d \tau=3/\theta$. Define
$\tilde{\tau}(t)=\int_{t_{1}}^{t}\theta(s)/3d s$,
where $t_{1}\in I_{-}$. Then $\tilde{\tau}:I_{-}\rightarrow 
\tilde{\tau}(I_{-})$ is a 
diffeomorphism and strictly monotone on $I_{-}$. Since $\theta$ 
is positive in $I_{-}$, $\tilde{\tau}$ increases with $t$.

Since $\theta$ is continuous beyond $t_{0}$, it is clear that 
$\tilde{\tau}(t)\rightarrow \tau_{-} \in\mathbb{R}$ as $t\rightarrow 
t_{0}$. 
To prove that $t\rightarrow t_{-}$ corresponds to $\tau\rightarrow
-\infty$, we make the following observation. One of the expressions 
$\theta$ and $d \theta/d t$ is unbounded on $(t_{-},t_{1}]$,
since if both were bounded the same would be true of $\sigma_{ij}$,
$\mu$ and $n_{ij}$ by (\ref{eq:raychaudhuri}) and (\ref{eq:dndt})
respectively. Then we would be able to extend the solution beyond
$t_{-}$, contradicting the fact that $I$ is the maximal existence 
interval (observe that $t_{-}>-\infty$ by Lemma
\ref{lemma:completeness}). If $\tilde{\tau}$ were bounded from below 
on $I_{-}$, then $\theta$ and $\theta'$ would be bounded on 
$\tilde{\tau}((t_{-},t_{1}])$ by Lemma
\ref{lemma:existence}, and thus $\theta$ and $d \theta/d t$
would be bounded on $(t_{-},t_{1}]$. Thus $t\rightarrow t_{-}$
corresponds to $\tau\rightarrow -\infty$.
Similar arguments yield the same conclusion concerning $I_{+}$. $\Box$

\section*{Acknowledgments}
This research was supported in part by the National Science Foundation
under Grant No. PHY94-07194. Part of this work was carried out
while the author was enjoying the hospitality of the Institute for
Theoretical Physics, Santa Barbara. The author also wishes to 
acknowledge the support of Royal Swedish Academy of Sciences. 
Finally, he would like to express his gratitude to Lars Andersson
and Alan Rendall, whose suggestions have improved the article.

\end{document}